\documentclass[11pt,preprint]{aastex}
\usepackage{color}
\usepackage{ulem}
\usepackage{natbib}

\usepackage{lscape}
\usepackage{rotating}
\usepackage{float}
\usepackage{graphicx}
\usepackage{epstopdf}

\slugcomment{ Ap J Suppl 187, 460 (2010)}

\shorttitle{First \textit{Fermi} LAT Pulsar Catalog}
\shortauthors{Fermi-LAT Collaboration}

\begin{document}
%\documentclass[preprint]{aastex}
%\pagestyle{empty}
%\begin{document}
\title{The First Fermi Large Area Telescope Catalog of Gamma-ray Pulsars
%\\Author list created Thursday 17 Sep 2009 01:15 PDT
}
\author{
A.~A.~Abdo\altaffilmark{2,3}, 
M.~Ackermann\altaffilmark{4}, 
M.~Ajello\altaffilmark{4}, 
W.~B.~Atwood\altaffilmark{5}, 
M.~Axelsson\altaffilmark{6,7}, 
L.~Baldini\altaffilmark{8}, 
J.~Ballet\altaffilmark{9}, 
G.~Barbiellini\altaffilmark{10,11}, 
M.~G.~Baring\altaffilmark{12}, 
D.~Bastieri\altaffilmark{13,14}
% ...
}
% ...
\altaffiltext{2}{Space Science Division, Naval Research Laboratory, Washington, DC 20375, USA}
\altaffiltext{3}{National Research Council Research Associate, National Academy of Sciences, Washington, DC 20001, USA}
\altaffiltext{4}{W. W. Hansen Experimental Physics Laboratory, Kavli Institute for Particle Astrophysics and Cosmology, Department of Physics and SLAC National Accelerator Laboratory, Stanford University, Stanford, CA 94305, USA}
\altaffiltext{5}{Santa Cruz Institute for Particle Physics, Department of Physics and Department of Astronomy and Astrophysics, University of California at Santa Cruz, Santa Cruz, CA 95064, USA}
\altaffiltext{6}{Department of Astronomy, Stockholm University, SE-106 91 Stockholm, Sweden}
\altaffiltext{7}{The Oskar Klein Centre for Cosmoparticle Physics, AlbaNova, SE-106 91 Stockholm, Sweden}
\altaffiltext{8}{Istituto Nazionale di Fisica Nucleare, Sezione di Pisa, I-56127 Pisa, Italy}
\altaffiltext{9}{Laboratoire AIM, CEA-IRFU/CNRS/Universit\'e Paris Diderot, Service d'Astrophysique, CEA Saclay, 91191 Gif sur Yvette, France}
\altaffiltext{10}{Istituto Nazionale di Fisica Nucleare, Sezione di Trieste, I-34127 Trieste, Italy}
\altaffiltext{11}{Dipartimento di Fisica, Universit\`a di Trieste, I-34127 Trieste, Italy}
\altaffiltext{12}{Rice University, Department of Physics and Astronomy, MS-108, P. O. Box 1892, Houston, TX 77251, USA}
\altaffiltext{13}{Istituto Nazionale di Fisica Nucleare, Sezione di Padova, I-35131 Padova, Italy}
\altaffiltext{14}{Dipartimento di Fisica ``G. Galilei", Universit\`a di Padova, I-35131 Padova, Italy}
% ...
%\section{}
%\end{document}

%\title{The First \textit{Fermi} Large Area Telescope Catalog of Gamma-ray Pulsars}

%\author{\textit{Fermi} LAT Collaboration and \textit{Fermi} Pulsar Timing Consortium \altaffilmark{1}}

%\altaffiltext{1}{Contact authors: G.A. Caliandro (\myemail), 
%E.C. Ferrara (Elizabeth.C.Ferrara@nasa.gov),
%D. Parent (parent@cenbg.in2p3.fr), R.W. Romani (rwr@astro.stanford.edu)}

\begin{abstract}

% \section{0.0 Abstract}
%
% PeterH. - Dave T. 31 July 2009
% ECF - 1 Aug 2009, 2 Aug 2009
% Incorporated comments from F. Camilo - 22 Aug 2009
% RWR -- 27 Aug 2009 re-ordered, incorporated the rest of Fernando's suggestions.
% ECF - 16 Sep 2009 (Corrected mis-wording noted by DJT email 9/11/09)
%
% Comments from FC 
% I have two issues with this: (1) you're conflating two different
% populations (_all_ gamma-ray pulsars, including MSPs, with just non-MSP
% radio pulsars), which makes this a misleading statement; (2) in any case,
% what's the point?  This statement might be useful only in the context of
% a model (are you perhaps alluding, or thinking in the back of your mind,
% that radio beams are supposed to be narrow, collimated, from relatively
% low in the magnetosphere, while gamma-ray beams are broader?) - but
% none of that is clear here, and the abstract is not the place for it.
% To be clear: I don't object to stating how many gamma-ray pulsars have
% two peaks; I do object to bringing the radio pulsars in at this point.
%
% -- I agree with Fernando here. In any case the statement as it was 
% in the abstract was wrong; many PSR have multiple radio peaks -- the
% phase spread is, however, typically much smaller. Without a sharper model
% comparison, this statement wasn't really helpful, anyway, so I amended it -- Roger.
% _______________
% \begin{document}

The dramatic increase in the number of known gamma-ray pulsars since the
launch of the \textit{Fermi Gamma-ray Space Telescope} (formerly
GLAST) offers the first opportunity to study a sizable population of these
high-energy objects.  This catalog summarizes 46 high-confidence pulsed detections
using the first six months of data taken by the Large Area Telescope
(LAT), \textit{Fermi}'s main instrument.  Sixteen previously unknown
pulsars were discovered by searching for pulsed signals at the
positions of bright gamma-ray sources seen with the LAT, or at the
positions of objects suspected to be neutron stars based on
observations at other wavelengths. The dimmest observed flux among these
gamma-ray-selected pulsars is $6.0 \times 10^{-8}$ ph cm$^{-2}$\,s$^{-1}$
(for $E>$100 MeV). Pulsed gamma-ray emission was discovered from 
twenty-four known pulsars by using
ephemerides (timing solutions) derived from monitoring radio
pulsars. Eight of these new gamma-ray pulsars are millisecond pulsars.  The dimmest observed
flux among the radio-selected pulsars is $1.4 \times 10^{-8}$ ph cm$^{-2}$\,s$^{-1}$
(for $E>$100 MeV).  The remaining six gamma-ray pulsars
were known since the \textit{Compton Gamma Ray Observatory} mission, or before.
The limiting flux for pulse detection is non-uniform over the sky owing to different
background levels, especially near the Galactic plane.
The pulsed energy spectra can be described by a power law with an
exponential cutoff, with cutoff energies in the range $\sim 1-5$ GeV.
The rotational energy loss rate ($\dot{E}$) of these neutron stars spans 5 decades,
from $\sim$$3 \times 10^{33}$ erg s$^{-1}$ to 5 $\times$ 10$^{38}$ erg s$^{-1}$,
and the apparent efficiencies for conversion to gamma-ray emission range from
$\sim 0.1$\% to $\sim$ unity, although distance uncertainties complicate efficiency
estimates. The pulse shapes show substantial diversity, but
roughly 75\% of the gamma-ray pulse profiles have two peaks,
separated by $\ga 0.2$ of rotational phase.  For most of the
pulsars, gamma-ray emission appears to come mainly from the outer
magnetosphere, while polar-cap emission remains plausible for a
remaining few.  Spatial associations imply that many of these pulsars power
pulsar wind nebulae. Finally, these discoveries suggest that gamma-ray-selected
young pulsars are born at a rate comparable to that of their radio-selected
cousins and that the birthrate of all young gamma-ray-detected pulsars is
a substantial fraction of the expected Galactic supernova rate.

% \end{document}

\end{abstract}

\keywords{catalogs -- gamma rays: observations -- pulsars: general -- stars: neutron}

{\em Version corrected for an Erratum sent to the Ap J, December 2010:}
In the original paper, an error was made in accounting for the delay due to interstellar dispersion in the
radio phasing of PSR J1124$-$5916.  This changes the measured gamma-ray to radio lag ($\delta$) to 0.11 $\pm$
0.01.  An error was also made in the off-pulse phase range in Table 3 for that pulsar.  This error did not
affect the spectral results. Corrected versions of Table 3 (with the revised numbers in bold face), Figure 4,
and Figure A20 are included here. In addition there was an error in the caption to Figure 9.  The figure with corrected caption is included here,
with the changed word in bold face.

\section{Introduction}
% \section{1.0 Intoduction}
%
% Andrea and  Dave T. -  22 July 2009
% PeterH. - 29 July 2009 
% ECF - 1 Aug 2009
%
% Incorporated comments from F. Camilo - 22 Aug 2009
% Comments from FC not yet addressed are:
% _______________
% l40: radio-quiet pulsars - should you perhaps define what this means?
% das elects not to, 27 august 2009.
% \begin{document}

Following the 1967 discovery of pulsars by Bell and
Hewish \citep{Hewish68}, \citet{Gold68} and \citet{Pacini68}
identified these objects as rapidly rotating neutron stars whose
observable emission is powered by the slow-down of the rotation. With
their strong electric, magnetic, and gravitational fields, pulsars
offer an opportunity to study physics under extreme conditions.  As
endpoints of stellar evolution, these neutron stars, together with
their associated supernova remnants (SNRs) and pulsar wind nebulae (PWNe), help
probe the life cycles of stars.

Over 1800 rotation-powered pulsars are now listed in the ATNF pulsar
catalog \citep{ATNFcatalog}\footnote{http://www.atnf.csiro.au/research/pulsar/psrcat},
as illustrated in Figure \ref{SkyMap}.
The vast majority of these pulsars were discovered by radio
telescopes.  Small numbers of pulsars have also been seen in the optical band,
with more in the X-ray bands \citep[see e.g.][]{NeutronStarReview}.

In the high-energy gamma-ray domain ($\geq30$ MeV) the first
indications for pulsar emission were obtained for the Crab pulsar by
balloon-borne detectors \citep[e.g.][]{Browning71}, and confirmed by the
SAS-2 satellite \citep{Kniffen74}, which also found gamma radiation
from the Vela pulsar \citep{Thompson75}.  The \textit{COS-B} satellite provided
additional details about these two gamma-ray pulsars, including a
confirmation that the Vela pulsar gamma-ray emission was not in phase
with the radio nor did it have the same emission pattern (light curve)
as seen in the radio \citep[see e.g.][]{Kanbach80}.

The \textit{Compton Gamma Ray Observatory} (\textit{CGRO}) expanded the number of
gamma-ray pulsars to at least seven, with six clearly seen by the
\textit{CGRO} high-energy instrument, EGRET. This gamma-ray pulsar
population allowed a search for trends, such as the increase in
efficiency $\eta = L_\gamma/\dot E$ with decreasing
values of the open field line voltage of the pulsar, first noted
by \citet{Arons96}, for gamma-ray luminosity  $L_\gamma$
and spin-down luminosity $\dot E$. A summary of gamma-ray pulsar results in the 
\textit{CGRO} era is given by \citet{Thompson04}.

The third EGRET catalog \citep[3EG;][]{3rdCat} included 271 sources of
which $\sim$170 remained unidentified. Determining the nature of these
unidentified sources is one of the outstanding problems in high-energy
astrophysics. Many of them are at high Galactic latitude and are most
likely active galactic nuclei or blazars. However, most of the sources at low Galactic
latitudes ($|b|\leq 5 \degr$) are associated with star-forming
regions and hence may be pulsars, PWNe, SNRs, 
winds from massive stars, or high-mass X-ray binaries
\citep[e.g.][]{Kaaret96,Yadigaroglu1997,Romero1999}. 
A number of radio pulsars were subsequently discovered in EGRET error
boxes \citep[e.g.][]{Kramer03}, but gamma-ray pulsations
in the archival EGRET data were never clearly seen.
Solving the puzzle of the
unidentified sources will constrain pulsar emission
models: pulsar population synthesis studies, such as those
by \citet{Cheng98}, \citet{Gonthier02}, and \citet{McLaughlin00},
indicate that the number of detectable pulsars in either EGRET or 
\textit{Fermi} data, as well as the expected ratio of radio-loud and
radio-quiet pulsars \citep{Harding07}, strongly depends on the assumed
emission model.

The Large Area Telescope (LAT) on the \textit{Fermi Gamma-ray Space
Telescope} has provided a major increase in the known gamma-ray pulsar
population, including pulsars discovered first in gamma-rays \citep{BSP} 
and millisecond pulsars (MSPs) \citep{MSP}. The first aim
of this paper is to summarize the properties of the gamma-ray pulsars
detected by \textit{Fermi}-LAT during its first six months of data
taking. The second primary goal is to use this gamma-ray pulsar
catalog to address astrophysical questions such as:

\begin{enumerate}
  \item Are all the gamma-ray pulsars consistent with one type of
  emission model?  
  \item How do the gamma-ray pulsars compare to the
  radio pulsars in terms of physical properties such as age, magnetic
  field, spin-down luminosity, and other parameters?  
  \item Are the trends suggested by the \textit{CGRO} pulsars 
  confirmed by the LAT gamma-ray pulsars?  
  \item Which of the LAT pulsars are associated with SNRs, 
  PWNe, unidentified EGRET sources, or TeV sources?  
%  \item Were any expected pulsars not seen by LAT, and if so what do 
%  those results imply?
\end{enumerate}
  
The structure of this paper is as follows: Section 2 describes the LAT
and the pulsar data analysis procedures; Section 3 presents the
catalog and derives some population statistics from our sample; Section 4 studies
the LAT sensitivity for gamma-ray pulsar detection, while in Section 5
the implications of our results are briefly discussed. Finally, our conclusions are
summarized in Section 6.
  
% \end{document}

\section{Observations and Analysis}
% \section{2.0 Observations and Analysis}
%
% Andrea \& David - 24 July 2009
% PeterH. - 29 July 2009 
% Incorporates additional PdH updates from 29 July 2009
% ECF - 1 Aug 2009, 2 Aug 2009
% ECF - 16 Sep 2009 (Changed final paragraph per PSR email 9/14/09)
% 
% Incorporated comments from F. Camilo - 22 Aug 2009
% Comments from FC not yet addressed are:
% 
% l78: within theta_68, where -> within theta_68 of the zenith, where ECF - is this correct??)
%  NO IT IS NOT. I put "of a point source" -- das.

% 
% \begin{document}

The \textit{Fermi Gamma-ray Space Telescope} was successfully launched 
on 2008 June 11, carrying two gamma-ray instruments: the LAT and the 
Gamma-ray Burst Monitor (GBM). The LAT, 
\textit{Fermi}'s main instrument, is described in detail in \citet{LATinstrument}, 
with early on-orbit performance reported in \citet{OnOrbit}.  It is a 
pair-production telescope composed of a $4\times4$ grid of towers. Each 
tower consists of a silicon-strip detector and a tungsten-foil tracker/converter, 
mated with a hodoscopic cesium-iodide calorimeter. This grid of towers is 
covered by a segmented plastic scintillator anti-coincidence detector. The 
LAT is sensitive to gamma  rays with energies in the range from 20 MeV to 
greater than 300 GeV, and its on-axis effective area is $\sim8000$ cm$^{2}$ 
for $E>1$ GeV. The gamma-ray point spread function (PSF) is energy dependent, and 68\% of 
photons have reconstructed directions within $\theta_{68} \simeq 0\fdg8 E^{-0.8}$
of a point source, with $E$ in GeV, leveling off to $\theta_{68} \lesssim 0\fdg1$ for $E>10$ GeV .
Effective area, PSF, and energy resolution are tabulated into bins of photon energy and
angle of incidence relative to the LAT axis. The tables are called ``instrument response functions,Ó 
and are described in detail in \citet{OnOrbit}. This work uses the version called {\tt P6\_v3\_diffuse}.

Gamma-ray events recorded with the LAT have time stamps that are derived 
from a GPS-synchronized clock on board the \textit{Fermi} satellite. The 
accuracy of the time stamps relative to UTC is $<1$ $\mu$s \citep{OnOrbit}. The 
timing chain from the GPS-based satellite clock through the barycentering and 
epoch folding software has been shown to be accurate to better than a few 
$\mu$s for binary orbits, and significantly better for isolated pulsars \citep{DAS08}.

The LAT field-of-view is about 2.4 sr. Nearly the entire first year in orbit has been 
dedicated to an all-sky survey, imaging the entire sky every two orbits, i.e. every 
3 hours. Any given point on the sky is observed roughly $1/6^{\rm th}$ of the 
time.
The LAT's large effective area and excellent source localization coupled with improved 
cosmic-ray rejection led to the detection 
of 46 gamma-ray pulsars in the first six months of LAT observations. These include 
the six gamma-ray pulsars clearly seen with EGRET \citep{Thompson04}, 
two young pulsars seen marginally with EGRET \citep{EGRETB0656p14,EGRETB1046m58},
the MSP seen marginally with EGRET  \citep{Kuiper2000}, PSR J2021+3651 discovered in 
gamma-rays by \textit{AGILE} \citep{Halpern2008}, and some of the other pulsars also studied
by \textit{AGILE} \citep{AgilePulsars}.

During the LAT commissioning period, several configuration settings  
were tested that affected the LAT energy resolution and  
reconstruction. However, these changes had no effect on the LAT  
timing.  Therefore, for the spectral analyses, the data were collected  
from the start of the \textit{Fermi} sky-survey observations (2008  
August 4, shortly before the end of the commissioning period) until  
2009 February 1, while the light curve and periodicity test analysis  
starts from the first events recorded by the LAT after launch (2008  
June 25) and also extends through 2009 February 1.

% \end{document}

s
	\subsection{Timing Analysis}
	% \section{2.1 Timing Analysis}
%
% David S. \& Andrea - 29 July 2009
% Incorporates updates by PdH from 29 July 2009
% Cleaned by ECF - 1 Aug 2009, 2 Aug 2009
% Camilo, Manchester, Caraveo comments by David S, 27 August 2009.
%
% Incorporated comments from F. Camilo - 22 Aug 2009
% Comments from FC not yet addressed are:
% 
% % - l108:   Do the errors in DM for the ephemerides
%	      used here result% in ~0.1P delta uncertainties?  No.  ~0.01P?  
%	      Yes, quite often.% 
%% \begin{document}

We have conducted two distinct pulsation searches of \textit{Fermi} LAT data. 
One search uses the ephemerides of known pulsars, obtained from radio and X-ray observations. 
The other method searches for periodicity in the arrival times of gamma-rays coming from the direction
of neutron star candidates (``blind period searches''). Both search strategies have advantages. 
The former is sensitive to lower gamma-ray fluxes, and the comparison of phase-aligned pulse profiles 
at different wavelengths is a powerful diagnostic of beam geometry.  The blind period search allows for 
the discovery of new pulsars with selection biases different from those of radio searches, such as, for 
example, favoring pulsars with a broader range of inclinations between the rotation and magnetic axes. 
%The number 
%of observed radio-quiet (Geminga-like) pulsars will constrain beaming models and population studies.

For each gamma-ray event (index $i$), the topocentric gamma-ray arrival time recorded by the LAT is
transferred to times at the solar-system barycenter $t_i$ by correcting for the position of \textit{Fermi} 
in the solar-system frame. The rotation phase $\phi_i(t_i)$ of the neutron star is calculated from a timing 
model, such as a truncated Taylor series expansion,
\begin{equation}
\phi_i(t_i) = \phi _0 + \sum _{j = 0}^{j = N} { {f_{j} \times (t_i - T_0)^{j+1}}\over {(j+1)!}}.
\end{equation}
Here, $T_0$ is the reference epoch of the pulsar ephemeris and $\phi _0$ is the pulsar phase at $t = T_0$.
The coefficients $f_j$ are the rotation frequency derivatives of order $j$. The rotation period is $P = 1/f_0$. 
Different timing models, described in detail in \citet{Edwards06}, can take into account various
physical effects. Most germane to the present work is accurate $\phi_i(t_i)$ computations, even in
the presence of the rotational instabilities of the neutron star called ``timing noise''.
``Phase-folding'' a light curve, or pulse profile, means filling
a histogram with the fractional part of the $\phi_i$ values. An ephemeris includes the pulsar coordinates necessary for
barycentering, the $f_{j}$ and $T_0$ values, 
and may include parameters describing the pulsar proper motion, glitch epochs, and more. 
The radio dispersion measure (DM) is used to extrapolate 
the radio pulse arrival time to infinite frequency, and uncertainties in the DM translate to an uncertainty 
in the phase offset between the radio and gamma-ray peaks. 

% \end{document}

		\subsubsection{Pulsars with Known Rotation Ephemerides}
		% \section{2.1.1 Pulsars with Known Rotation Ephemerides}
%
% David S. - 29 July 2009
% Incorporates updates from PdH - 29 July 2009
% ECF - 1 Aug 2009, 2 Aug 2009
% PdH 5 Aug
% ECF - 16 Sep 2009 (Updates to address Fernando2)
%
% Incorporated comments from F. Camilo - 22 Aug 2009
% Comments from FC not yet addressed are:
%
% \begin{document}

The ATNF database \citep[version 1.36,][]{ATNFcatalog} lists 1826 pulsars, and more have
been discovered and await publication (Figure \ref{SkyMap}).  The LAT observes them
continuously during the all-sky survey.  Phase-folding the gamma rays
coming from the positions of all of these pulsars (consistent with the
energy-dependent LAT PSF) requires only modest computational
resources. 
However, the best candidates for gamma-ray emission are the pulsars with
high $\dot E$, which often have substantial timing noise.
Ephemerides accurate enough to allow phase-folding into, at a minimum,
25-bin phase histograms can degrade within days to months.
The challenge is to have \textit{contemporaneous} ephemerides. 

We have
obtained 762 contemporaneous pulsar ephemerides from radio observatories, and 5 from
X-ray telescopes, in two distinct groups.
The first consists of 218 pulsars with high spin-down
power ($\dot E>10^{34}$ erg s$^{-1}$) timed regularly
as part of a campaign by a consortium of astronomers
for the \textit{Fermi} mission, as described
in \citet{DAS08}.  With one exception (PSR J1124$-$5916,
which is a faint radio source with large timing noise), all of the 218 targets
of the campaign have been monitored since shortly
before \textit{Fermi} launch. Some results from the timing campaign
can be found in \citet{ParkesFermiTiming}.

The second group is a sample of 544 pulsars from nearly the entire
$P-\dot P$ plane (Figure \ref{PPdot}) being timed for other
purposes for which ephemerides were shared with the LAT team. These
pulsars reduce possible bias of the LAT pulsar searches created by our
focus on the high $\dot E$ sample requiring frequent monitoring. Gamma-ray pulsations from six radio pulsars
with $\dot E < 10^{34}$ erg s$^{-1}$ were discovered in this manner, all of
which are MSPs.

Table \ref{tbl-char} lists properties of the 46 detected gamma-ray pulsars. 
Five of the 46 pulsars, all MSPs, are in binary systems.
The period first derivative $ \dot{P}$ is corrected for the  
kinematic Shklovskii effect \citep{Shklovskii}: 
$\dot{P}=\dot{P}_{\rm obs}-\mu^2P_{\rm obs}d/c$, where $\mu$ is the pulsar proper 
motion, and $d$ the distance. The correction is small except for a few MSPs \citep{MSP}. 
The characteristic age is $ \tau_{\rm c} = P/2\dot{P}$ and the
spin-down luminosity is
\begin{equation}
 \dot{E} =  4\pi^2I\dot{P}P^{-3},
\end{equation}
taking the neutron star's moment of inertia $I$ to be $10^{45}$g cm$^2$.
The magnetic field at the light cylinder (radius $R_{\rm LC} = cP/2\pi$) is
\begin{equation}
 B_{\rm LC} = \left( \frac{3I8\pi^4\dot{P}}{c^3P^5} \right)^{1/2} \approx 2.94 
 \times 10^{8}(\dot{P}P^{-5})^{1/2}\, {\rm G}.
\end{equation}

Table \ref{tbl-det} lists which observatories provided ephemerides
for the gamma-ray pulsars. An ``L'' indicates that the pulsar was
timed using LAT gamma rays, as described in the next Section.

``P'' is the Parkes radio telescope \citep{Parkes,ParkesFermiTiming}.
The majority of the Parkes observations were carried out at intervals of 4 --
6 weeks using the 20 cm Multibeam receiver \citep{ParkesReceiver}
with a 256 MHz band centered at 1369 MHz. At 6 month intervals
observations were also made at frequencies near 0.7 and 3.1~GHz with
bandwidths of 32 and 1024 MHz respectively.
The required frequency resolution to avoid dispersive smearing across the
band was provided by a digital spectrometer system.

``N'' is the Nan\c cay radio telescope \citep{Nancay}. 
Nan\c cay observations are carried out every three weeks on average. The 
recent version of the BON backend is a GPU-based coherent dedispersor 
allowing the processing of a 128 MHz bandwidth over two complex 
polarizations \citep{Desvignes}. The majority of the data are 
collected at 1398 MHz, while for MSPs in particular, observations are 
duplicated at 2048 MHz to allow DM monitoring.

``J'' is the 76-m Lovell radio telescope at Jodrell Bank in the United
Kingdom. Jodrell Bank observations \citep{Jodrell} were carried
out at typical intervals of between 2 days and 10 days in a 64-MHz band
centered on 1404 MHz, using an analog filterbank to provide the frequency
resolution required to remove interstellar dispersive broadening.
Occasionally, observations are also carried out in a band centered on 610
MHz to monitor the interstellar dispersion delay.

``G'' is the 100-m NRAO Green Bank Telescope (GBT).  PSR~J1833--1034
was observed monthly at 0.8\,GHz with a bandwidth of 48\,MHz using the
BCPM filter bank \citep{GBT_FilterBank}.  The other pulsars monitored at
the GBT were observed every two weeks at a frequency of 2\,GHz across
a 600\,MHz band with the Spigot spectrometer \citep{GBT}.
Individual integration times ranged between 5 minutes and 1 hour.

Arecibo ("A") observations of the very faint PSRs J1930+1852 and
J2021+3651 are carried out every two weeks, with the L-wide receiver 
(1100 to 1730 MHz). The back-ends used are the Wideband Arecibo 
Pulsar Processor (WAPP) correlators (Dowd et al. 2000), each with
a 100 MHz-wide band. The antenna voltages are 3-level digitized and 
then auto-correlated with a total of 512 lags, accumulated every 128 $\mu$s, 
and written to disk as 16-bit sums. Processing includes
Fourier-transforms to obtain power spectra, which are then dedispersed
and phase-folded. The average pulse profiles are cross-correlated
with a low-noise template profile to obtain topocentric times of arrival.

% Arecibo (``A'') timing observations of two pulsars (PSRs
% J1930+1852 and J2021+3651, among the faintest within the Arecibo
% declination range) are carried out every two weeks, with the L-wide
% receiver (which is sensitive to radio waves from 1100 to 1730 MHz).
% The back-ends used are the Wideband Arecibo Pulsar Processor (WAPP)
% correlators \citep{Arecibo}, each capable of processing a
% 100 MHz-wide band. These are normally centered at 1170, 1410, 1510 and
% 1610 MHz to minimize the adverse effects of radio frequency
% interference.
%    For the purposes of this program, the antenna voltages are 3-level
% digitized and then auto-correlated with a total of 512 lags. The
% results are accumulated every 128 $\mu$s and written to disk as 16-bit
% sums. During processing these are Fourier-transformed to obtain power
% spectra, which are then dedispersed at the pulsar's DM and folded at
% the pulsar's spin period. The resultant average pulse profiles are
% cross-correlated with a low-noise template profile to obtain
% topocentric times of arrival.

``W'' is the Westerbork Synthesis Radio
Telescope, with which observations were made 
approximately monthly at central frequencies of 328, 382 and 1380 MHz with
bandwidths of 10 MHz at
the lower frequencies and 80 MHz at the higher frequencies. The PuMa
pulsar backend \citep{Westerbrok} was used to record all the observations. 
Folding and dedispersion were performed offline.  

The rms of the radio timing residuals for most of the solutions used in
this paper is $<0.5$\% of a rotation period, but ranges as high as
$1.2$\% for five pulsars. This is adequate for the 50- or 25-bin phase 
histograms used in this paper. The ephemerides used for this catalog will
be available on the \textit{Fermi} Science Support Center data
servers\footnote{http://fermi.gsfc.nasa.gov/ssc/}.

% \end{document}

		\subsubsection{Pulsars Discovered in Blind Periodicity Searches}
		% \section{2.1.2 Pulsars Discovered in Blind Periodicity Searches}
%
% P.S.R - 30 June 2009
% Dave S - 29 July 2009
% ECF - 1 Aug 2009
% P.S.R - 24 Aug 2009
% ECF - 15 Sep 2009 (updated per PSR email 9/14/09)
% ECF - 16 Sep 2009 (Added dataset info per PSR email 9/14/09)
% ECF - 16 Sep 2009 (Updates to address Fernando2)
%
% Incorporated comments from F. Camilo - 22 Aug 2009
% Comments from FC not yet addressed are:
%
% - l206: "In several cases" - how many? [can they be indicated in a table?]
% 
% \begin{document}

For all 16 of the pulsars found in the blind searches of the LAT data, we determined the 
timing ephemerides used in this catalog directly from the LAT data as described below.  
In addition, for two other pulsars the LAT data provided the best available timing model. 
The first is the radio-quiet pulsar Geminga. Since
Geminga is such a bright gamma-ray pulsar, it is best timed directly using
gamma-ray observations.  During the period between EGRET and \textit{Fermi}, occasional 
\textit{XMM-Newton} observations maintained the timing model \citep{jh05} but
a substantially improved ephemeris has now been derived from the LAT data
\citep{Geminga1}.  The second is PSR J1124$-$5916, which is extremely faint in the
radio (see Table \ref{tbl-char}) and exhibits a large amount of 
timing noise \citep{Camilo2002b}. In this Section, we briefly describe the blind pulsar
searches and how the timing models for these pulsars are created.
These pulsars have an ``L'' in the ``ObsID'' column of Table \ref{tbl-det}. 

Even though $L_\gamma$ of a young pulsar can be several percent of $\dot{E}$, the 
gamma-ray counting rates are low. As an 
example, the LAT detects a gamma ray from the Crab pulsar approximately every 500 
rotations, when the Crab is well within the LAT's field-of-view.
Such sparse photon arrivals make periodicity searches difficult. Extensive searches 
for pulsations performed on EGRET data \citep{Chandler2001,Ziegler2008} were just sensitive 
enough to detect the very bright pulsars Vela, Crab, and Geminga in a blind search, had they not already been known as pulsars. Blind periodicity searches of all other EGRET sources proved fruitless.

By contrast, the improvements afforded by the LAT have enabled highly successful blind searches for pulsars. In the first six months of operation, we discovered a total of 16 new pulsars in direct pulsation searches of the LAT data \citep[see e.g.][]{CTA1,BSP}.
A computationally efficient time-difference search technique made these searches possible 
\citep{Atwood2006}, enabling searches of hundreds of \textit{Fermi} sources to be performed 
on a small computer cluster with only a modest loss in sensitivity compared 
to fully coherent search techniques.  Still, owing to the large number of frequency and 
frequency derivative trials required to search a broad parameter space, the minimum 
gamma-ray flux needed for a statistically significant detection is considerably higher than the 
minimum flux needed for the phase-folding technique using a known ephemeris (as in Section 2.1, Eq. 1).

We performed these blind searches on $\sim100$ candidate sources 
identified before launch and on another $\sim200$ newly detected LAT sources. 
The parameter space covered by the blind searches included frequencies 
from 0.5 Hz to 64 Hz (periods of 156.25 ms to 2 s), and a frequency 
derivative from zero to the spin-down of the young Crab pulsar 
($f_1 = -3.7\times10^{-10}$), which covers $\sim86$\% of the pulsars 
contained in the ATNF database \citep{BSP}.
Of the 16 pulsars detected in these searches, 13 are associated 
with previously known EGRET sources.  The discoveries include several long-suspected 
pulsars in SNRs and PWNe.

These 16 pulsars are gamma-ray selected, as they were discovered by the 
LAT and thus the population is subject to very different selection effects than the general 
radio pulsar population. However, this does not necessarily imply that they are radio 
quiet. For several cases, deep radio searches have already been 
performed on known PWNe or X-ray point sources suspected of harboring pulsars.  
In most cases, new radios searches are required to investigate whether there is a radio 
pulsar counterpart down to a meaningful luminosity limit.  These searches are 
now being undertaken and are yielding the first results \citep{Camilo2009}.  

For these 18 pulsars (16 new plus Geminga and PSR J1124$-$5916), we derived timing 
models from the LAT data using the procedure 
summarized here. A more detailed description of pulsar timing using LAT 
data can be found in \citet{BlindTiming}.  

The LAT timing analysis starts from the first events recorded by the 
LAT after launch (2008 June 25) and extends through about 2009 May 1. 
During the commissioning period, several configuration settings were 
tested that affected the LAT energy resolution and reconstruction. 
However, these changes had no effect on the LAT timing.  We selected photons from a 
small radius of interest (ROI) around the pulsar of $<0.5\degr$ or $<1\degr$ (see further 
Section 2.1.3 and 
Table~\ref{tbl-det}). For this pulsar timing analysis, we used \textit{diffuse}  class photons 
with energies above a cutoff (typically $E>300$ MeV) selected to optimize the 
signal-to-noise ratio for that particular pulsar. We converted the photon arrival times to 
the geocenter using the \textsc{gtbary} science tool\footnotemark[2]. %% Warning!  This footnote number won't get changed automatically!! It must be updated if footnotes are added or deleted before this point!
This correction removes the effects of the spacecraft motion about the Earth, resulting in 
times as would be observed by a virtual observatory at the geocenter.

%The LAT timing analysis starts from the Þrst events recorded by the Ê
%LAT after launch (2008 June 25) and extends through about 2009 May 1. Ê
%During the commissioning period, several conÞguration settings were Ê
%tested that affected the LAT energy resolution and reconstruction. 
%However, these changes had no effect on the LAT timing. 
Using an initial 
timing model for the pulsar, we then used \textsc{Tempo2} \citep{Hobbs2006} 
in its predictive mode to generate polynomial coefficients describing the pulse phase as a 
function of time for an observatory at the geocenter.  Using these predicted phases, we 
produced folded pulse profiles over segments of the LAT observation.  The length of the 
segments depends on the brightness of the pulsar but are typically 10--20 days. We then 
produced a pulse time of arrival (TOA) for each data segment by Fourier domain 
cross-correlation with a template profile \citep{tay92}. The template profile for most of the 
pulsars is based on a multi-Gaussian fit to the observed LAT pulse profile.  However, in 
the case of Geminga, which has very high signal-to-noise and a complex profile not well 
described by a small number of Gaussians, we used a template profile that was the full 
mission light curve itself.

Finally, we used \textsc{Tempo2} to fit a timing model to each pulsar. For
most of the pulsars, the model includes pulsar celestial coordinates,
frequency and frequency derivative. In several cases, the fit also required
a frequency second derivative term to account for timing noise. In the case
of PSR J1124$-$5916, we required three sinusoidal terms \citep{Hobbs2006}
to model the effects of the strong timing noise in this source.  Two pulsars
(J1741$-$2054 and J1809$-$2332) have positions too close to the ecliptic
plane for the Declination to be well-constrained by pulsar timing and thus
we fixed the positions based on X-ray observations of the presumed
counterparts\footnote{Pulsar timing positions are measured by fitting the sinusoidal delays of
the pulse arrival times associated with the Earth moving along its orbit. For
pulsars very close to the ecliptic plane the derivative of this delay with respect to ecliptic latitude is greatly reduced and thus such pulsars have one spatial dimension poorly constrained, as
discussed in \citet{BlindTiming}.}.
For Geminga and PSR J1124$-$5916 we also used known, fixed positions 
\citep{Caraveo1998,Faherty2007,Camilo2002b} because
they were of much higher precision than could be determined from less than
one year of \textit{Fermi} timing. The rms of the timing residuals are
between $0.5$ and $2.9$\% of a
rotation period, with the highest being for PSR J1459$-$60. The \textsc{Tempo2} 
timing models used for the catalog analysis will be made available online at 
the FSSC web site\footnotemark[2].
% \end{document}

		\subsubsection{Light Curves}
		% \section{2.1.3 Light Curves}
%
% Andrea - 23 June 2009
% Dave S - 29 July 2009
% ECF - 1 Aug 2009, 2 Aug 2009
% ECF - 16 Sep 2009 (Updates to address Fernando2)
%
% Incorporated comments from F. Camilo - 22 Aug 2009
% Comments from FC not yet addressed are:
% 
% - l221: "Phase-aligned radio profiles for the radio selected pulsars
% are in the bottom panel".  I notice that for two pulsar that are gamma-ray
% selected you also include the radio profiles (cf. Camilo et al. 2009b),
% so please modify this.
% _______________
% - l222: "requiring a) at least 50 counts per bin in the peak" - this
% is not strictly true; see, e.g., J2124-3358.
% 
% \begin{document}

The light curves of 46 gamma-ray pulsars detected by the LAT are appended to the end of this paper, in Figures 
\ref{fig:J0007p7303_lightcurve} to \ref{fig:J2238p59_lightcurve}. The gray light curve 
in the top panel includes all photons with $E>0.1$ GeV, while the other panels show 
the profiles in exclusive energy ranges: $E>1.0$ GeV (with $E>3.0$ GeV in black) in 
the second panel from the top; $0.3$ to $1.0$ GeV in the next panel; and $0.1$ to 
$0.3$ GeV in the fourth panel. Phase-aligned radio profiles for the radio-selected 
pulsars are in the bottom panel. The light curves are plotted with $N = 25$ or 50 bins, with 25 bins used when required to keep at least 50 counts per bin in the peak of the light
curve or to prevent undue smearing due to the accuracy of the timing model.

Table~\ref{tbl-pulse} lists light curve parameters, taken from the $>100$ MeV profiles 
(top panel of Figures A-1--A-46). For some pulsars (e.g. PSR J1420$-$6048) the peak multiplicity is unclear:
the data are consistent with both a single, broad peak and with two closely spaced narrow peaks.
For the Crab (PSR J0534+2200) the phase offset of the first gamma-ray peak is relative to radio ``precursor'' peak,
not visible in Figure A-8.

Table \ref{tbl-det} lists the $Z^2_2$ \citep{Buccheri1983} and $H$ \citep{DeJager1989} 
periodicity test values for $E>0.3$ GeV. Only one trial is made for each pulsar, and the 
significance calculations do not take into account the trials factor for the $\sim 800$ 
pulsars searched. Detection of gamma-ray 
pulsations are claimed when the significance of the periodicity test exceeds 5$\sigma$ 
(i.e.~a chance probability of $<6\times10^{-7}$). We have used the $Z$-test with $m=2$ 
harmonics ($Z^2_2$) which provides an analytical distribution function for the null 
hypothesis described by a $\chi^2$ distribution with $2m$ degrees of freedom. 
The $H$-test uses Monte-Carlo simulations to calculate probabilities, limited to a minimum 
of $4\times 10^{-8}$ (equivalent to 5.37$\sigma$). Each method is sensitive to different pulse profile shapes. Four 
pulsars in the catalog fall short of the 5$\sigma$ significance threshold in the six-month 
data set with the selection cuts applied here: the 3 MSPs J0218+4232,
J0751+1807, and J1744$-$1134 reported in \citet{MSP}, and the radio pulsar 
PSR J2043+2740. The characteristic pulse shape as well as the trend of the significance 
versus time lead us to include these four in the catalog.

Table \ref{tbl-det} also lists ``maxROI'', the maximum angular radius around the pulsar 
position within which gamma-ray events were kept, generally 
$1\fdg0$, but $0\fdg5$ in some cases. The choice was made by using the energy 
spectrum for the phase-averaged source, described in Section 2.2, to maximize $S^2/N$ 
over a grid of maximum radii and minimum energy thresholds (where $S$ is the number 
of counts attributed to the point source, and $N$ is the number of counts due to the diffuse
background and neighboring sources). We selected photons within a radius
 $\theta_{68}$ of the pulsar position, requiring a radius of at least $0\fdg35$, but no 
 larger than the reported maxROI.

%The background level drawn in the gray light curves of Figures \ref{fig:J0007p7303_lightcurve} 
% to \ref{fig:J2238p59_lightcurve} (top panels) was computed from the 
%diffuse emission model fitted by the likelihood spectral analysis described in Section 2.2. 
%
We estimated the background level represented by the dashed horizontal lines
in Figures \ref{fig:J0007p7303_lightcurve} to \ref{fig:J2238p59_lightcurve}
from an annulus between 1\degr $< \theta < $2\degr~surrounding the source. 
Nearby sources were removed, and we normalized to the same solid angle 
as the source ROI. 
The poor spatial resolution of the LAT at low energies can blur
structured diffuse emission and bias this background estimate.
The levels shown are intended only to guide the eye. Detailed analyses
of off-pulse emission will be discussed in future work.

% \end{document}

	\subsection{Spectral Analysis}

The pulsar spectra were fitted with an exponentially cutoff power-law model
of the form
\begin{equation}
\frac{{\rm d} N}{{\rm d} E} = K E_{\rm GeV}^{-\Gamma}
                              \exp \left(- \frac{E}{E_{\rm cutoff}} \right)
\label{expcutoff}
\end{equation}
in which the three parameters are 
the photon index at low energy $\Gamma$, the cutoff energy $E_{\rm cutoff}$,
and a normalization factor $K$ (in units of ph cm$^{-2}$ s$^{-1}$ MeV$^{-1}$),
in keeping with the observed spectral shape of bright pulsars \citep{LATVELA}.
The energy at which the normalization factor $K$ is defined is arbitrary.
We chose 1 GeV because it is, for most pulsars, close to the energy
at which the relative uncertainty on the differential flux is minimal.

We wish to extract the spectra down to 100 MeV in order to constrain the power-law 
part of the spectrum, and to measure the flux above 100 MeV directly.
Because the spatial resolution of \textit{Fermi} is not very good at low energies
($\sim 5\degr$ at 100~MeV), we need to account for all neighboring sources
and the diffuse emission together with each pulsar. This was done using the framework 
used for the LAT Bright Source List \citep{BSL}. A 6-month source list was generated 
in the same way as the 3-month source list described in \citet{BSL}, but covering the 
extended period of time used for the pulsar analysis.
We have added the source Cyg X-3 \citep{CygX3}, although it was not detected 
automatically as a separate source, because it is very close to PSR J2032+4127,
and impacts the spectral fit of the pulsar. Cyg X-3 was fit with a simple power-law
as were all other non-pulsar sources in the list.

We used a Galactic diffuse model designated \texttt{54\_77Xvarh7S} 
calculated using GALPROP\footnote{http://galprop.stanford.edu/},
an evolution of that used in \citet{BSL}. A similar model, \texttt{\texttt{gll\_iem\_v02}},
is publicly available$^2$.

We kept events with $E>100$ MeV belonging to the 
\textit{diffuse} event class, which has the tightest cosmic-ray background rejection 
\citep{LATinstrument}. 
To avoid contamination by gamma-rays produced by cosmic-ray interactions in the Earth's atmosphere, 
we select time intervals when the entire ROI, 
of radius $10\degr$ around the source, has a zenith angle $<105\degr$.
We extracted events in a circle of radius $10\degr$ around each pulsar,
and included all sources up to $17\degr$ into the model (sources outside
the extraction region can contribute at low energy). Sources further away 
than $3\degr$ from the pulsar were assigned fixed spectra, taken from the all-sky 
analysis. Spectral parameters for the pulsar and sources within $3\degr$ of it were 
left free for the analysis.

The fit was performed by maximizing unbinned likelihood (direction and energy
of each event is considered) as described in \citet{BSL} and using the 
\textsc{minuit} fitting engine\footnote{http://lcgapp.cern.ch/project/cls/work-packages/mathlibs/minuit/doc/doc.html}. 
The uncertainties on the parameters were estimated from the quadratic 
development of the log(likelihood) surface around the best fit.
In addition to the index $\Gamma$ and the cutoff energy $E_{\rm cutoff}$
which are explicit parameters of the fit, the important physical quantities are 
the photon flux $F_{100}$ (in units of ph cm$^{-2}$ s$^{-1}$) and 
the energy flux $G_{100}$ (in units of erg cm$^{-2}$ s$^{-1}$),
\begin{eqnarray}
F_{100} = \int_{\rm 100 \, MeV}^{\rm 100 \, GeV} \frac{{\rm d} N}{{\rm d} E} {\rm d} E,\,\rm{and} \\
G_{100} = \int_{\rm 100 \, MeV}^{\rm 100 \, GeV} E \frac{{\rm d} N}{{\rm d} E} {\rm d} E.
\end{eqnarray}
These derived quantities are obtained from the primary fit parameters.
Their statistical uncertainties are obtained using their derivatives with respect
to the primary parameters and the covariance matrix obtained
from the fitting process.

For a number of pulsars, an exponentially cutoff power-law spectral model
is not significantly better than a simple power-law. We identified these by 
computing $TS_{\rm cutoff} = 2\Delta$log(likelihood)
(comparable to a $\chi^2$ distribution with one degree of freedom)
between the models with and without the cutoff. 
Pulsars with $TS_{\rm cutoff} < 10$ have poorly measured
cutoff energies. $TS_{\rm cutoff}$ is reported in Table \ref{tbl-spec}.

The above analysis yields a fit to the overall spectrum, including both the pulsar
and any unpulsed emission, such as from a PWN. 
To do better we split the data into on-pulse and off-pulse samples and 
modeled the off-pulse spectrum by a simple power-law. The off-pulse 
window used for this background estimation is defined in the last column 
of Table \ref{tbl-pulse}.

In a second step we re-fitted the on-pulse emission to the exponentially
cutoff power-law as before, with the off-pulse emission
(scaled to the on-pulse phase interval) added to the model and fixed to the off-pulse result. 
In many cases the off-pulse emission was not significant
at the $5 \sigma$ or even $3 \sigma$ level, but we kept the formal
best fit anyway, in order to not bias the pulsed emission upwards. The 
results summarized in Table \ref{tbl-spec} come from this
on-pulse analysis. 

Using an off-pulse pure power law is not ideal for the Crab or any other PWN 
with synchrotron and inverse Compton components within the \textit{Fermi} energy range.
Judging from the Crab pulsar, using a simple
power-law to model the off-pulse emission mainly affects the value of the cutoff 
energy. The analysis specific to the Crab, with a model adapted to the pulsar synchrotron
component low energies and to the high energy nebular component, yields
$E_{\rm cutoff} \sim6$ GeV \citep{FermiCrab}. 
This is the value listed in Table \ref{tbl-spec}.
The cutoff value obtained with the simplified model applied to most pulsars in this paper
is higher ($>$ 10 GeV). The photon and energy fluxes given by the two analyses are
within 10\% of each other.  Additional exceptions in Table \ref{tbl-spec} 
are for PSRs J1836+5925 and J2021+4026. The off-pulse phase definition for these 
pulsars is unclear, so the spectral parameters reported in the Table are from the initial, 
phase-averaged spectral analysis.

We have checked whether our imperfect knowledge of the Galactic diffuse
emission may impact the pulsar parameters by applying the same analysis
with a different diffuse model, as was done in \citet{BSL}.
The phase-averaged emission is affected. Seven (relatively faint) pulsars
see their flux move up or down by more than a factor of 1.5.
On the other hand, the pulsed flux is much more robust, because the off-pulse
component absorbs part of the background difference, and the 
source-to-background ratio is better after on-pulse phase selection. Only two pulsars
see their pulsed flux move up or down by more than a factor of 1.2, and
none shift by more than  a factor of 1.4 when changing the diffuse model.
Overall, the systematic uncertainties due to the diffuse model on the fluxes $F_{100}$ and $G_{100}$, 
on the photon index $\Gamma$, and on $E_{\rm cutoff}$ scale with the statistical uncertainty. 
Adding the statistical and systematic errors in quadrature amounts, to a good approximation, 
to multiplying the statistical errors on $F_{100}$, $G_{100}$, and $\Gamma$ by $1.2$.
%The uncertainties listed in Table \ref{tbl-spec} and plotted in the figures \ref{IndexvsEdot} and \ref{FluxVsSensitivity} 
The uncertainties listed in Table \ref{tbl-spec} and plotted in the figures include this correction. 
The increase in the
uncertainty on $E_{\rm cutoff}$ due to the diffuse model is $<5\%$ and is neglected.

Systematic uncertainties on the LAT effective area are of order $5$\% near 1 GeV, 
10\% below $0.1$ GeV, and 20\% above 10 GeV. To propagate their effect on the spectral
parameters in Equation 4, we modify the instrument response functions to bracket
the nominal values, and repeat the likelihood calculations. This is reported
in detail for most of the individual LAT pulsars already referenced. 
The bias values reported in \citet{MSP} well describe our current knowledge of the effect
of the uncertainties in the instrument response functions on the spectral parameters:
they are $\delta\Gamma = (+0.3,\,-0.1)$, $\delta E_{\rm cutoff} = (+20\%,\,-10\%)$, 
$\delta F_{100} = (+30\%,\,-10\%)$, and $\delta G_{100} = (+20\%,\,-10\%)$. 
The bias on the integral energy flux is somewhat less than that of the integral photon flux, 
due to the weighting by photons in the energy range where the effective area uncertainties are smallest.
We do not sum these uncertainties in quadrature with the others, since a change in instrument response will
tend to shift all spectral parameters similarly.

% \end{document}

%		\subsubsection{Unfolding analysis}
		% \section{2.2_part 2 Unfolding}
%
% Francesco L. - 30 June 2009
% das - 29 July 2009
% ECF - 1 Aug 2009
%
% Incorporated comments from F. Camilo - 22 Aug 2009
% das - 26 August 2009
%
% Comments from FC not yet addressed are:
% % _______________
% - l336-337: "The results obtained from the unfolding analysis were found to
% be consistent with the likelihood analysis results"
% 
% Always?  and what about errors, how do those compare?
% 
% 
% 
% \begin{document}

The pulsar spectra were also evaluated using an unfolding 
method \citep{dagostini1995, mazziotta2009}, that takes into 
account the energy dispersion introduced by the instrument 
response function and does not assume any model for the 
spectral shapes. ``Unfolding'' is essentially a deconvolution 
of the observed data from the instrument response functions. 
For each pulsar we selected photons within 68\% of the PSF 
with a minimum radius of $0\fdg35$ and a maximum of $5\degr$.

The observed pulsed spectrum was built by selecting the
events in the on-pulse phase interval and subtracting the
events in the off-pulse interval, properly scaled for the 
phase ratio. The instrument response function, expressed as
a smearing matrix, was evaluated 
using the LAT \textit{Geant4}-based\footnote{http://geant4.web.cern.ch/geant4/} 
Monte Carlo simulation package called \textit{Gleam} \citep{GLEAM03}, 
taking into account the pointing history of the source.

The true pulsar energy spectra were then reconstructed 
from the observed ones using an iterative procedure based on 
Bayes' theorem \citep{mazziotta2009}. Typically, 
convergence is reached after a few iterations. 
When the procedure has converged, both statistical and 
systematic errors on the observed energy distribution 
can be easily propagated to the unfolded spectra. The results 
obtained from the unfolding analysis were  
consistent within errors with the likelihood analysis results.

% \begin{thebibliography}{}
% 
% \bibitem[D'Agostini 1995]{dagostini1995} D'Agostini, G. 1995, NIM, A362, 487
% 
% \bibitem[Mazziotta 2009]{mazziotta2009} Mazziotta, M.~N. 2009,
% Proc. of the 31st ICRC, Lodz 7-15 July 2009
% 
% \end{thebibliography}
% 
% 
% \end{document}

%\section{Catalog Description and Sample Population Statistics}
\section{The LAT Pulsar Sample}
We describe here the astronomical context of the observed LAT pulsars,
including our current best understanding of the source distances, the Galactic distribution and
possible associations. We also note correlations among some observables which may help probe the
origin of the pulsar emission.

	\subsection{Distances}
	\label{Distances}
	% \section{3.1 Distances}
%
% Teresa M. \& Peter D. - 28 July 2009 
% PeterH - 31 July 2009
% ECF - 1 Aug 2009
% PeterH - 5 Aug 2009
% David S - 7 August 
% ECF - 16 Sep 2009 (Updates to address Fernando2)
% 
% 
% Incorporated comments from F. Camilo - 22 Aug 2009
% 
% 
% \begin{document}

Converting measured pulsar fluxes to radiated power requires
reliable distance estimates.
Annual trigonometric parallax measurements are the most
reliable, but are generally only available for a few relatively nearby
pulsars.

The most commonly used technique to obtain radio pulsar distances  
exploits the pulse delay as a function of wavelength 
by free electrons along the path to Earth. A distance
can be computed from the DM coupled to an
electron density distribution model. 
We use the NE2001 model \citep{Cordes2002} unless noted otherwise.
It assumes uniform electron densities in and 
between the Galactic spiral arms, with smooth transitions between
zones, and spheres of greater density for specific regions such
as the Gum nebula, or surrounding Vela.
Specific lines-of-sight can traverse unmodeled regions of 
over- (or under-) density, as, for example, along the tangents
of the spiral arms, causing 
significant discrepancies between the true pulsar distances and
those inferred from the electron-column density.

A third method, kinematic, associates the pulsar with
objects whose distance can be measured from the Doppler shift of absorption or emission
lines in the neutral hydrogen (HI) spectrum, together with
a rotation curve of the Galaxy. 
It breaks down where the velocity gradients are very small or where
the distance-velocity relation has double values.
The associations are often uncertain, and
these distance measurements can be controversial. 

In a small number of cases, the distance is evaluated either from X-ray
measurements of the absorbing column at low energies (below 1 keV),
or from consideration of the X-ray flux
assuming some standard parameters for the neutron star.

Table \ref{tab:dist} presents the best known distances of 37 pulsars
detected by \textit{Fermi}, the methods used to obtain them, and the references.  
For distances obtained from the NE2001 model and the DM, the reference 
indicates the DM measurement. 
We assume a minimum DM distance uncertainty of 30\%.  
When distances from different methods disagree
and no method is more convincing than the other, a range is
given, and 30\% uncertainties on the upper and lower
values are used. 
For the remaining 9 \textit{Fermi}-discovered pulsars no distance
estimates have been established so far.
Here follow comments for some of the distance values reported in Table \ref{tab:dist}:

%\begin{itemize}

%\item 
{\it PSR J0205+6449} -- The pulsar is in the PWN 3C 58. 
  NE2001 gives 4.5 kpc for DM=141 cm$^{-3}$ pc in this direction \citep{Camilo2002c}.
%\citep{Malofeev2003}.  
  Using HI absorption and emission lines from the PWN yields
from 2.6 kpc \citep{Green1982} to 3.2 kpc \citep{Roberts1993}. 
The lower V-band reddening \citep{Fesen1988,Fesen2008}
  compared to the Galactic-disk edge \citep{Schlegel1998} suggests
  that the PWN is in the range 3--4 kpc.  
  Table~\ref{tab:dist} quotes
  the distance range found by \citet{Green1982}
  and \citet{Roberts1993}.

%\item
{\it PSR J0218+4232} -- 
%The distance to the only millisecond pulsar which was marginally 
%detected by EGRET \citep{Kuiper2000} is rather uncertain. 
The DM measurements from \citet{Navarro1995} together with
NE2001 yield 2.7 $\pm$ 0.8 kpc.
Comparing the pulsar characteristic age with the cooling models of its white-dwarf companion
gives a distance range of 2.5 to 4 kpc \citep{Bassa2003}.

{\it PSR J0248+6021 } -- The DM of 376 cm$^{-3}$
 pc \citep{Cognard2009} puts this pulsar beyond the edge of the Galaxy
 for this line-of-sight.  The line-of-sight, however, borders the
 giant HII region W5 in the Perseus Arm.
%  and the distance estimate
% could be affected by a dense local environment. 
 We bracket the pulsar
 distance as being between W5 (2 kpc) and the Galaxy edge (9 kpc).
  
%\item
%{\it PSR J0534+2200} -- The Crab pulsar and its nebula belong to the
%  best studied sources in the sky. Despite many instruments over the
%  entire electromagnetic spectrum have observed the system, the
%  distance is poorly known. According to \citet{Kaplan2008} neither
%  timing parallax, radio interferometric parallax, nor optical
%  parallax measurements are likely to significantly improve our
%  knowledge of the pulsar's distance in the near future.  The estimate
%  of the distance reported in Table~\ref{tab:dist} is performed
%  by \citet{Trimble1973} using several different methods.

%\item
{\it PSR J0631+1036} -- The  DM $= 125.3$ 
  cm$^{-3}$\,pc \citep{Zepka1996} is large for a source in the direction
  of the Galactic anticenter.  The dark cloud LDN 1605, part of the
  active star-forming region 3 Mon, is in the line-of-sight.
  The pulsar could be inside the cloud, at $\sim$0.75 kpc.
  Ionized material in the cloud could cause NE2001 to overestimate
  the distance.

%\item
{\it PSR J1124$-$5916} -- It lies towards the Carina arm where NE2001 biases are acute.
  The DM distance is 5.7 kpc \citep{Camilo2002b}.  
  The kinematic distance of the associated SNR (G292.0+1.8) indicates a
  lower limit of 6.2$\pm$0.9\,kpc \citep{Gaensler2003}.
%   which is higher
%  than the previous evaluation of 3.2 kpc \citep{Caswell1975}
%  performed with the same method.  
  The value in Table~\ref{tab:dist} is derived by \citet{Gonzalez2003} linking the
  X-ray absorption column with the extinction along the
  pulsar direction.

%\item
{\it PSR J1418$-$6058} -- This pulsar is likely associated with the
  PWN G313.3+0.1, near the Kookaburra complex.   
  A nearby HII region is at 13.4~kpc \citep{Caswell1987} but
  could easily be in the background.
  Such a large distance implies an unreasonably large gamma-ray
  efficiency.   
  Table~\ref{tab:dist} lists a crude estimate of
  the distance range with the lower limit \citep{Yadigaroglu1997}
  taking the pulsar to be related to one of
  the near objects (Clust 3, Cl Lunga 2 or SNR G312.4$-$0.4), and the higher
  limit \citep{Ng2005} determined by applying the relation found
  by \citet{Possenti2002} and the correlation between pulsar X-ray
  photon index and luminosity given by \citet{Gotthelf2003}.
  
%\item
{\it PSR J1709$-$4429} -- The NE2001 DM 
distance is 2.3$\pm$0.7 kpc \citep{TaylorDM1993}. Kinematic
distances give upper and lower limits
of 3.2$\pm$0.4 kpc and 2.4$\pm$0.6 kpc,
respectively \citep{Koribalski1995}. The X-ray flux from the neutron
star detected by {\it Chandra} \citep{Romani2005} and \textit{XMM-Newton} 
\citep{McGowan2004} is compatible with a distance of 1.4--2.0
kpc. We assume the range 1.4--3.6 kpc.

%\item
{\it PSR J1747$-$2958} -- The pulsar is associated with the PWN
  G359.23$-$0.82. HI measurements
  yield a distance upper limit of 5.5~kpc \citep{Uchida1992}, but
  the DM (101 pc\,cm$^{-3}$) suggests 
  2.0$\pm$0.6\,kpc \citep{Camilo2002a}. The X-ray absorbing column
  detected by {\it Chandra} is
  between 4 and 5 kpc, while the closer value of 2 kpc would
  imply that an otherwise unknown molecular cloud lies in front of the
  pulsar \citep{Gaensler2004}. A range of 2--5 kpc is used in our analysis.
    
%\item
%{\it PSR J1836+5925} -- The pulsar is likely associated with
% RX~J1836.2+5925 (1RXS~J183613.6+59253) whose upper limit on the
% distance, quoted in Table~\ref{tab:dist}, is based on a 'reasonable'
% proper motion \citet{Halpern2007}.

%\item
%{\it PSR J1952+3252} -- It is better known as B1951+32 and it is
%  associated with SNR CTB80. The distance evaluated from DM is
%  3.1$\pm$0.2\,kpc \citep{Strom2000}, but the kinematic distance is
%  rather 2 kpc \citep{Greidanus1990}.

%\item 
{\it PSR J2021+3651} -- The DM 
%(369 pc\,cm$^{-3}$) locates it at a
  distance of $\sim$12 kpc implies a high gamma-ray
  conversion efficiency \citep{Roberts2002,PSR2021LAT}.  
  The open cluster Berkeley 87 near the line-of-sight could be responsible for 
  an electron column density higher than modeled by NE2001.  
  The distance in Table~\ref{tab:dist} comes from a 
  {\it Chandra} X-ray observation of the pulsar and its surrounding
  nebula \citep{VanEtten2008}. A similar range (1.3--4.1 kpc) was
  obtained for the X-ray flux detected from the associated PWN.

%\item 
{\it PSR J2032+4127} -- The DM value (115 pc cm$^{-3}$)
  gives an NE2001 distance of 3.6 kpc. If the pulsar belongs to
  the star cluster Cyg OB2, it would be located at approximately 1.6
  kpc \citep{Camilo2009}. In this text we use a range of 1.6--3.6
  kpc for this source.

%\item
{\it PSR J2229+6114} -- The distance derived from the X-ray
  absorption  
%  before the discovery of
%  pulsed emission from the source 
  is $\sim$3 kpc \citep{Halpern2001b},
  between the values from the DM \citep[6.5
  kpc;][]{Halpern2001a} and from the kinematic
  method \citep[0.8 kpc;][]{Kothes2001}. 
%  These two values are quoted
%  as a range in Table~\ref{tab:dist}.
  
%\end{itemize}

Figure \ref{SkyProjection} shows 
%how these pulsars are distributed around us, 
a polar view of the distribution of known pulsars over the Galactic plane. 
When two different distances are listed, we plot the closer one.

% \end{document}

	\subsection{Spatial Distributions, Luminosity, and Other Pulse Properties}
	% \section{3.0 Catalog Description and Sample Population Statistics}
%
% Andrea - 1 July 2009
% Incorporates comments by PdH - 31 July 2009
% Comments from DAS - 1 Aug 2009
% ECF - 1 Aug 2009
% ECF - 16 Sep 2009 (Updates to address Fernando2)
%
% Incorporated comments from F. Camilo - 22 Aug 2009
% Comments from FC not yet addressed are:
% 
% - l346: I believe the first people to address the "Shklovskii effect"
% (really, acceleration biases more generally) for MSPs were Camilo et
% al. (1994).
% _______________
% - ll396-399: I don't understand why L_x for the Crab is being brought to
% this discussion.  Although you start the paragraph by saying L_tot~L_gamma
% for most pulsars... in figure 8 you're really plotting L_gamma (except
% for the Crab).  But many pulsars don't have detected gamma-rays.  So are
% you talking about all pulsars?  Surely not.  Some very selected sample.
% What's the purpose of this?  It's not clear to me.
% 
%
% 
% \begin{document}

Figure \ref{SkyMap} shows the pulsars projected on the sky. A Gaussian fit to the
Galactic latitude distribution for those with $|b|<10^\circ$ and having
distance estimates yields a standard deviation of $\sigma_{\rm b} = 3.5 \pm
0.8$ degrees. The distances range from $d=0.25$ to $5.6$ kpc, and we can use
the most distant to place an upper limit on the scale height, 
obtaining $h<d\sin{3\fdg5} = 340$ pc, close to the typical scale height for all radio
pulsars (the value used in \citet{ParkesSurveyII} is 450 pc).

The light curve peak separations $\Delta$ and the radio lags $\delta$
from Table \ref{tbl-pulse} are summarized in Figure \ref{RadioSep}. 
As we will discuss in Section 5, outer magnetosphere 
emission models predict correlations between these parameters.
Figure \ref{BlcvsAge} shows 
$B_{\rm LC}$ versus the characteristic age ($\tau_{\rm c}$). 
The magnetic fields at the light cylinder for the detected MSPs are comparable to those of the 
other gamma-ray pulsars, suggesting that the emission mechanism for the two families 
may be similar.
%In Figure \ref{EcvsBlc} we plot the cutoff energy versus $B_{\rm LC}$, 
%with the energy cutoff histogram on the right Y axis. This plot seems 
%almost flat until at least $5\times10^4$~G.
%
%In Figures \ref{IndexvsEdot} and \ref{DeltavsEdot} we plot the photon index, 
%and the gamma-ray peak separation versus $\dot E$, respectively. 
%The histogram of the photon indices is distributed around $\sim1.5$. 
%The $\Delta$ distribution is bimodal, 
%with gamma-ray peak separations peaking at  $\sim0.15$ and $\sim0.5$ in phase.

Table \ref{tbl-spec} lists $L_\gamma$  and $\eta$, while
Figure \ref{LEdot} plots $L_\gamma$ vs.\ $\dot{E}$. The dashed line indicates 
$L_\gamma=\dot{E}$, while the dot-dashed line indicates $L_\gamma\propto\dot{E}^{1/2}$, where 
\begin{equation}
  L_\gamma \equiv 4\pi d^2 f_\Omega G_{100}.
\end{equation}
%We use a 30\% systematic error on $G_{100}$, stemming from the uncertainties
%in the instrument response and the diffuse emission model. 
The  flux correction factor $f_\Omega$ \citep{Watters09} is 
model-dependent and depends on the magnetic inclination and observer angles $\alpha$ 
and  $\zeta$. 
%For instance, a larger $f_\Omega$ is needed for pulsars with large impact 
%angles $\beta = \zeta - \alpha$ if a particular model predicts low-level off-beam emission.
Both the outer gap and slot gap models predict $f_\Omega\sim1$, 
in contrast to earlier use of $f_\Omega=1/4\pi\approx0.08$ \citep[in e.g.][]{Thompson94}, 
or $f_\Omega=0.5$ for MSPs \citep[in e.g.][]{Fierro95}. For simplicity, we use 
$f_\Omega=1$ throughout the paper, which presumably induces
an artificial spread in the quoted $L_\gamma$ values. 
%This 
%is also true in the case of the outer gap model of \citet{Zhang04}, where 
%$L_\gamma = f^3(P,B,\alpha)\dot{E}$ is dominated by the fractional gap size $f$.
%
%This may be written as:
%\begin{equation}
%  \eta_\gamma \approx 0.0486f_\Omega d_1^2 G_{100} I_{45}^{-1}\dot{P}_{-15}^{-1}P_{0.1}^3,
%\end{equation}
%where $G_{100}$ is measured in $10^{-5}$~MeV cm$^{-2}$ s$^{-1}$, $I_{45} = I/10^{45}$ 
%g\,cm$^2$, $\dot{P}_{-15} = \dot{P}/10^{-15}$~s\,s$^{-1}$, $P_{0.1} = P/0.1$~s and 
%$d_1=d/1$~kpc. 
However, it is the quadratic distance dependence for $L_\gamma$ that dominates
the uncertainty in $L_\gamma$
in nearly all cases. 
%Distances are discussed above.

Gamma-rays dominate the total power $L_{\rm tot}$ radiated by most known high-energy pulsars, 
that is, $L_{\rm tot}\approx L_\gamma$. The Crab is a notable exception, 
with X-ray luminosity $L_X\sim10L_\gamma$. 
In Figure \ref{LEdot} we plot both $L_\gamma$ and $L_X + L_\gamma$. 
$L_X$  for $E<100$ MeV is taken from Figure 9 of \citet{Kuiper2001}.
%
%A break is clearly seen around $\dot{E}_{\rm break}\sim10^{35}$ erg s$^{-1}$. 
%While the MSPs seem to follow $L_\gamma\propto\dot{E}$, the pulsars with higher $\dot{E}$ seem to follow a 
%trend which is flatter than the expected $L_\gamma\propto\dot{E}^{1/2}$.

% \end{document}

	\subsection{Associations}
	% \section{3.2 Associations}
%
% David S. - 3 July 2009
% Elizabeth F. and Dave T. - 16 July 2009 
% PeterH - 31 July 2009
% Comments from DAS - 1 Aug 2009
% ECF - 1 Aug 2009, 2 Aug 2009
% ECF - 17 Sep 2009 (Corrected numbers in first paragraph to match the table)
% 
% Incorporated comments from F. Camilo - 22 Aug 2009
% Comments from FC not yet addressed are:
% 
% - l491: I think this paragraph would be clearer if you started
% with stating that 31 of the 46 pulsars are associated with EGRET sources,
% including the 6 confirmed and 3 marginal EGRET pulsar detections.
% 
% \begin{document}
 
Table \ref{tbl-assoc} provides some alternate names and positional associations 
of the pulsars in this catalog with other astrophysical sources.  For the EGRET 3EG,  
EGR, and GEV and AGILE AGL catalogs, the uncertainties in the localization of the 
counterparts is worse than for the LAT sources. In these cases, we consider a source 
is a possible counterpart to a LAT pulsar when the separation between the two positions 
is less than the quadratic sum of their 95\% confidence error radii.

We see that 25 of the 46 pulsars are associated with sources in the 3EG, EGR and GEV 
catalogs of EGRET sources, though  19 were seen only as unidentified unpulsed sources.  
A number of these unidentified EGRET sources had previously been associated with SNRs, 
PWNe, or other objects \citep[e.g.][]{Walker2003,DeBecker2005}. In all cases, the gamma-ray 
emission seen with the LAT is dominated by the pulsed emission.  Of the 25 EGRET sources, 
14 are gamma-ray-selected pulsars,  and 11 are radio-selected, including 2 MSPs. All 6 
high-confidence EGRET pulsars \citep{Nolan96} are detected, and the 3 marginal EGRET 
detections are confirmed as pulsars \citep{EGRETB0656p14,EGRETB1046m58,Kuiper2000}.  
The 21 sources without 3EG, EGR, or GEV counterparts include 18 previously detected radio 
pulsars (6 of which are MSPs) and 3 gamma-ray selected pulsars.  
%However, we note that two of these 5 gamma-ray selected pulsars (PSRs J1813$-$1246 and J1907+06) 
%are associated with sources in the catalog of point sources above 1 GeV by \citet{Lamb1997}.

Not surprisingly, many of the young pulsars have SNR or PWN associations. At least 
19 of the 46 pulsars are associated with a PWN and/or SNR 
\citep{Roberts2005,Green2009}. 
We do not test here whether the gamma-ray flux from any of these pulsars includes a
non-magnetospheric component, as might be indicated by spatially extended emission or
a spectrum at pulse minimum not characteristic of a pulsar. Such studies are underway.

%A key test of whether any of these associations 
%include any gamma-ray component other than the pulsar will depend on seeing 
%spatially-extended emission or off-pulse emission with a different energy spectrum 
%from that produced by the pulsar.

At least 12 of the pulsars are associated with TeV sources, 9 of which
 are also associated with PWNe. Those pulsars with both 
TeV and PWN associations are typically young, with ages less than 20 kyr.

% \end{document}

\section{Pulsar Flux Sensitivity}
% \section{4.0 Pulsar Flux Sensitivity}
%
% smith, 1 July
% Andrea - 15 July 2009
% J Eric/Roger - 30 July 09
% Comments from DAS - 1 Aug 2009
% ECF - 1 Aug 2009, 2 Aug 2009
% ECF - 16 Sep 2009 (Updates to address Fernando2)
%
% 
% Incorporated comments from F. Camilo - 23 Aug 2009
% Comments from FC not yet addressed are:
% 
% - l541: "histograms are well matched" - have you done any kind of (KS?) tests,
% or is this by eye?
% 
% You mention a flux threshold to which pulsars are detected, but by
% this you mean the faintest pulsars detected, right?  You don't mean
% the flux below which you're incomplete, right?  Can't you get this
% from the change in slope in Log N-Log S?
% 
% 
% \begin{document}

In order to interpret the population of gamma-ray pulsars discovered with the LAT, we need to evaluate the sensitivity of our searches for pulsed emission.  While the precise sensitivity at any location is a function of the local background flux, the pulsar spectrum, and the pulse shape, we can derive an approximate pulsed sensitivity by calculating the \textit{unpulsed} flux sensitivity for a typical pulsar spectrum at all locations in the sky and correlating with the observed $Z^2_2$ test statistic for the ensemble of detected pulsars.

Figures \ref{EcvsBlc} and \ref{IndexvsEdot} show the distributions of the cutoff energy and the photon index, respectively, for all the LAT-detected pulsars.
The distribution of photon indices peaks in the range $\Gamma = 1-2$, and the distribution of cutoffs peaks at $E_{\rm cutoff} = 1 - 3$ GeV.  For a typical spectrum, we used $\Gamma = 1.4$ and $E_{\rm cutoff} = 2.2$ GeV, values approximately equal to their respective weighted averages.

We then generated a sensitivity map for unpulsed emission for the six-month data set used here.  For each \textit{(l,b)}
location in the sky, we computed the DC flux sensitivity at a threshold likelihood test statistic TS $=$ 25 integrated
above 100 MeV, assuming the typical pulsar spectrum within the source PSF and an underlying diffuse gamma-ray flux from
the  \texttt{rings\_Galaxy\_v0} model \citep{LATVelaLike}. This is an earlier version of 
the publicly available$^2$ model \texttt{\texttt{gll\_iem\_v02}}, similar to the model used for the spectral analysis.
%, but more closely matches the observed Galactic structure.
We note that the likelihood calculation assumes that the
source flux is small compared to the diffuse background flux within the PSF, which is appropriate for a source just at
the detection limit. Finally, we converted this map to pulsed sensitivity by a simple scale factor that accounts for the
correspondence between the $Z^2_2$ periodicity test confidence level and the unpulsed likelihood TS for the detected
pulsars.

The resulting $5\sigma$ sensitivity map for pulsed emission is shown in Figure \ref{SensitivitySkyMap}. 
Comparing the measured fluxes with the predicted sensitivities at the pulsar locations (Figure
\ref{FluxVsSensitivity}), we see that this 5$\sigma$ limit indeed provides a reasonable lower envelope to the
pulsed detections in this catalog. Thus the effective sensitivity for high latitude (e.g.~millisecond) pulsars
with known rotation ephemerides is $1-2 \times 10^{-8}$ cm$^{-2}$ s$^{-1}$; at low latitude there is large
variation, with typical detection thresholds $3-5\times$ higher. We expect the threshold to be somewhat higher
for pulsars found in blind period searches. Figure \ref{FluxVsSensitivity} suggests that this threshold is
$2-3\times$ higher than that for pulsars discovered in folding searches, with resulting values as high as
$2\times 10^{-7}$ cm$^{-2}$ s$^{-1}$ on the Galactic plane.

The Log\,N--Log\,S plot is shown in Figure \ref{LogN-LogS}. 
%The dashed line is for all the detected pulsars, the radio-selected gamma-ray pulsars (including MSPs) are
% colored gray, and the blue histogram is for the gamma-ray-selected pulsars. 
The approximate $N \propto 1/S$ dependence expected for a disk population is apparent for the higher flux objects.
This shows that
while radio-selected pulsars are detected down to a threshold of $2 \times 10^{-8}$ cm$^{-2}$
s$^{-1}$, the faintest gamma-ray-selected pulsar detected has a flux $\sim 3\times$ higher at $6
\times 10^{-8}$ cm$^{-2}$ s$^{-1}$. It is interesting to note that, aside from the lower flux
threshold for the former, the radio-selected and gamma-ray-selected histograms are well matched,
suggesting similar underlying populations.

% \end{document}

\section{Discussion}
% \section{5.0 Discussion}
%
% Roger \& Alice - 16 July 2009
% RWR 29 July 2009
% ECF - 1 Aug 2009 (still need lots of references)
%  iRWR 24/8/09 
% ECF - 15 Sep 2009 (corrected several Figure and Table references)
% ECF - 16 Sep 2009 (Updates to address Fernando2)
% ECF - 16 Sep 2009 (Added text to support detectability figure)
%
%
% Incorporated comments from F. Camilo - 23 Aug 2009
% Comments from FC not yet addressed are:
% 
% - "there is a weak correlation of E_c with B_LC, as shown in Figure 5".
% What's the correlation coefficient?  And what about if you plotted this
% only for those pulsars with TS_cutoff>10?  would it strengthen the
% correlation?  weaken it?  
% 
% 
% \begin{document}

The striking results of the early \textit{Fermi} pulsar discoveries demonstrate the LAT's 
excellent power for pulsed gamma-ray detection. By increasing the gamma-ray pulsar
sample size by nearly an order of magnitude and by firmly establishing the
gamma-ray-selected (radio-quiet Geminga-type) and millisecond gamma-ray
pulsar populations, we have promoted GeV pulsar astronomy to a major probe
of the energetic pulsar population and its magnetospheric physics. Our large
pulsar sample allows us both to establish patterns in the pulse emission
possibly pointing to a common origin of pulsar gamma-rays and to find anomalous systems
that may point to exceptional pulsar geometries and/or unusual emission physics.
In this Section we discuss some initial conclusions drawn from the sample,
recognizing that the full exploitation of these new results will flow from the
detailed population and emission physics studies to follow.

\subsection{Pulsar Detectability}

A widely-cited predictor of gamma-ray pulsar detectability
is the spin-down flux at Earth ${\dot E}/d^2$ \citep[see e.g.][]{DAS08}. 
However, as argued by 
\citet{Arons06} \citep[see also][]{Harding02}, it is natural in many 
models for the gamma-ray emitting gap to maintain
a fixed voltage drop. This implies that $L_\gamma$ is simply proportional to the
particle current \citep{Harding1981}, which gives $L_\gamma \propto {\dot E}^{1/2}$,
i.e. gamma-ray efficiency increases with decreasing spin-down power
down to ${\dot E} \sim 10^{34}- 10^{35}$ erg s$^{-1}$ where the gap saturates at
large efficiency. In Figure \ref{DetectMetric} we show how our
detected pulsars rank in ${\dot E}^{1/2}/d^2$ against the set of searched 
pulsars. We see that for both MSPs and young pulsars, the detected objects have
among the largest values of this metric. The presence of missing objects
among the detected pulsars is interesting, but must be treated with caution,
as the detectability metric may be inflated by poor DM distances, or the
sensitivity of the pulse search might be anomalously low due to
high local background or unfavorable pulsar spectrum or pulse profile.
Alternatively, some missing objects may be truly gamma-ray faint for the 
Earth line-of-sight. A more complete study of the implications of the pulsar 
non-detections and upper limits is in progress.

	To study the luminosity evolution in the observed pulsar population,
we plot in Figure \ref{LEdot} our present best estimate of $L_\gamma$ 
against ${\dot E}$, based on the pulsed flux
measured for each pulsar.  Two important caveats
must be emphasized here. First, the inferred luminosities are quadratically sensitive
to the often large distance uncertainties. Indeed, for many radio selected
pulsars (green points) we have only DM-based distance estimates. For
many gamma-ray-selected pulsars we have only rather tenuous SNR or birth cluster
associations with rough distance bounds. Only a handful of pulsars have
secure parallax-based distances. Second, we have assumed here uniform
phase-averaged beaming across the sky ($f_\Omega$=1). This is not realized
for many emission models, especially for low ${\dot E}$
pulsars \citep{Watters09}.

To guide the eye, Figure \ref{LEdot} shows lines for 100\% conversion
efficiency ($L_\gamma = {\dot E}$) and a heuristic constant voltage
line $L_\gamma =(10^{33}$ erg s$^{-1} {\dot E})^{1/2}$. In view
of the large luminosity uncertainties, we must conclude that it is not
yet possible to test the details of the luminosity evolution. However,
some trends are apparent and individual objects highlight possible
complicating factors.  For the highest ${\dot E}$ pulsars, there does seem
to be rough agreement with the ${\dot E}^{1/2}$ trend. However, large
variance between different distance estimates for the Vela-like PSRs
J2021+3651 and J1709$-$4429 complicate the interpretation. In the
range $10^{35}$ erg s$^{-1} < {\dot E} < 10^{36.5}$ erg s$^{-1}$,
the $L_\gamma$ seems nearly constant, although the lack of
precise distance measurements limits our ability to draw conclusions.
For example, the very large nominal DM distance of PSR J0248+6021
would require $>100$\% efficiency, and so is unlikely to be correct.
Two other pulsars with apparent high efficiency (J1836+5925 and J2021+4026) 
are plotted including relatively bright unpulsed emission; this may
be magnetospheric, but may also be a surrounding  or nearby source.
The association distances for the gamma-ray-selected pulsars must
additionally be treated with caution. For example
PSR J2021+4026 has a $\tau_{{\rm c}} \sim 10\times$ larger than the
age of the putative associated SNR $\gamma$ Cygni.
Improved distance estimates in this range are the key to probing
luminosity evolution.

From $10^{34}$ erg s$^{-1} < {\dot E} < 10^{35}$ erg s$^{-1}$
we have several nearby pulsars with reasonably accurate parallax
distance estimates. However we see a wide range of gamma-ray
efficiencies. This is the range over which, for both slot gap 
and outer gap models, the gap is expected to `saturate' and use
most of the available potential to maintain the pair cascade.  In slot gap models
\citep{Muslimov2003}, the break occurs at about $10^{35}$ erg s$^{-1}$, 
when the gap is limited by screening of the accelerating 
field by pairs. The efficiency below this saturation is predicted to be 
$\sim$ 10\%.  In outer gap models \citep{Zhang04}, the break is
predicted to occur at somewhat lower $\dot E \sim 10^{34}$ erg s$^{-1}$.  With
the present statistics and uncertainties, it is not possible to discriminate
between these model predictions except to note that both are 
consistent with the observed results.
In some models the gap saturation dramatically affects the shape of
the beam on the sky and accordingly the flux conversion factor $f_{\Omega}$;
for outer gap models \citet{Watters09} estimate
$f_{\Omega} \sim 0.1-0.15$ for Geminga (similar values are obtained for
J1836+5925), driving down the rather high inferred luminosity of these pulsars
by an order of magnitude. In contrast, another pulsar with an
accurate parallax distance, PSR J0659+1414, has an inferred
luminosity $30\times$ {\it lower} than the ${\dot E}^{1/2}$ prediction.
Clearly, some parameter in addition to ${\dot E}$ controls the
observed $L_\gamma$. Finally, for $< 10^{34}$ erg s$^{-1}$ the sample
is dominated by the MSPs. These nearby, low luminosity objects
clearly lie below the ${\dot E}^{1/2}$ trend, and in fact seem more
consistent with $L_\gamma \propto {\dot E}$.

	Upper limits on radio pulsars with high values
of the spin-down flux at Earth or large ${\dot E}^{1/2}/d^2$
can help constrain viable efficiency models. In practice, the
modest present exposure, the large background in the Galactic
plane and the need to rely on uncertain dispersion-based distance
estimates limit the value of such constraints. Still, a few pulsars
are already interesting; for example, using the DM-based distance,
the sensitivity in Figure \ref{SensitivitySkyMap}, and an assumed $f_\Omega=1$, we find that 
PSR J1740+1000 shows less than $1/5$ of the flux expected from 
the ${\dot E}^{1/2}$ (constant voltage) line in Figure \ref{LEdot}. Similarly,
PSRs J1357$-$6429 and J1930+1852 have upper limits just below the 
expected fluxes. Further, some detected pulsars, e.g. PSRs J0659+1414 
and J0205+6449, lie significantly below the constant voltage trend.
We expect that as LAT exposure and the significance of such limits
increase, we should obtain additional constraints on the factors 
controlling pulsar detectability.

One likely candidate for the additional factor affecting
gamma-ray detectability is beaming. For PSR J1930+1852 \citep{Camilo2002d}, X-ray
torus fitting \citep{Ng2008} suggests a small viewing angle
$|\zeta|\sim 33\degr$. In outer gap models this makes
it highly unlikely that the pulsar will produce strong emission
on the Earth line-of-sight. Similarly it has been argued
that PSR J0659+1414 has a small viewing angle
$\zeta < 20 \degr$ \citep{Everett2001} (but see 
\citet{Weltevrede2009} for a discussion of uncertainties).
Again, strong emission from above the null charge surface is
not expected for this $\zeta$. One possible interpretation is
that we are seeing slot gap or even polar cap emission from
this pulsar, which is expected at this $\zeta$. The unusual
pulse profile and spectrum of this pulsar may allow us to
test this idea of alternate emission zones.

In discussing non-detections, we should also note
that the only binary pulsar systems reported in this paper
are the radio-timed MSPs. In particular, our blind searches
are not, as yet, sensitive to pulsars that are undergoing strong
acceleration in binary systems.  However, we do expect such objects to exist. Population
syntheses \citep{phal02} % ApJ 574, 364
suggest that several percent of the young pulsars are born while
retained in massive star binary systems. A few such systems are known in the radio
pulsar sample (e.g. the TeV-detected PSR B1259$-$63); we expect that with the
gamma-ray signal immune to dispersion effects an appreciable number of pulsar
massive-star binaries will eventually be discovered. Indeed, it is entirely
possible that the bright gamma-ray binaries LSI $+61^\circ$ 303 \citep{LATLSI+61}
and LS 5039 \citep{LATLS5039} may host pulsed GeV signals that have not yet been found.

\subsection{Pulsar Population}
% Reference figure 8

With the above caveats about missing binary systems in mind, we can already
draw some conclusions about the {\it single} gamma-ray pulsar population.
For example, there are 17 gamma-ray selected pulsars with a faintest
flux of $\sim 6 \times 10^{-8} {\rm cm^{-2} s^{-1}}$; there are 16
non-millisecond radio-selected pulsars to this flux limit.
%For example, we have 21 non-millisecond radio-selected pulsars and 17 gamma-ray selected
%pulsars to the shallower flux limit ($\sim 6 \times 10^{-8}$ cm$^{-2}$ s$^{-1}$)
%of the latter. 
Of course, some gamma-ray-selected objects can indeed be detected 
in the radio \citep{Camilo2009}. Indeed, the detection of PSR J1741$-$2054 at
$L_{1.4 {\rm GHz}} \approx 0.03$ mJy kpc$^{2}$ underlines the fact that the radio
emission can be very faint. Deep searches for additional radio counterparts are
underway. However, with deep radio observations of several objects,
e.g.  Geminga, PSR J0007+7303, PSR J1836+5925
\citep{ka99,Halpern2004,Halpern2007}, %ka99 = apJ 527, 101; ha02=ApJ 612, 398; ha07=ApJ 668, 1154
providing no convincing detections,
it is clear that some objects are truly radio faint. The substantial number of radio
faint objects suggests that gamma-ray emission has an appreciably larger extent
than the radio beams, such as expected in the outer gap (OG) and slot-gap/two pole caustic (SG/TPC)
models.

Population synthesis studies for normal (non-millisecond) pulsars predicted that 
LAT would detect from 40--80 radio loud pulsars and comparable numbers of radio 
quiet pulsars in the first year \citep{Gonthier04,Zhang07}.
The ratio of radio-selected to gamma-ray-selected gamma-ray pulsars has been noted as
a particularly sensitive discriminator of models, since the outer magnetosphere
models predict much smaller ratios than polar cap models \citep{Harding07}.
Studies of the MSP population \citep{Story07} predicted that
LAT would detect around 12 radio-selected and 33--40 gamma-ray-selected 
MSPs in the first year, in rough agreement with the number of 
radio-selected MSPs seen to date (searches
for gamma-ray selected MSPs have not yet been conducted).
Thus, in the first six months the numbers of LAT pulsar detections are consistent
with the predicted range, and the large number of gamma-ray selected pulsars
discovered so early in the mission points towards the outer magnetosphere models.

We can in fact use our sample of detected gamma-ray pulsars to
estimate the Galactic birthrates.  For each object with an available distance 
estimate, we compute the maximum distance for detection from 
$D_{\rm max}=D_{\rm est} (F_\gamma/F_{\rm min})^{1/2}$,
where $D_{\rm est}$ comes from Table \ref{tab:dist}, the photon flux $F_{100}$ from Table \ref{tbl-spec}
and $F_{\rm min}$ from Figure \ref{SensitivitySkyMap}. We limit $D_{\rm max}$ to 15\,kpc, 
and compare $V$, the volume enclosed within the estimated source distance,
to $V_{\rm max}$, that enclosed within the maximum distance, for a Galactic disk with radius
10\,kpc and thickness 1\,kpc. 
%These are NOT true PSR population values.
%In principle we should use detailed pop models seperately for the young and MSP. 
If we assume a blind search threshold $2\times$ higher than that
for a folding search at a given sky position, the inferred
values of $\langle V/V_{\rm max}\rangle $ are 0.49, 0.59 and 0.55 for
the radio-selected young pulsars, millisecond pulsars and
gamma-ray-selected pulsars, respectively. 
%These are close to the expected value of 0.5 \citep{Schmidt1968}; 
These are close to 0.5, the value expected for a population uniformly
filling a given volume \citep{Schmidt1968}; the MSP value is somewhat high as our sample
includes three objects detected at $<5\sigma$. The value for the
gamma-ray-selected pulsars is also high but is controlled by 
PSR J2032+4127. If we exclude this object from the sample, 
we get $\langle V/V_{\rm max}\rangle = 0.5$ at an effective threshold of 
$3\times$ the ephemeris-folding value.

Although we do not attempt a full population synthesis here, if we 
assume that the pulsar characteristic ages are the true ages,
our sample can give rough estimates for local volume birthrates:
$8 \times 10^{-5}$ kpc$^{-3}$ yr$^{-1}$ (young radio-selected),
$4 \times 10^{-5}$ kpc$^{-3}$ yr$^{-1}$ (young gamma-ray-selected, $2\times$ threshold)
and $2 \times 10^{-8}$ kpc$^{-3}$ yr$^{-1}$ (MSP). Note that
only half of the gamma-ray-selected objects have distance estimates.
If we assume that the set without distance information has comparable
luminosity, the gamma-ray-selected birthrate is $\sim2\times$ larger. 
These estimates retain appreciable uncertainty; for example if
the effective blind search detection threshold is $3\times$ that for folding, the
inferred gamma-ray-selected birthrate increases by an additional $\sim 65\%$. 
If we extrapolate these local birthrates to a full disk with an effective radius 
of 10\,kpc we get $1/100$ yr (radio-selected young pulsars), 
$1/100$ yr (gamma-ray-selected pulsars) 
and $1/(6 \times 10^5$ yr) (radio-selected MSPs).  
Normally in estimating radio pulsar birthrates one would
correct for the radio beaming fraction. However if young gamma-ray-selected
pulsars are simply similar objects viewed outside of the radio beam, this
would result in double-counting. In any case one infers a total
Galactic birthrate for energetic pulsars of $\sim 1/50$ yr,
with gamma-ray-selected objects representing half.
This represents a large fraction of the estimated Galactic supernova
rate, so clearly more careful population syntheses will be
needed to see if these numbers are compatible.

\subsection{Trends in Light Curves and Other Observables}

The pulse shape properties can also help us to probe the geometry and physics of
the emission region. The great majority of the pulsars show two dominant,
relatively sharp peaks, suggesting that we are seeing caustics from the edge
of a hollow cone.  When a single peak is seen, it tends to be broader, suggesting
a tangential cut through an emission cone.  This picture is realized in the OG
and the high altitude portion of the SG models.

For the radio-emitting pulsars, we can compare the
phase lag between the radio and first gamma-ray peak $\delta$ with the separation
of the two gamma-ray peaks $\Delta$. As first pointed out in \citet{Romani1995}, 
these should be correlated in outer magnetosphere models --- this is
indeed seen (Figure \ref{RadioSep}). The distribution can be compared with predictions of
the TPC and OG models shown in \citet{Watters09}. The $\delta - \Delta$ distribution
and in particular the presence of $\Delta \sim 0.2-0.3$ values appear
to favor the OG picture. However, there are a greater number having $\Delta \sim 0.4-0.5$,
which favors TPC models. In Figure \ref{DeltavsEdot} we show the peak
separation as a function of pulsar spin-down luminosity --- the $\Delta$ distribution appears
to be bimodal, with no strong dependence on pulsar $\dot E$ (or age). 
A full comparison will require detailed population models,
which are being created. It may also be hoped that the precise distribution of measured
values can help probe details of the emission geometry. In particular, whenever
we have external constraints on the viewing angle $\zeta$ (typically from X-ray
images of the PWN) or magnetic inclination $\alpha$ (occasionally measured from
radio polarization), then the observed values of $\delta$ and $\Delta$ become
a powerful probe of the precise location of the emission sheet within the
magnetosphere. This can be sensitive to the field perturbations
from magnetospheric currents and hence can probe the global electrodynamics
of the pulsar magnetosphere.

If one examines the energy dependence of the light curves of
both the radio-selected and gamma-ray-selected pulsars, a decrease in the
P1/P2 ratio with increasing energy seems to be a common feature.
However, the P1/P2 ratio evolution does not occur for all pulsars, notably
J0633+0632, J1028$-$5819, J1124$-$5916, J1813$-$1246, J1826$-$1256,
J1836+5925, J2021+3651, and J2238+59.  Most of these pulsars have two peaks with phase 
separation of $\sim$0.5 and little or no bridge emission between the peaks.  
Perhaps the lack of P1/P2 energy evolution is connected with
an overall symmetry of the light curve.

The LAT pulsar sample also shows evidence of trends in other observables
that may offer additional clues to the pulsar physics. While the detected 
objects have a wide range of surface magnetic fields, their inferred light cylinder
magnetic fields $B_{\rm LC}$ are uniformly relatively large ($\ga 10^3$ G). Indeed, 
the LAT-detected MSPs are those with the highest light cylinder fields with
values very similar to those of the detected normal pulsars.  Comparison of the 
spectral cut-off $E_{\rm cutoff}$ with surface magnetic field shows no significant
correlation. This evidence argues against classical low altitude polar cap
models supported by $\gamma$-B cascades.  However, there is a weak
correlation of $E_{\rm cutoff}$ with $B_{\rm LC}$, as shown in Figure \ref{EcvsBlc}.  It is
interesting that the values of $E_{\rm cutoff}$ have a range of only about a decade,
from 1 to 10 GeV, and that all the different types of pulsars seem to follow
the same correlation.  This strongly implies that the gamma-ray emission
originates in similar locations in the magnetosphere relative to the light 
cylinder.  Such a correlation of $E_{\rm cutoff}$ with $B_{\rm LC}$ is
actually expected in all outer magnetosphere models where the gamma-ray 
emission primarily comes from curvature radiation of electrons whose acceleration
is balanced by radiation losses.  In this case, 
\begin{equation}
E_{\rm cutoff} = 0.32 \lambda_c \left(E_{\parallel}\over e \right)^{3/4} \rho_c^{1/2}
\end{equation}
in $m_{{\rm e}}c^2$, where $\lambda_c$ is the electron Compton wavelength, $E_{\parallel}$ is the 
electric field that accelerates particles parallel to the magnetic field and
$\rho_c$ is the magnetic field radius of curvature.  In both SG \citep{Muslimov04} and OG \citep{Zhang04,Hirotani08} models, $E_{\parallel}
\propto B_{\rm LC}\,w^2$, where $w$ is the gap width.  All these models
give values of $E_{\rm cutoff}$ that are roughly consistent with those measured for the
LAT pulsars.  Although $\rho_c \sim R_{\rm LC}$, 
the gap widths are expected to decrease with increasing $B_{\rm LC}$, so that
$E_{\rm cutoff}$ is predicted to be only weakly dependent on $B_{\rm LC}$ in most outer magnetosphere
models, as observed.  
%Spectral trends

%
% MGB addition 08/06/09
%

The detection of pulsed flux at $E_{\rm max}>$ a few GeV provides 
additional, physical motivation for high altitude emission, since one expects
strong (hyper-exponential) attenuation from $\gamma$-B$\longrightarrow e^+e^-$
absorption in the high magnetic fields at low altitudes of a near-surface
polar gap. For a polar cap model with spin period $P$ and surface field $10^{12}B_{12}$G, 
Equation 1 of \citet{Baring2004AdSpR..33..552B} gives
$r\gtrsim (E_{\rm max} B_{12}/1.76\hbox{GeV})^{2/7}\, P^{-1/7}\, 10^6$cm.
While the largest minimum $r$ is derived from the 25 GeV detection of the Crab 
by MAGIC \citep{Albert2008ApJ...674.1037A}, significant minimum altitudes of 2 to 3 stellar radii 
are found for many LAT pulsars with large $B_{12}$ and high $E_{\rm cutoff}$, assuming 
maximum energies   set at around $2.5E_{\rm cutoff}$. Such altitudes are inconsistent
with the $r \la 1-2$ stellar radii of polar cap models, hence implicating outer 
magnetosphere (e.g. TPC or OG) models where the bulk of the emission occurs at tens to hundreds
of stellar radii.

% {\bf ADD THESE REFERENCES:
% \obeylines
% Albert, J. et al., 2008, ApJ, 674, 1037
% Baring, M.G. 2004, Adv. Sp. Res., 33, 552
% }
%%%

In Figure \ref{IndexvsEdot}, we see a general trend for the young pulsars to show a
softer spectrum at large ${\dot E}$, although there is a great deal 
of scatter; a similar trend was noted in
\citet{Thompson99}.  This may be indicative of higher pair multiplicity,
which would steepen the spectrum for the more energetic pulsars, 
either by steepening the spectrum of the curvature radiation-generating 
primary electrons \citep{Romani96} or by inclusion of 
an additional soft spectral component associated with robust pair formation
\citep{Takata07,Harding08}.  In either case,
one would expect steepening from the simple monoenergetic
curvature radiation spectrum $\Gamma=2/3$ for the higher $\dot E$ pulsars. 
Interestingly, the MSPs do not extend the trend to
lower ${\dot E}$. Of course EGRET (and now the LAT) find strong variations
of photon index with phase for the brighter pulsars. A full understanding
of photon index trends will doubtless require phase-resolved modeling.

% I say skip this stuff -- its relevance is not clear.
%
%Weltevrede and Johnston 2008 argue that high linear polarization in the radio
%strongly correlates with strong $\gamma$-ray emission. We plot in figure (x)
%the scatter between the L-band linear polarization fraction and the inferred
%$\gamma$-ray luminosity.  It is interesting that the transition from low to high linear
%polarization roughly coincides with the break at $\dot E \sim 10^{35}\,\rm erg s^{-1}$
%in the $L_\gamma ({\dot E})$ plot (Fig X).  This confirms that the high $\dot E$ $\gamma$-ray pulsars have
%high linear polarization, suggesting that their radio pulses may be somewhat
%different to those of less energetic pulsars. One likely possibility is
%that the radio emission arises from relatively high altitude cones (Kasteg.
%\& Johnston 2007), with high linear polarization. Another high-energy-radio
%connection has been noted by Romani \& Johnston, who find that `Giant pulses'
%seen from the Crab and a handful of other radio pulsars tends to occur at
%phases showing hard power-law X-ray pulse components. We find that this persists
%to the $\gamma$-ray band, where the pulsed emission
%[Actually, this case would be best made with MPSRs 1937 and B1821, neither of which
%made the final catalog...]

% \end{document}

\section{Conclusion}
% \section{6.0 Conclusion}
%
% Roger \& Alice - 30 June 2009
% RWR - 29 July  2009
% ECF - 1 Aug 2009, 2 Aug 2009
% ECF - 16 Sep 2009 (Updates to address Fernando2)
%
% Incorporated comments from F. Camilo - 23 Aug 2009
% Comments from FC not yet addressed are:
% 
%
% - l733: "we are detecting the millisecond pulsars with the highest
% spin-down flux at Earth" - where have you mentioned this before? (if
% you're using it in the conclusions, it must have been mentioned before?  perhaps the Edot/d^2 figure I suggested would show this?)
% _______________
% - "and some of the remaining unidentified sources also contain
%spin-powered pulsars" - do you mean remaining EGRET or Fermi unidentified
% sources (from 3EG, or from BSL?)  It is a virtual certainty that
%many
% Fermi unidentified sources contain spin-powered pulsars (I'd say most
% of the 50 unidentified, non-variable ones among the 205 in the
%BSL to
% start with).  So can you make this statement clearer/stronger?
% _______________
% - l756: "The photon emission also occupies a
%large fraction of the
% spin-down luminosity" - do you mean eta is large?  The sentence as
% written can appear awkward, I think.
% 
%"... increasing as the pulsars approach Edot ~1e33-34 erg/s" - where
% was this shown to be the case?
% _______________
% 
% 
% \begin{document}

\begin{figure}[!ht]
\centering
\includegraphics[width=1.0\textwidth]{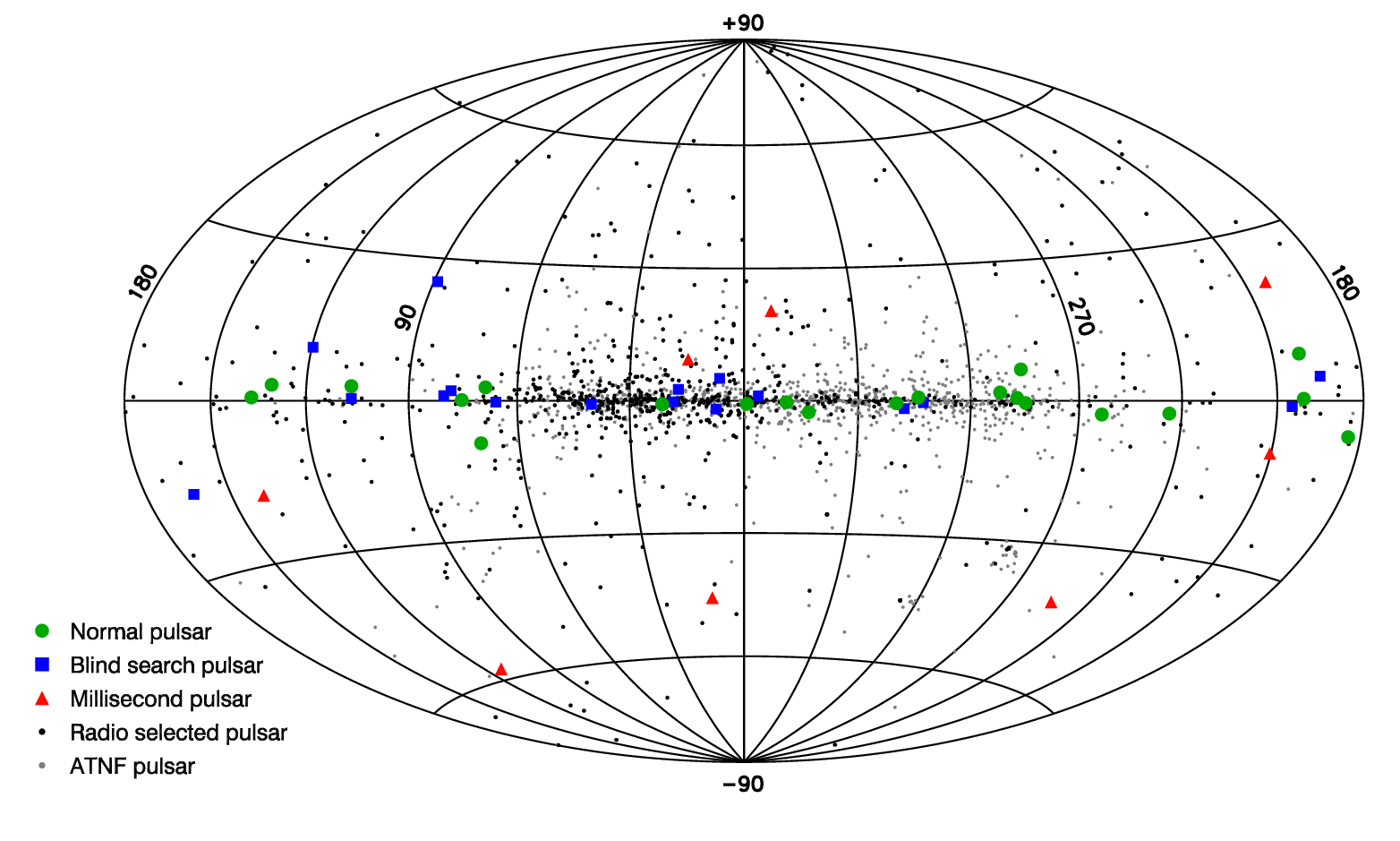}
\caption{Pulsar sky map in Galactic coordinates. 
Blue squares: gamma-ray-selected pulsars. 
Red triangles: millisecond gamma-ray pulsars. 
Green circles: all other radio loud gamma-ray pulsars. 
Black dots: Pulsars for which gamma-ray pulsation searches were conducted using rotational ephemerides.
Gray dots: Known pulsars which were not searched for pulsations.
\label{SkyMap}}
\end{figure}

\begin{figure}[!ht]
\includegraphics[width=0.8\textwidth]{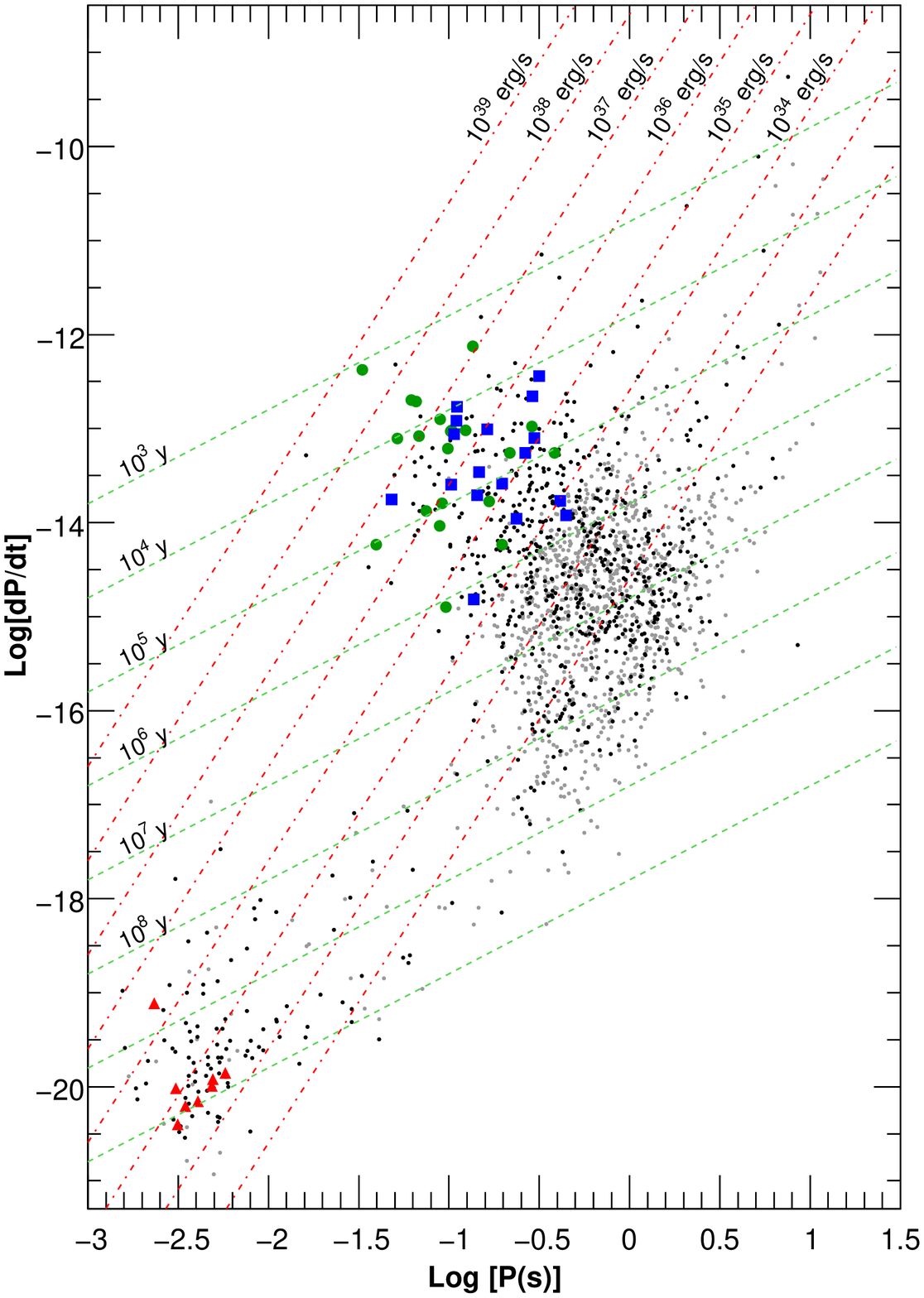}
\caption{ $P-\dot P$ diagram. 
Dashed lines: characteristic age $\tau_{\rm c}$. Dot-dashed lines: rotational energy loss rate $\dot E$. 
Blue squares: gamma-ray-selected pulsars. 
Red triangles: millisecond gamma-ray pulsars. Green circles: all other radio loud gamma-ray pulsars. 
Black dots: Pulsars for which gamma-ray pulsation searches were conducted using rotational ephemerides.
Gray dots: Known pulsars which were not searched for pulsations.
\label{PPdot}}
\end{figure}

\begin{figure}[!ht]
\includegraphics[width=1.0\textwidth]{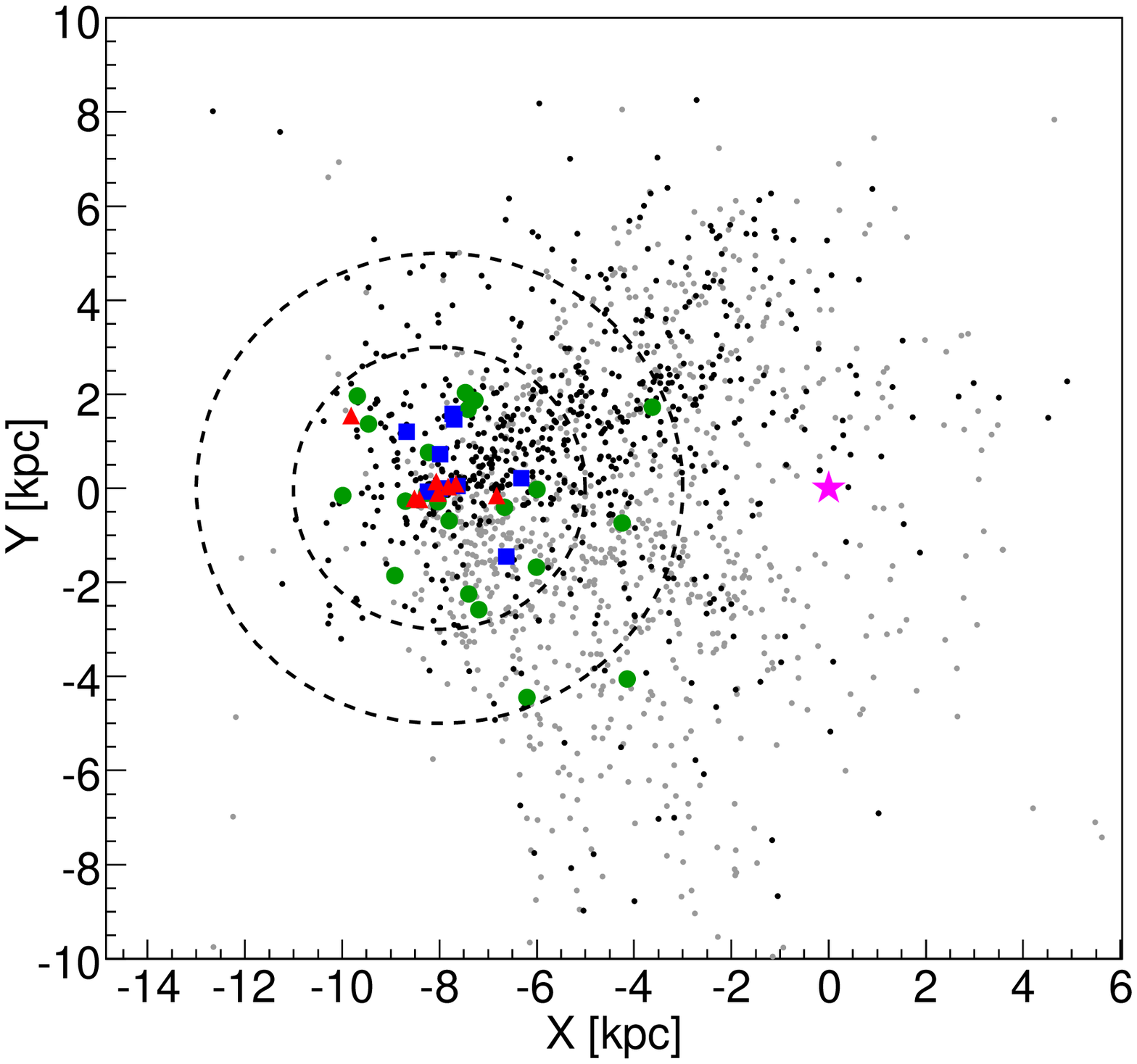}
\caption{Galactic plane pulsar distribution (polar view). The star represents the Galactic center. 
The two circles centered at the Earth's position have radii of 3 kpc and 5 kpc.  
For pulsars with different possible distances, the nearer values from Table \ref{tab:dist} are used. 
Note that the millisecond pulsars (MSPs), while having a significantly lower $\dot E$ than the other pulsars 
(see Figure \ref{IndexvsEdot}), are detectable due to their close proximity. The one exception (PSR J0218+4232) also 
exhibits a significantly higher $\dot E$ than the other MSPs.
Blue squares: gamma-ray-selected pulsars.
Red triangles: millisecond gamma-ray pulsars. Green circles: all other radio loud gamma-ray pulsars. 
Black dots: Pulsars for which gamma-ray pulsation searches were conducted using rotational ephemerides.
Gray dots: Known pulsars which were not searched for pulsations.
%The farthest pulsars is J0248+6021 for which there is only a lower limit for its distance of 10kpc.
\label{SkyProjection}}
\end{figure}

\begin{figure}[!ht]
\includegraphics[width=0.6\textwidth]{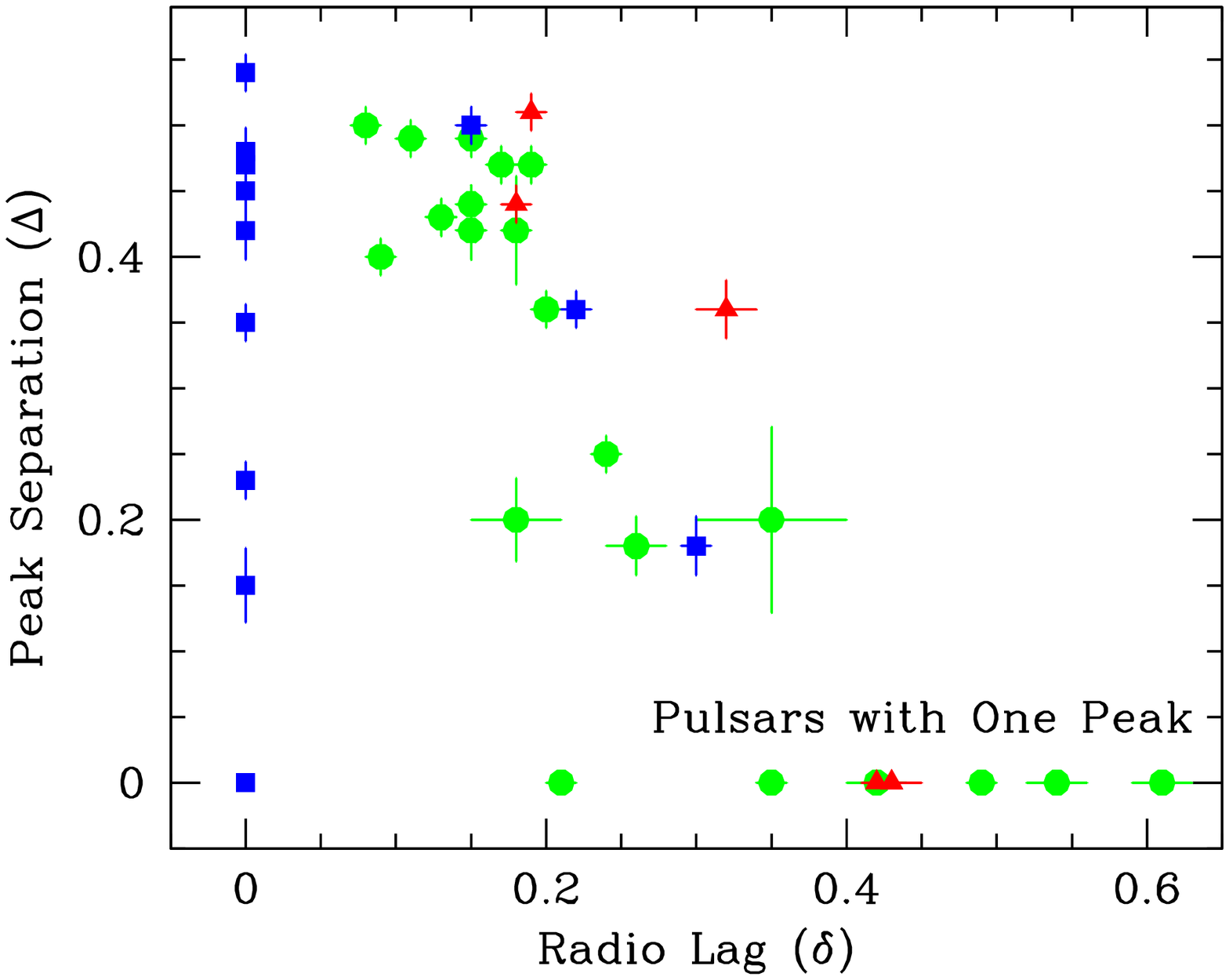}
\caption{Phase difference $\Delta$ between the gamma-ray peaks, versus
the phase lag $\delta$ between the main radio peak and the nearest gamma-ray
peak. Pulsars without a radio detection are plotted with $\delta=0$. With
present light curves we cannot generally measure $\Delta < 0.15$; objects
classified as single-peaked are plotted with $\Delta$=0. Two such
objects, both MSPs, are off the plot at $\delta>0.8$. 
Blue squares: gamma-ray-selected pulsars. 
Red triangles: millisecond gamma-ray pulsars. 
Green circles: all other radio loud gamma-ray pulsars. The plot has been corrected for the
updated value of $\delta$ for PSR J1124-5916, as per an Erratum sent to the ApJ (December 2010).
\label{RadioSep}}
\end{figure}

\begin{figure}[!ht]
\includegraphics[width=0.6\textwidth]{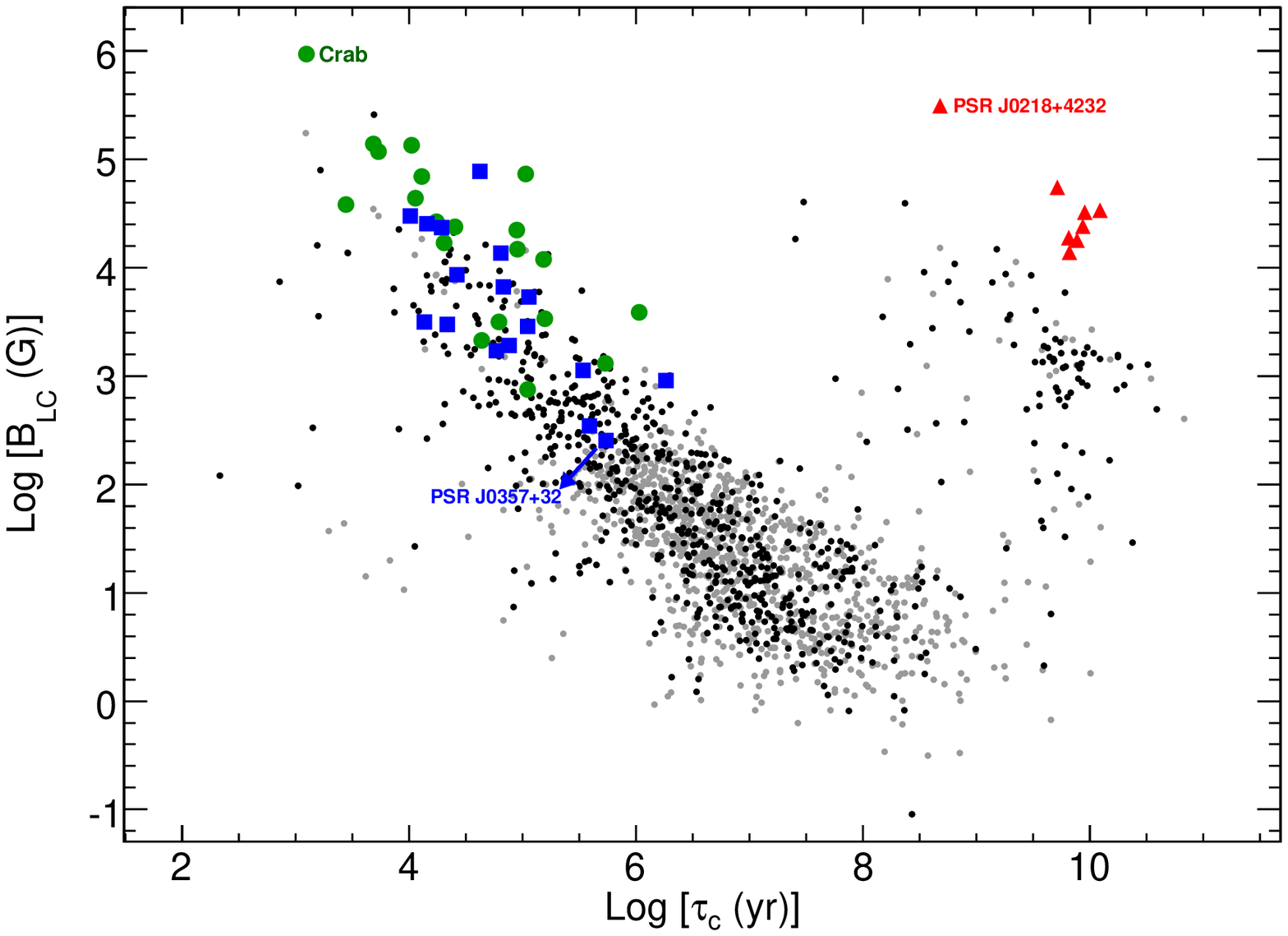}
\caption{Magnetic field strength at the light cylinder $B_{\rm LC}$ versus pulsar characteristic age $\tau_{\rm c}$.
Blue squares: gamma-ray-selected pulsars.
Red triangles: millisecond gamma-ray pulsars. Green circles: all other radio loud gamma-ray pulsars. 
Black dots: Pulsars for which gamma-ray pulsation searches were conducted using rotational ephemerides.
Gray dots: Known pulsars which were not searched for pulsations.
\label{BlcvsAge}}
\end{figure}

\begin{figure}[!ht]
\centering
\includegraphics[width=1.0\textwidth]{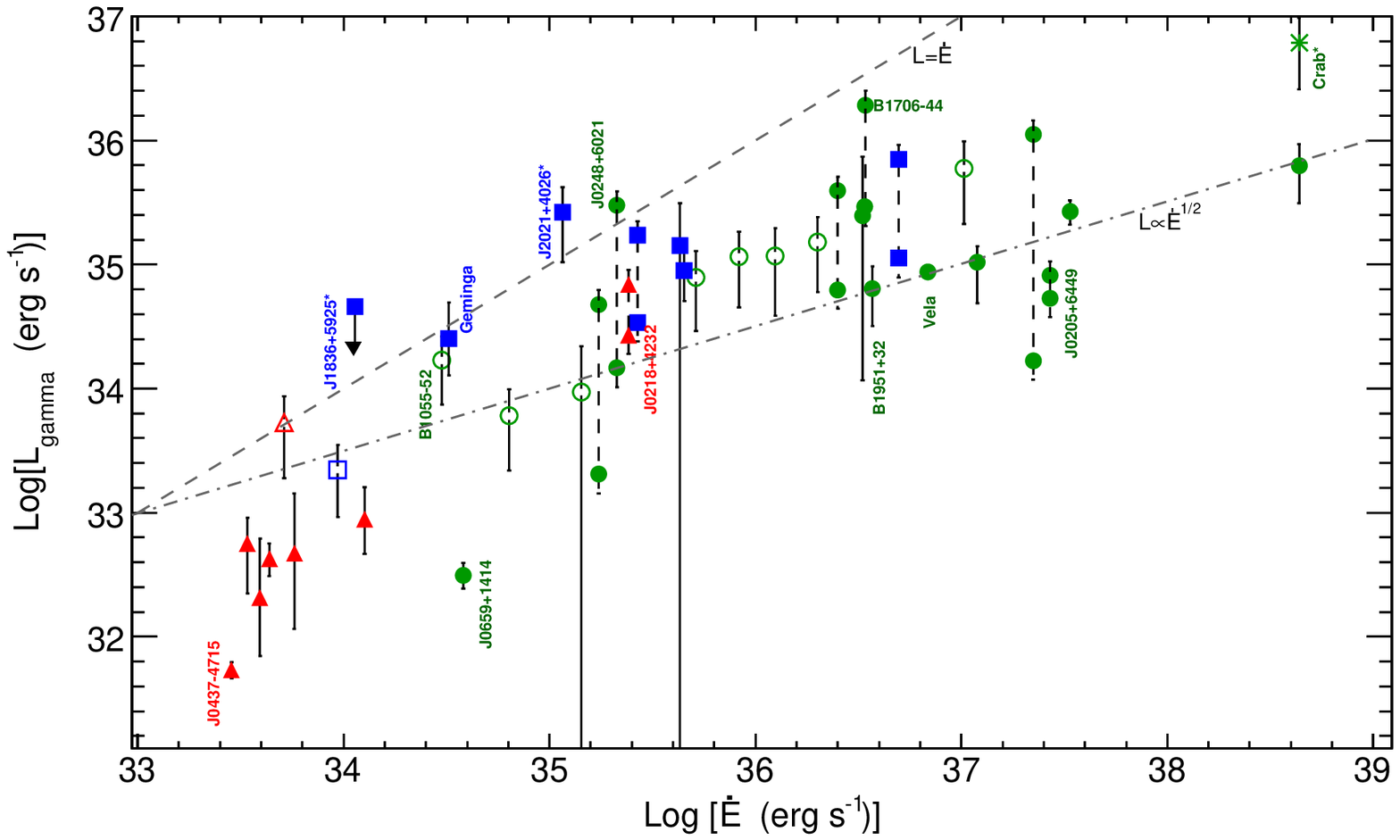}
\caption{Gamma-ray luminosity $L_\gamma$ versus the rotational energy loss rate $\dot E$. 
Dashed line: $L_\gamma$ equal to $\dot E$. 
Dot-dashed line: $L_\gamma$ proportional to the square root of $\dot E$. 
$L_\gamma$ is calculated using a beam correction factor $f_\Omega=1$ for all pulsars 
and the integral energy flux $G_{100}$ from the on-pulse spectral analysis (see Section 2.2),
except for PSRs J1836+5925 and J2021+4026 which use the total background-corrected phase-averaged
flux, including a relatively bright unpulsed component (see Section 2.2). 
For the Crab we also plot the total high energy luminosity, $L_{\rm tot} = L_X + L_\gamma$, indicated by *.
Several notable pulsars have been labeled.
Blue squares: gamma-ray-selected pulsars.
Red triangles: millisecond gamma-ray pulsars. Green circles: all other radio loud gamma-ray pulsars. 
Unfilled markers indicate pulsars for which only a DM-based distance estimate is available (see Table \ref{tab:dist}).
Pulsars with two distance estimates have two markers connected with dashed error bars.
\label{LEdot}}
\end{figure}

\begin{figure}[!ht]
\includegraphics[width=0.5\textwidth]{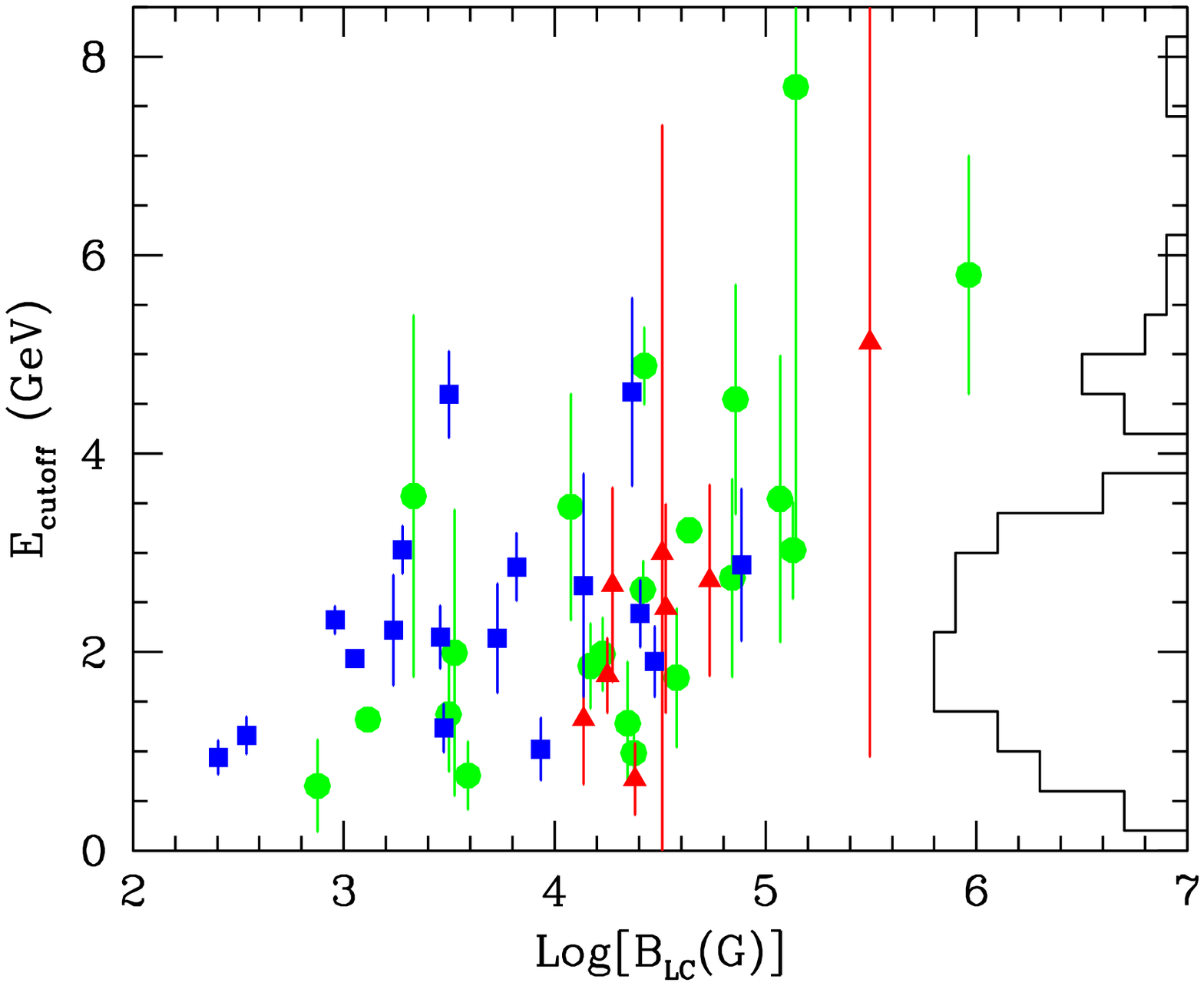}
\caption{Value of the exponential cutoff $E_{\rm cutoff}$ versus the magnetic field at the light cylinder, $B_{\rm LC}$.
The statistical uncertainties on $E_{\rm cutoff}$ are shown.
An additional systematic bias of ($+20\%,\,-10\%$) may affect $E_{\rm cutoff}$ (see text).
The histogram of $E_{\rm cutoff}$ values is projected along the right-hand axis.
Blue squares: gamma-ray-selected pulsars.
Red triangles: millisecond gamma-ray pulsars. Green circles: all other radio loud gamma-ray pulsars. 
\label{EcvsBlc}}
\end{figure}

\begin{figure}[!ht]
\includegraphics[width=0.5\textwidth]{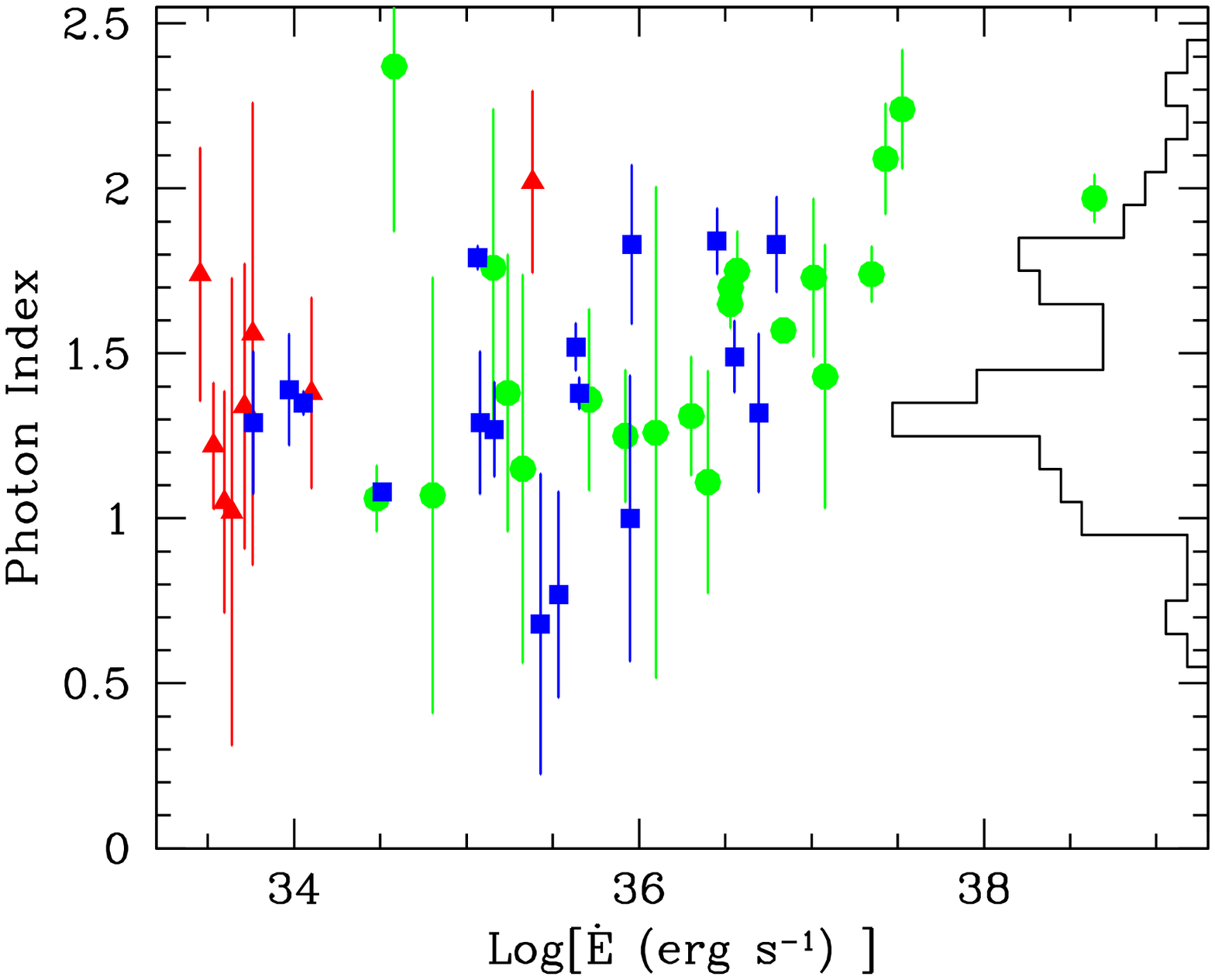}
\caption{Photon index $\Gamma$ versus the rotational energy loss rate, $\dot E$. 
For $\Gamma$, the statistical uncertainties combined
with the systematic uncertainties due to the diffuse emission model are shown. 
An additional systematic bias of ($+0.3,\,-0.1$) affects $\Gamma$ (see text).
The histogram of the photon indices is projected along the right-hand axis.
Blue squares: gamma-ray-selected pulsars.
Red triangles: millisecond gamma-ray pulsars. Green circles: all other radio loud gamma-ray pulsars. 
\label{IndexvsEdot}}
\end{figure}

\begin{figure}[!ht]
\includegraphics[width=0.5\textwidth]{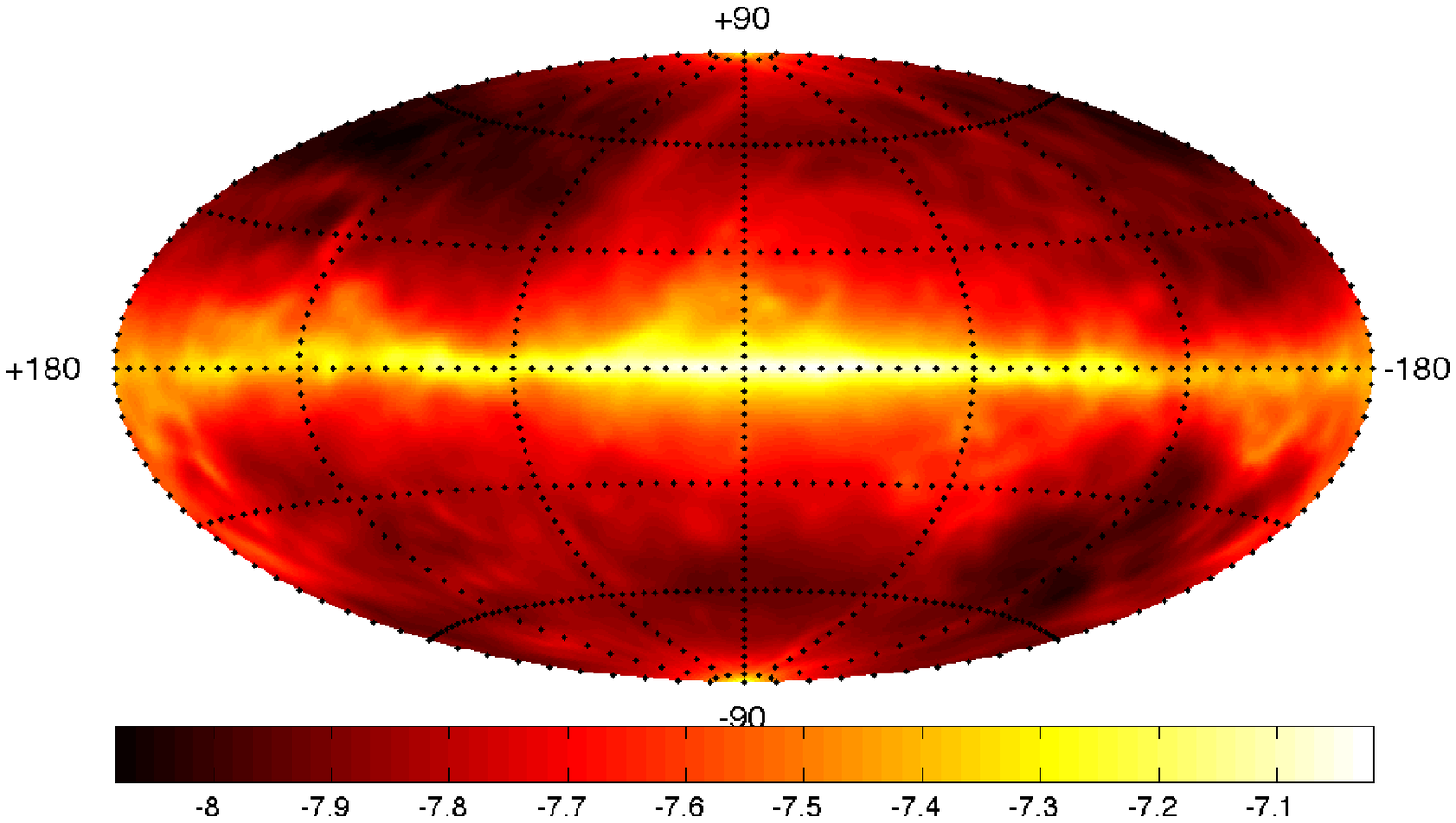}
\caption{Aitoff projection sky map of the $5\sigma$ sensitivity in units of logarithmic {\bf photon}
flux (Log($L_\gamma$) ph cm$^{-2}$\,s$^{-1}$) for six months of \textit{Fermi} LAT sky-survey
data. The sensitivity analysis uses the model of the diffuse gamma-ray background described 
in the text (Section 4), and pulsar spectra with differential photon indices of $\Gamma = 1.4$ 
with an exponential cutoff energy of $E_{\rm cutoff} = 2.2$ GeV. 
\label{SensitivitySkyMap}}
\end{figure}

\begin{figure}[!ht]
\includegraphics[width=0.5\textwidth]{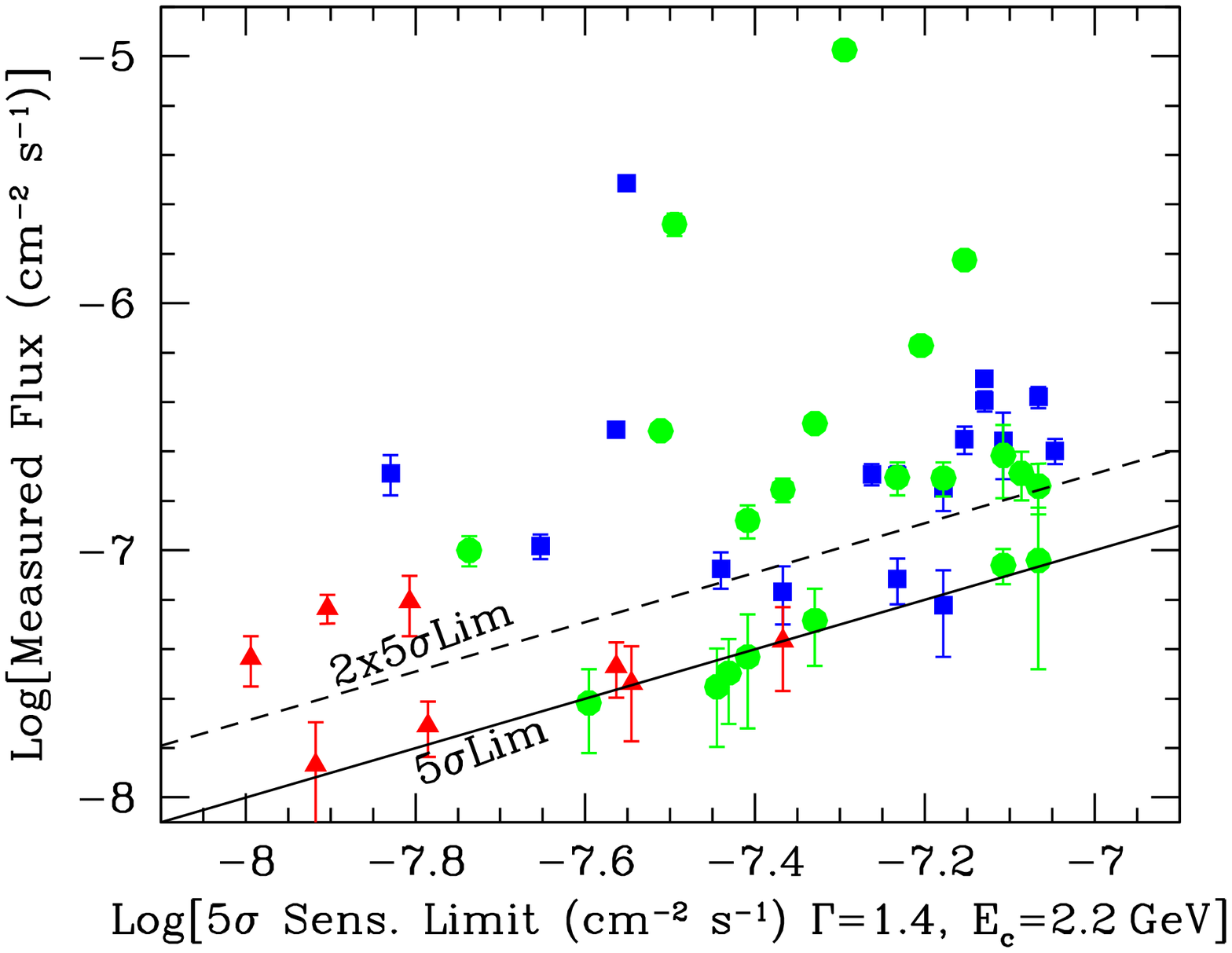}
\caption{Measured integral photon flux above 100 MeV, $F_{100}$, versus the $5\sigma$ flux 
sensitivity described in Figure \ref{SensitivitySkyMap}.  
For $F_{100}$, the statistical uncertainties combined
with the systematic uncertainties due to the diffuse emission model are shown. 
An additional systematic bias of ($+30\%,\,-10\%$) affects $F_{100}$ (see text).
The effective blind search
sensitivity is comparable to the $2\times 5\sigma$ line, although a few pulsars are 
discovered at lower flux, presumably due to favorable pulse profiles, spectra or
local backgrounds.
Blue squares: gamma-ray-selected pulsars.
Red triangles: millisecond gamma-ray pulsars. Green circles: all other radio loud gamma-ray pulsars. 
\label{FluxVsSensitivity}}
\end{figure}

\begin{figure}[!ht]
\includegraphics[width=0.6\textwidth]{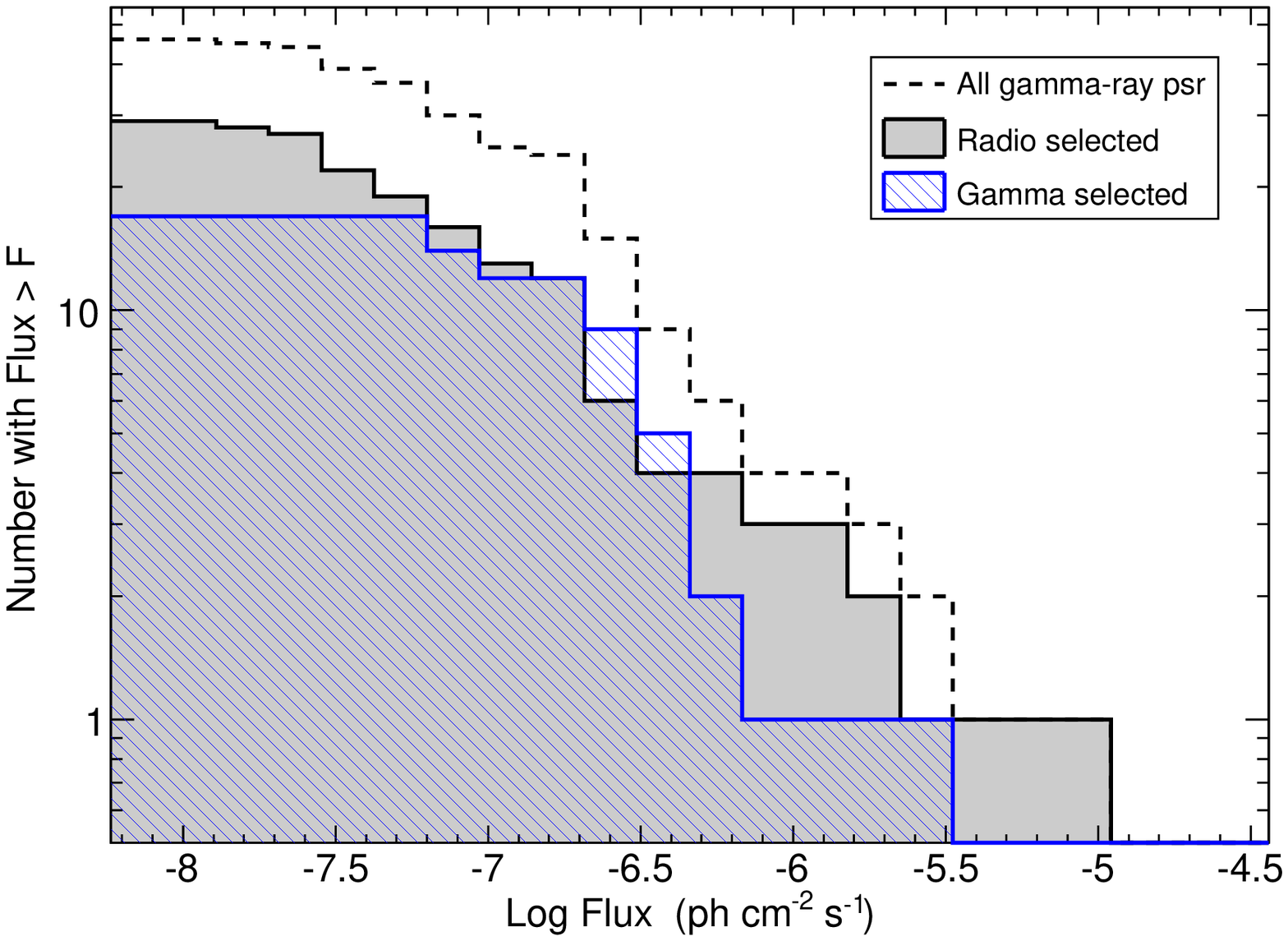}
\caption{Log\,N--Log\,S distribution as described in Section 4 for all the detected pulsars 
(black dashed line), the radio-selected gamma-ray pulsars including MSPs (grey histogram), 
and the gamma-ray-selected pulsars (blue hatched histogram).
\label{LogN-LogS}}
\end{figure}

\begin{figure}[!ht]
\includegraphics[width=0.7\textwidth]{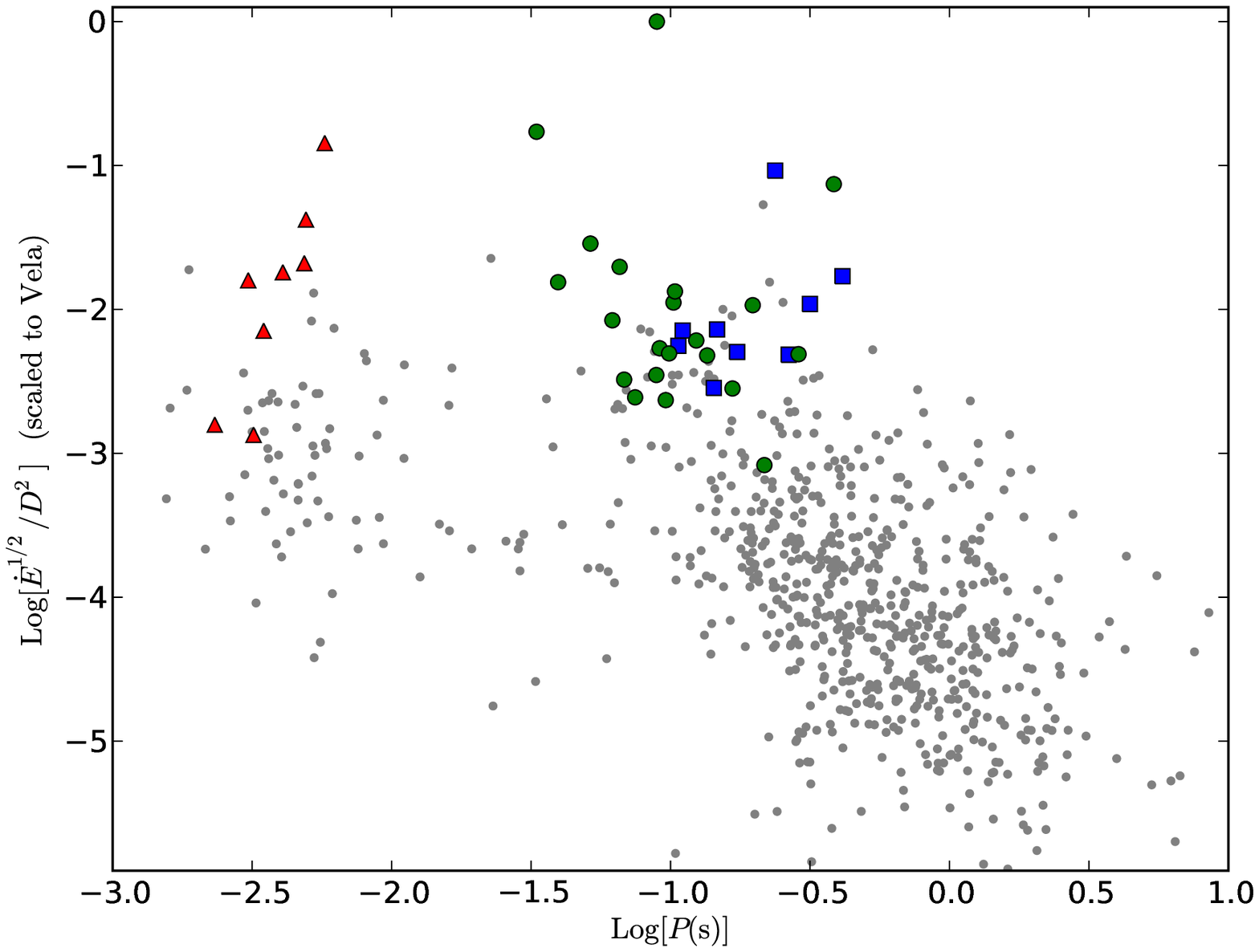}
\caption{Pulsar ``detectability'' metric (${\dot E}^{1/2}/d^2$, normalized to Vela)
% use this if we agree to normalize the y-axis!!!!  WE DO!!!
vs. spin period. Detected MSPs (red triangles) and young pulsars 
(radio-selected, green circles; gamma-ray-selected, blue squares) all have high values 
of this metric. For objects with a distance range in Table \ref{tab:dist}, we use here the 
geometric mean of the maximum and minimum values. Searched, but presently 
undetected objects (gray dots) are plotted using DM-derived distances. For 
the possible causes of non-detection see Section 5.1.
\label{DetectMetric}}
\end{figure}

\begin{figure}[!ht]
\includegraphics[width=0.7\textwidth]{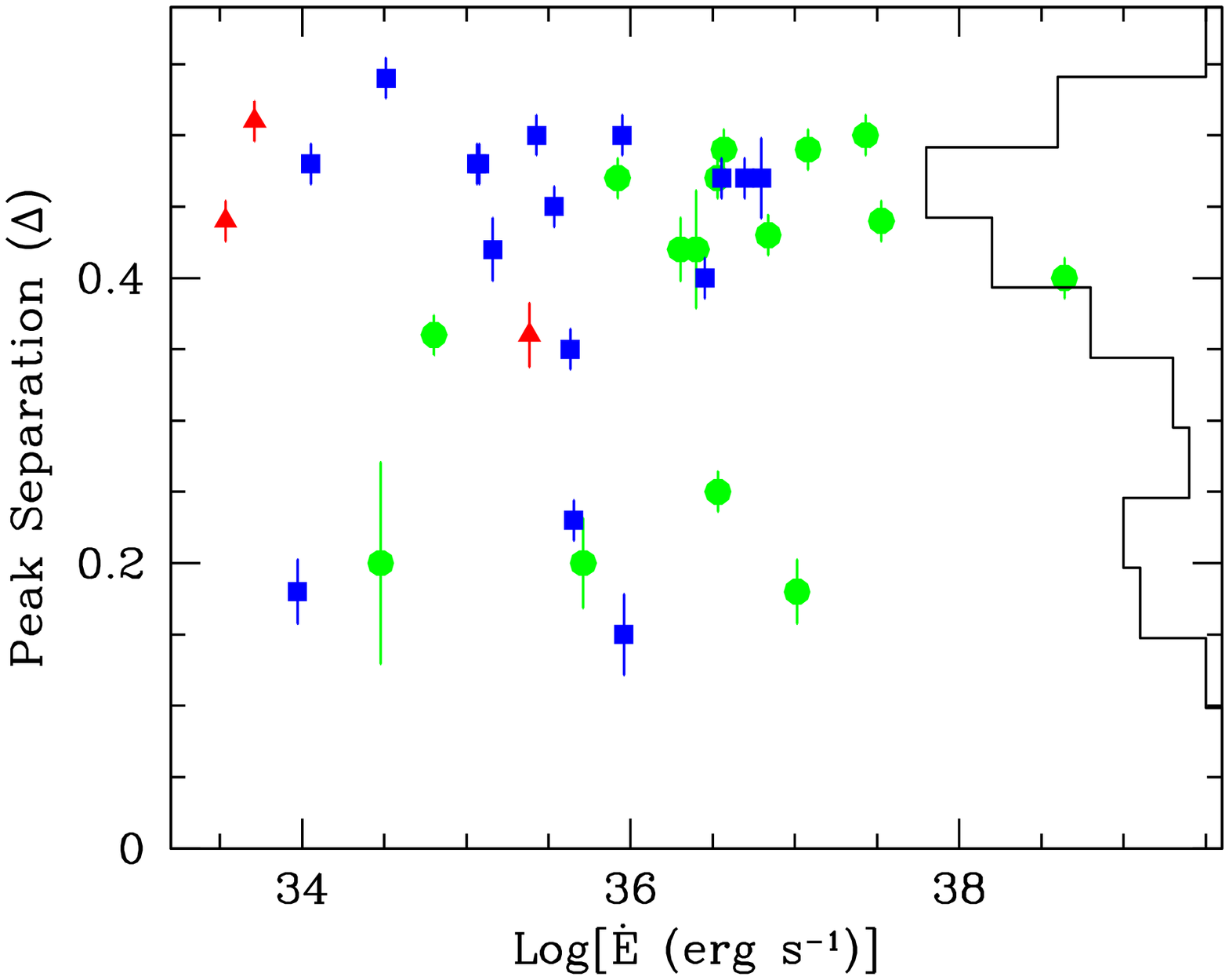}
\caption{Separations $\Delta$ between the gamma-ray peaks,
for those pulsars with two identified peaks, versus the spin-down power $\dot E$.
The histogram of peak separations is projected along the right-hand axis.
Blue squares: gamma-ray-selected pulsars.
Red triangles: millisecond gamma-ray pulsars. Green circles: all other radio loud gamma-ray pulsars. 
\label{DeltavsEdot}}
\end{figure}

The new gamma-ray pulsar populations established by early LAT observations
show that we are detecting many nearby young pulsars.
In addition we are detecting the millisecond pulsars with the highest spin-down flux
at Earth.  Thus we see that the LAT is providing a new, local, but relatively 
unbiased view of the energetic pulsar population (see Figures \ref{SkyMap}, \ref{PPdot}, and \ref{SkyProjection}). 
These detections provide a new window into pulsar demographics and physics. 

We conclude that a large fraction of the local energetic pulsars are GeV
emitters. There is also a significant correlation with X-ray and TeV bright pulsar
wind nebulae. Conversely, we have now uncovered the pulsar origin of a large fraction 
of the bright unidentified Galactic EGRET sources, as proposed by several authors 
\citep{Kaaret96, Yadigaroglu1997}. We have also found plausible
pulsar counterparts for several previously detected TeV sources. In this sense
the ``mystery'' of the unidentified EGRET sources is largely solved. It is possible that
the two LAT-detected massive binaries (LSI $+61^\circ$ 303, LS 5059) and some of the remaining unidentified
sources also contain spin-powered pulsars. Thus we expect that the LAT pulsar population
will increase, with both the detection of binary gamma-ray pulsars and fainter
and more distant pulsars. 
	
The light curve and spectral evidence summarized above suggests that these
pulsars have high altitude emission zones whose fan-like beams scan over a large portion
of the celestial sphere. This means that they should provide a relatively unbiased census of
energetic neutron star formation. A rough estimate of the young gamma-ray pulsar birthrate
extrapolating from our local sample suggests a Galactic birthrate as high as $\sim 1/50$ yr,
a large fraction of the estimated Galactic supernova rate. Gamma-ray detectable MSPs 
in the Galactic field are born rarely, $\sim 1/6\times 10^5$ yr, but with their long lifetimes are
inferred to contribute comparably to the number of (in principle) detectable Galactic
gamma-ray pulsars.

The data also advance our understanding of emission zone physics. It is now clear
that the gamma-ray emission from the brightest pulsars arises largely in the outer 
magnetosphere. The photon emission also accounts for a large fraction of the spin-down luminosity,
increasing as the pulsars approach ${\dot E} \sim 10^{33-34}$ erg s$^{-1}$. While these wide, bright
beams are a boon for population studies, as noted above, they represent a challenge for
theorists trying to understand pulsar magnetospheres. Further LAT pulsar observations
and, in particular, the high quality, highly phase-resolved spectra now being obtained
for the brightest LAT pulsars will surely sharpen this challenge.

% \end{document}

\acknowledgments
The \textit{Fermi} LAT Collaboration acknowledges the generous
support of a number of agencies and institutes that
have supported the Fermi LAT Collaboration. These
include the National Aeronautics and Space Administration
and the Department of Energy in the United States,
the Commissariat \`a l'Energie Atomique and the Centre
National de la Recherche Scientifique / Institut National
de Physique Nucl\'eaire et de Physique des Particules
in France, the Agenzia Spaziale Italiana and the Istituto
Nazionale di Fisica Nucleare in Italy, the Ministry
of Education, Culture, Sports, Science and Technology
(MEXT), High Energy Accelerator Research Organization
(KEK) and Japan Aerospace Exploration Agency
(JAXA) in Japan, and the K. A. Wallenberg Foundation
and the Swedish National Space Board in Sweden.

Additional support for science analysis during the operations phase is gratefully
acknowledged from the Istituto Nazionale di Astrofisica in Italy and the 
Centre National d'\'Etudes Spatiales in France.

The Parkes radio telescope is part of the Australia Telescope which is funded by the Commonwealth
Government for operation as a National Facility managed by CSIRO.  The Green Bank Telescope is operated
by the National Radio Astronomy Observatory, a facility of the National Science Foundation operated
under cooperative agreement by Associated Universities, Inc. The Arecibo Observatory is part of the
National Astronomy and Ionosphere Center (NAIC), a national research center operated by Cornell
University under a cooperative agreement with the National Science Foundation. The Nan\c cay Radio
Observatory is operated by the Paris Observatory, associated with the French Centre National de la
Recherche Scientifique (CNRS). The Lovell Telescope is owned and operated by the University of
Manchester as part of the Jodrell Bank Centre for Astrophysics with support from the Science and
Technology Facilities Council of the United Kingdom. The Westerbork Synthesis Radio Telescope is
operated by Netherlands Foundation for Radio Astronomy, ASTRON.

%\begin{thebibliography}{}
%\input{Bibliography}
%\end{thebibliography}

\bibliographystyle{apj}
%\bibliography{Bibliography/Pulsar_Catalog_Intro_Refs}
\bibliography{Bibliography/Pulsar_Catalog_ALL_Refs}
%\bibliography{Bibliography/2-ObservationAnalysis}

% \table{tbl-char}
%
% Incorporated comments from F. Camilo - 22 Aug 2009
% Comments from FC not yet addressed are:
% 
%-  I don't know why the Pdot column is necessary - you give all the
% relevant derived parameters one might have wished for.
% 
% ECF - 15 Sep 2009 (updated to correct table caption)
% 
%%%%%%%%%%%%%%%%%%%%%%%%%%%%inizio variazioni

\clearpage
%\begin{landscape}
\tabletypesize{\scriptsize}
\begin{deluxetable}{llrrrrrrrr}
\tablewidth{0pt}
\tablecaption{Measured and intrinsic parameters of LAT-detected pulsars
\label{tbl-char}}

\tablehead{
\colhead{PSR} & \colhead{Type,} &  \colhead{$l$} & \colhead{$b$} & \colhead{$P$} & \colhead{$\dot P$} &\colhead{age $\tau_{{\rm c}}$} & \colhead{$\dot E$}&
\colhead{$B_{LC}$} & \colhead{$S_{1400}$}  \\
& \textit{Ref.} & \colhead{($^\circ$)} & \colhead{($^\circ$)} & \colhead{(ms)} & \colhead{($10^{-15}$)} & \colhead{(kyr)} & \colhead{($10^{34}$ erg s$^{-1}$)}
& \colhead{(kG)} & \colhead{(mJy)} 
}

\startdata
%PSR				LAT?			l		b			P			Pdot					age				Edot			B_LC			S1400													
%			  					(deg)		(deg)		(ms)			(10^-15)				(kyr)			(10^34 ergs/s)		(1000 G)			(mJy)															
J0007$+$7303	 &	   g  $^{a,b}$ &	       119.7   &	       10.5    &	   316\phd\phn     &	361\phd\phn	&	  14		    &		   45.2    &    	 3.1     &	$^1\,<0.1$\phn\phn		\\
J0030$+$0451	 &	   m  $^{c,d}$ &	       113.1   &	       $-$57.6   &	       4.9    &  10$\times 10^{-6}$   &	7.7$\times 10^{6}$ &		     0.3   &	  	17.8    &		  0.6\phn\phn		\\
J0205$+$6449	 &	   r  $^e$     &	       130.7   &	       3.1     &	     65.7    &	       	194\phd\phn	&	    5		    &	       2700\phd\phn &	115.9   &		  0.04\phn			\\
J0218$+$4232	 &	   mb $^d$     &	       139.5   &	       $-$17.5   &	       2.3    &  77$\times 10^{-6}$   &	0.5$\times 10^{6}$ &		  24\phd\phn &	313.1   &		  0.9\phn\phn		\\
J0248$+$6021	 &	   r  $^f$     &	       137.0   &	       0.4     &	   217\phd\phn     &	  55.1		&	  63		    &		  21\phd\phn &	3.1	  &		  9\phd\phn\phn\phn	\\
J0357$+$32	 &	   g  $^b$     &	       162.7   &	       $-$16.0   &	   444\phd\phn     &	  12.0		&	590		    &		    0.5     &	  	0.2	  &		\nodata			\\
J0437$-$4715	 &	   mb $^d$     &	       253.4   &	       $-$42.0   &	       5.8    &  14$\times 10^{-6}$   &	6.6$\times 10^{6}$ &		    0.3    &	  	13.7    &	       140\phd\phn\phn\phn	\\
J0534$+$2200	 &	   r  $^h$     &	       184.6   &	       $-$5.8    &	     33.1    &	      	423\phd\phn	&	     1		    &	    46100\phd\phn &	950.0   &            14\phd\phn\phn\phn	\\
J0613$-$0200	 &	   mb $^d$     &	       210.4   &	       $-$9.3    &	       3.1    &    9$\times 10^{-6}$    &	5.3$\times 10^{6}$ &		    1.3    &	 	 54.3    &		  1.4\phn\phn     	\\
J0631$+$1036	 &	   r  $^i$     &	       201.2   &	       0.5     &	   288\phd\phn     &	105\phd\phn	&	  44		    &		  17.3    &	  	2.1	  &		  0.8\phn\phn     	\\
J0633$+$0632 &	   g  $^b$     &	       205.0   &	       $-$1.0    &	   297\phd\phn     &	  79.5		&	  59		    &		  11.9    &	  	1.7	  &	   $^2\,<0.2$\phn\phn	\\
J0633$+$1746	 &	   g  $^h$     &	       195.1   &	       4.3     &	   237\phd\phn     &	  11.0		&	340 		    &		    3.3    &	  	1.1	  &	$<1$\phd\phn\phn\phn	\\
J0659$+$1414	 &	   r  $^i$     &	       201.1   &	       8.3     &	   385\phd\phn     &	  55.0		&	110 		    &		    3.8    &	  	0.7	  &		  3.7\phn\phn     	\\
J0742$-$2822	 &	   r  $^i$     &	       243.8   &	       $-$2.4    &	   167\phd\phn     &	  16.8		&	160 		    &		  14.3    &	  	3.3	  &		  15\phd\phn\phn\phn     \\
J0751$+$1807	 &	   mb $^d$     &	       202.7   &	       21.1    &	       3.5    &    6$\times 10^{-6}$    &	 8.0$\times 10^{6}$ &	    0.6    &	  	32.3    &		  3.2\phn\phn     	\\
J0835$-$4510	 &	   r  $^k$     &	       263.6   &	       $-$2.8    &	     89.3    &	       	124\phd\phn	&	  11		    &		688\phd\phn &	43.4    &		  1100\phd\phn\phn\phn     \\	
J1028$-$5819	 &	   r  $^l$     &	       285.1   &	       $-$0.5    &	     91.4    &	       	  16.1		&	  90  		    &		  83.2    &	  	14.6    &		  0.36\phn     		\\
J1048$-$5832	 &	   r  $^m$     &	       287.4   &	       0.6     &	   124\phd\phn     &	  96.3		&	  20		    &		201\phd\phn &	16.8    &		  6.5\phn\phn     	\\
J1057$-$5226	 &	   r  $^n$     &	       286.0   &	       6.6     &	   197\phd\phn     &	    5.8		&	540 		    &		    3.0    &	  	1.3	  &		  11\phd\phn\phn\phn	   \\
J1124$-$5916	 &	   r	       &	       292.0   &	       1.8     &	   135\phd\phn     &	747\phd\phn	&	    3		    &	       1190\phd\phn &	37.3    &		  0.08\phn     		\\
J1418$-$6058	 &	   g  $^b$     &	       313.3   &	       0.1     &	   111\phd\phn     &	170\phd\phn	&	  10		    &		495\phd\phn &	29.4    &	  $^{2,3}\,<0.06$\phn   	\\
J1420$-$6048	 &	   r  $^i$     &	       313.5   &	       0.2     &	     68.2    &	       	  83.2		&	  13		    &	       1000\phd\phn &	69.1    &		  0.9\phn\phn     	\\
J1459$-$60	 &	   g  $^b$     &	       317.9   &	       $-$1.8    &	   103\phd\phn     &	  25.5		&	  64		    &		  91.9    &	  	13.6    &	    $^2\,<0.2$\phn\phn	\\
J1509$-$5850	 &	   r  $^i$     &	       320.0   &	       $-$0.6    &	     88.9    &	       	    9.2		&	150 		    &		  51.5    &	  	11.8    &		  0.15\phn     		\\
J1614$-$2230	 &  	   mb $^d$     &	       352.5   &	       20.3    &	       3.2    &  4$\times 10^{-6}$      &	1.2$\times 10^{	6}$ &	    0.5    &	 	33.7    &		  \nodata	   		\\
J1709$-$4429	 &	   r  $^n$     &	       343.1   &	       $-$2.7    &	   102\phd\phn     &	  93.0		&	  18		    &		341\phd\phn &	26.4    &		  7.3\phn\phn     	\\
J1718$-$3825	 &	   r  $^i$     &	       349.0   &	       $-$0.4    &	     74.7    &	       	  13.2		&	  90		    &		125\phd\phn &	21.9    &		  1.3\phn\phn     	\\
J1732$-$31	 &	   g  $^b$     &	       356.2   &	       0.9     &	   197\phd\phn     &	  26.1		&	120		    &		  13.6    &	  	2.7	  &	    $^2\,<0.2$\phn\phn	\\
J1741$-$2054	 &	   g  $^{b,t}$ &	       6.4     &	       4.6     &	   414\phd\phn     &	  16.9		&	390		    &		    0.9     &	  	0.3	  & 	    $^2\,0.16$\phn     		\\
J1744$-$1134	 &	   m  $^d$     &	       14.8    &	       9.2     &	       4.1    &  7$\times 10^{-6}$      &	 9$\times 10^{6}$ &		    0.4    &	  	24.0    &		  3\phd\phn\phn\phn     	\\
J1747$-$2958	 &	   r  $^o$     &	       359.3   &	       $-$0.8    &	     98.8    &	      	  61.3		&	  26		    &		251\phd\phn &	23.5    &		  0.08\phn     		\\
J1809$-$2332	 &	   g  $^b$     &	       7.4     &	       $-$2.0    &	   147\phd\phn     &	  34.4		&	  68		    &		  43.0    &	  	6.5	  &	 $^{2,3}\,<0.06$\phn    	\\
J1813$-$1246	 &	   g  $^b$     &	       17.2    &	       2.4     &	     48.1    &	       	  17.6		&	  43		    &		626\phd\phn &	76.2    &	$^2\,<0.2$\phn\phn	   	\\
J1826$-$1256	 &	   g  $^b$     &	       18.5    &	       $-$0.4    &	   110\phd\phn     &	121\phd\phn	&	  14		    &		358\phd\phn &	25.2    &	$^{2,3}\,<0.06$\phn	   	\\
J1833$-$1034	 &	   r  $^o$     &	       21.5    &	       $-$0.9    &	     61.9    &	      	202\phd\phn	&	    5		    &	       3370\phd\phn &	137.3   &		  0.07\phn     		\\
J1836$+$5925	 &	   g  $^b$     &	       88.9    &	       25.0    &	   173\phd\phn     &	    1.5		&      1800		    &		    1.2     &	  	0.9	  &	     $^4\,<0.007$  		\\
J1907$+$06	 &	   g  $^{b,r}$   &	       40.2    &	       $-$0.9    &	   107\phd\phn     &	  87.3		&	  19		    &		284\phd\phn &	23.2    &		 $<0.02$\phn   		\\
J1952$+$3252	 &	   r  $^n$     &	       68.8    &	       2.8     &	     39.5    &	       	    5.8		&	110 		    &		374\phd\phn &	71.6    &		  1\phd\phn\phn\phn    	\\
J1958$+$2846	 &	   g  $^b$     &	       65.9    &	       $-$0.2    &	   290\phd\phn     &	222\phd\phn	&	  21		    &		  35.8    &	  	3.0	  &		  \nodata	   		\\
J2021$+$3651	 &	   r  $^p$     &	       75.2    &	       0.1     &	   104\phd\phn     &	  95.6		&	  17		    &		338\phd\phn &	26.0    &		  0.1\phn\phn     	\\
J2021$+$4026	 &	   g  $^{b,s}$   &	       78.2    &	       2.1     &	   265\phd\phn     &	  54.8		&	  77		    &		  11.6    &	  	1.9	  &	 	  \nodata	   		\\
J2032$+$4127	 &	   g  $^{b,t}$   &	       80.2    &	       1.0     &	   143\phd\phn     &	  19.6		&	120		    &		  26.3    &	  	5.3	  &	   $^2\,$ 0.24\phn     		\\
J2043$+$2740	 &	   r  $^q$     &	       70.6    &	       $-$9.2    &	     96.1    &	       	    1.3		&       1200	    &		    5.6    &	  	3.6	  &	   $^5\,$ 3\phd\phn\phn\phn	\\
J2124$-$3358	 &	   m  $^d$     &	       10.9    &	       $-$45.4   &	       4.9    &  12$\times 10^{-6}$    &	0.6$\times 10^{6}$ &		    0.4    &	  	18.8    &		  1.6\phn\phn     	\\
J2229$+$6114	 &	   r  $^m$     &	       106.6   &	       2.9     &	     51.6    		&	  78.3		&	    11		    &	      2250\phd\phn &	134.5   &		  0.25\phn     		\\
J2238$+$59	 &	   g  $^b$     &	       106.5   &	       0.5     &	   163\phd\phn     &	  98.6		&	    26		    &		  90.3    &	  	8.6	  &		  \nodata	   		\\

\enddata

\tablecomments{The first two columns are pulsar names and types: \textit{r} for 
radio-selected, \textit{g} for gamma-ray-selected, \textit{m} for MSPs, 
and \textit{b} for binary pulsars. The 3$^{rd}$ and 4$^{th}$ columns are 
Galactic coordinates for each pulsar. The 5$^{th}$ and 6$^{th}$ columns list the period ($P$) 
and its first derivative ($ \dot{P}$), corrected for the Shklovskii effect (see text).
Following are
the characteristic age $\tau_{\rm c}$ (column 7), the spin-down 
luminosity $\dot E$ (column 8), and the magnetic field at the light cylinder $B_{\rm LC}$ 
(column 9) . The last column is the radio flux density at 1400 MHz, or an upper 
limit when one is available. These values are taken from the ATNF database \citep{ATNFcatalog} except for 
the noted entries where: (1) \citep{Halpern2004}; (2) \citep{Camilo2009}; (3) \citep{Roberts2002}; 
(4) \citep{Halpern2007}; (5) \citep{Ray1996}. Note that PSR J1509$-$5850 should not be confused with PSR B1509$-$58 observed by \textit{CGRO}.}

\tablerefs{References to \textit{Fermi} LAT publications specific to these pulsars:
$a$ \citep{CTA1} ; 
$b$ \citep{BSP} ; 
$c$ \citep{PSR0030} ; 
$d$ \citep{MSP} ; 
$e$ \citep{LATPSR0205} ; 
$f$ \citep{Cognard2009} ;
$h$ \citep{FermiCrab} ;
$i$ \citep{LAT6pulsars} ;
$j$ \citep{Geminga1} ;
$k$ \citep{LATVELA} ;
$l$ \citep{PSR1028} ;
$m$ \citep{LATVelaLike} ;
$n$ \citep{LAT3EGRETpulsars} ;
$o$ \citep{LATPSR1833and1747} ;
$p$ \citep{PSR2021LAT} ;
$q$ \citep{LATPSR2043} ;
$r$ \citep{LATPSR1907} ;
$s$ \citep{gCyg} ;
$t$ \citep{Camilo2009}.\\
}

\end{deluxetable}

\normalsize
\clearpage
\noindent

% \table{tbl-det}
%
% Incorporated comments from F. Camilo - 22 Aug 2009
% ECF - 16 Sep 2009 (Updates to address Fernando2)
% 
% 
%%%%%%%%%%%%%%%%%%%%%%%%%%%%inizio variazioni

\clearpage

\begin{deluxetable}{lrrcc}
\tabletypesize{\scriptsize}
\tablecaption{Pulsation detection significances for LAT-detected pulsars
\label{tbl-det}}
\tablewidth{0pt}

\tablehead{
\colhead{PSR} & \colhead{$Z^2_2$ value} & \colhead{$H$ value} & \colhead{maxROI($^{\circ}$)} & \colhead{ObsID}
}
\startdata

J0007+7303   &   2072.1 &   2371.8 &  1.0 & L     \\
J0030+0451   &    121.1 &    362.7 &  1.0 & N      \\
J0205+6449   &     90.9 &    206.0 &  1.0 & G, J  \\
J0218+4232   &     24.7 &     22.5 &  1.0 & N, W    \\
J0248+6021   &     57.5 &     75.1 &  0.5 & N     \\
J0357+32     &    422.7 &    450.7 &  1.0 & L     \\
J0437$-$4715   &    126.9 &    153.6 &  1.0 & P      \\
J0534+2200   &   4397.8 &  15285.0 &  1.0 & N, J  \\
J0613$-$0200   &     93.6 &    139.9 &  1.0 & N      \\
J0631+1036   &     48.6 &     44.8 &  1.0 & N, J  \\
J0633+0632   &    230.2 &    573.3 &  1.0 & L     \\
J0633+1746   &  10053.6 &  20346.4 &  1.0 & L     \\
J0659+1414   &     80.5 &     99.0 &  1.0 & N, J \\
J0742$-$2822   &     38.9 &     44.9 &  1.0 & N, J  \\
J0751+1807   &     29.7 &     26.5 &  1.0 & N      \\
J0835$-$4510   &  26903.9 &  74716.7 &  1.0 & P     \\
J1028$-$5819   &    291.5 &    915.9 &  0.5 & P     \\
J1048$-$5832   &    208.5 &    634.0 &  1.0 & P     \\
J1057$-$5226   &   1668.9 &   1772.4 &  1.0 & P     \\
J1124$-$5916   &     93.5 &    179.9 &  1.0 & L     \\
J1418$-$6058   &    230.1 &    343.7 &  1.0 & L     \\
J1420$-$6048   &    104.7 &    114.4 &  1.0 & P     \\
J1459$-$60     &    148.2 &    159.3 &  1.0 & L     \\
J1509$-$5850   &     71.6 &     73.3 &  0.5 & P     \\
J1614$-$2230   &     36.2 &     69.5 &  0.5 & G      \\
J1709$-$4429   &   4680.1 &   5612.1 &  1.0 & P     \\
J1718$-$3825   &    111.9 &    109.8 &  0.5 & N, P  \\
J1732$-$31     &    141.2 &    279.6 &  1.0 & L     \\
J1741$-$2054   &    332.6 &    355.9 &  1.0 & L     \\
J1744$-$1134   &     28.4 &     38.1 &  1.0 & N      \\
J1747$-$2958   &     47.2 &     69.0 &  0.5 & G     \\
J1809$-$2332   &    589.3 &   1562.5 &  1.0 & L     \\
J1813$-$1246   &    140.0 &    162.0 &  1.0 & L     \\
J1826$-$1256   &    442.4 &    979.0 &  1.0 & L     \\
J1833$-$1034   &     35.2 &     87.6 &  1.0 & G     \\
J1836+5925   &    349.2 &    385.3 &  1.0 & L     \\
J1907+06     &    257.1 &    521.0 &  1.0 & L     \\
J1952+3252   &    464.8 &   1008.8 &  1.0 & J, N  \\
J1958+2846   &    146.9 &    233.1 &  1.0 & L     \\
J2021+3651   &   1433.5 &   4603.7 &  1.0 & G, A     \\
J2021+4026   &    222.0 &    275.8 &  1.0 & L     \\
J2032+4127   &    224.9 &    485.9 &  0.5 & L     \\
J2043+2740   &     28.2 &     38.2 &  1.0 & N, J  \\
J2124$-$3358   &     77.8 &     80.9 &  1.0 & N      \\
J2229+6114   &   1026.0 &   1237.4 &  1.0 & G, J     \\
J2238+59     &    135.8 &    373.0 &  1.0 & L     \\
\enddata

\tablecomments{Columns 2 and 3 list the $Z^2_2$ \citep{Buccheri1983} and $H$ \citep{DeJager1989} 
periodicity test values for $E>0.3$ GeV respectively. Detection of gamma-ray 
pulsations are claimed when the significance of the periodicity test exceeds 5$\sigma$, 
with the exceptions of J0218+4232, J0751+1807, J1744$-$1134, and J2043+2740 as described in Section 2.1.3.
A significance greater than 5$\sigma$ corresponds to $Z^2_2 > 36$ and $H > 42$ ;
greater than 7$\sigma$ corresponds to $Z^2_2 > 61$ ; and
greater than 10$\sigma$ corresponds to $Z^2_2 > 114$ (see Section 2.1.3).
Column 4 gives the maximum angular radius (maxROI) around the pulsar 
position within which gamma-ray events were searched for pulsations.
The final column indicates the observatories that provided ephemerides (see
Section 2.1.1 for details):
``A'' -- Arecibo telescope;
``G'' -- Green Bank Telescope;
``J'' -- Lovell telescope at Jodrell Bank;
``L'' -- Large Area Telescope;
``N'' -- Nan\c cay Radio Telescope;
``P'' -- Parkes radio telescope;
``W'' -- Westerbork Synthesis Radio Telescope.}

\end{deluxetable}
\clearpage
%\tiny
\noindent

\normalsize
%\clearpage

% \table{tbl-pulse}
%
% Incorporated comments from F. Camilo - 22 Aug 2009
% Comments from FC not yet addressed are:
% 
% Peak multiplicity:  how was this obtained?  I don't believe it's
% ever mentioned in text; by eye? by a fit of Gaussians, Lorentzians, what
% have you?  how did you decide 2 were enough, and 3 were not needed?  were
% you looking at the chi sq. of the fit, at residuals?  Somewhere in text
% this should be explained.
% _______________
% Radio lag uncertainties: have they all taken into account DM uncertainty?
% I'm not sure that for J1741-2054 has.  If there's a 'b' footnote for a
% couple of pulsars where it's not clear whether 1 or 2 peaks describe the
% profile, how about a reference for J1741, where Camilo et al. (2009b) claim
% 3 peaks?  At least mention that it's not clear whether it has 2 or 3, and
% give that ref - lest the unsuspecting reader come to believe on the basis
% of this table that all pulsars has 1 or 2 peaks.
% _______________
% As I already mentioned, why only one off-pulse range for all pulsars,
% when this doesn't describe many of them?
% 
% 
%%%%%%%%%%%%%%%%%%%%%%%%%%%%

\clearpage
\tabletypesize{\scriptsize}
\begin{deluxetable}{llcccc}
\tablewidth{0pt}
\tablecaption{Pulse shape parameters of LAT-detected pulsars
\label{tbl-pulse}}

\tablehead{
\colhead{PSR} & \colhead{Type$^a$} & \colhead{Peak} & \colhead{Radio lag} & \colhead{$\gamma$-ray peak separation} & \colhead{Off-pulse definition} \\
\colhead{}    &                    & multiplicity   & $\delta$            & $\Delta$                               & $\phi$ }
\startdata
 J0007$+$7303	  &	    g  &  2  &  ...		 &  0.23 $\pm$ 0.01 & 0.29 -- 0.87 \\
 J0030$+$0451	  &	    m  &  2  &  0.18 $\pm$ 0.01  &  0.44 $\pm$ 0.01 & 0.68 -- 0.12 \\
 J0205$+$6449	  &	    r  &  2  &  0.08 $\pm$ 0.01  &  0.50 $\pm$ 0.01 & 0.64 -- 0.02 \\
 J0218$+$4232	  &	    m &  2  &  0.32 $\pm$ 0.02  &  0.36 $\pm$ 0.02 & 0.84 -- 0.16 \\
 J0248$+$6021	  &	    r  &  1  &  0.35 $\pm$ 0.01  &  ... 	    & 0.71 -- 0.19 \\
 J0357$+$32	  	&	    g  &  1  &  ...		 &  ... 	    & 0.34 -- 0.86 \\
 J0437$-$4715	  	&	    m &  1  &  0.43 $\pm$ 0.02  &  ... 	    & 0.60 -- 0.20 \\
 J0534$+$2200	  &	    r  &  2  &  0.09 $\pm$ 0.01  &  0.40 $\pm$ 0.01 & 0.62 -- 0.98 \\
 J0613$-$0200	  	&	    m &  1  &  0.42 $\pm$ 0.01  &  ... 	    & 0.56 -- 0.16 \\
 J0631$+$1036	  &	    r  &  1  &  0.54 $\pm$ 0.02  &  ... 	    & 0.80 -- 0.20 \\
 J0633$+$0632	 &	    g  &  2  &  ...		 &  0.48 $\pm$ 0.01 & 0.09 -- 0.45 \\
 J0633$+$1746	  &	    g  &  2  &  ...		 &  0.50 $\pm$ 0.01 & 0.24 -- 0.54 \\
 J0659$+$1414	  &	    r  &  1  &  0.21 $\pm$ 0.01  &  ... 	    & 0.40 -- 1.00 \\
 J0742$-$2822	  	&	    r  &  1  &  0.61 $\pm$ 0.02  &  ... 	    & 0.84 -- 0.44 \\
 J0751$+$1807	  &	    m &  1  &  0.43 $\pm$ 0.02  &  ... 	    & 0.63 -- 0.99 \\
 J0835$-$4510	  	&	    r  &  2  &  0.13 $\pm$ 0.01  &  0.43 $\pm$ 0.01 & 0.66 -- 0.06 \\
 J1028$-$5819	  	&	    r  &  2  &  0.19 $\pm$ 0.01  &  0.47 $\pm$ 0.01 & 0.76 -- 0.12 \\
 J1048$-$5832	  	&	    r  &  2  &  0.15 $\pm$ 0.01  &  0.42 $\pm$ 0.02 & 0.64 -- 0.04 \\
 J1057$-$5226	  	&	    r  &  2  &  0.35 $\pm$ 0.05  &  0.20 $\pm$ 0.07 & 0.72 -- 0.20 \\
 J1124$-$5916	  	&	    r  &  2  &  \textbf{0.11 $\pm$ 0.01} &  0.49 $\pm$ 0.01 & \textbf{0.70 -- 0.06} \\
 J1418$-$6058	  	&	    g  &  2  &  ...		 &  0.47 $\pm$ 0.01 & 0.54 -- 0.90 \\
 J1420$-$6048	  	&	    r  &  2$^b$ &  0.26 $\pm$ 0.02  &  0.18 $\pm$ 0.02 & 0.60 -- 0.10 \\
 J1459$-$60	  	&	    g  &  2  &  ...		 &  0.15 $\pm$ 0.03 & 0.34 -- 0.78 \\
 J1509$-$5850	  	&	    r  &  2$^b$ &  0.18 $\pm$ 0.03  &  0.20 $\pm$ 0.03 & 0.52 -- 1.00 \\
 J1614$-$2230	  	&	    m &  2  &  0.19 $\pm$ 0.01  &  0.51 $\pm$ 0.01 & 0.92 -- 0.14 \\
 J1709$-$4429	  	&	    r  &  2  &  0.24 $\pm$ 0.01  &  0.25 $\pm$ 0.01 & 0.66 -- 0.14 \\
 J1718$-$3825	  	&	    r  &  1  &  0.42 $\pm$ 0.02  &  ... 	    & 0.68 -- 0.20 \\
 J1732$-$31	  	&	    g  &  2  &  ...		 &  0.42 $\pm$ 0.02 & 0.49 -- 0.93 \\
 J1741$-$2054	  	&	    g  &  2  &  0.30 $\pm$ 0.01  &  0.18 $\pm$ 0.02 & 0.67 -- 0.19 \\
 J1744$-$1134	  	&	    m  &  1  &  0.83 $\pm$ 0.02  &  ... 	    & 0.08 -- 0.44 \\
 J1747$-$2958	  	&	    r  &  2  &  0.18 $\pm$ 0.01  &  0.42 $\pm$ 0.04 & 0.64 -- 0.10 \\
 J1809$-$2332	  	&	    g  &  2  &  ...		 &  0.35 $\pm$ 0.01 & 0.41 -- 0.89 \\
 J1813$-$1246	  	&	    g  &  2  &  ...		 &  0.47 $\pm$ 0.02 & 0.56 -- 0.90 \\
 J1826$-$1256	  	&	    g  &  2  &  ...		 &  0.47 $\pm$ 0.01 & 0.54 -- 0.94 \\
 J1833$-$1034	  	&	    r  &  2  &  0.15 $\pm$ 0.01  &  0.44 $\pm$ 0.01 & 0.68 -- 0.10 \\
 J1836$+$5925	  &	    g  &  2  &  ...		 &  0.48 $\pm$ 0.01 & ... \\
 J1907$+$06	  	&	    g  &  2  &  ...		 &  0.40 $\pm$ 0.01 & 0.46 -- 0.94 \\
 J1952$+$3252	  &	    r  &  2  &  0.15 $\pm$ 0.01  &  0.49 $\pm$ 0.01 & 0.68 -- 0.08 \\
 J1958$+$2846	  &	    g  &  2  &  ...		 &  0.45 $\pm$ 0.01 & 0.55 -- 0.95 \\
 J2021$+$3651	  &	    r  &  2  &  0.17 $\pm$ 0.01  &  0.47 $\pm$ 0.01 & 0.70 -- 0.04 \\
 J2021$+$4026	  &	    g  &  2  &  ...		 &  0.48 $\pm$ 0.01 & ... \\
 J2032$+$4127	  &	    g  &  2  &  0.15 $\pm$ 0.01  &  0.50 $\pm$ 0.01 & 0.60 -- 0.92 \\
 J2043$+$2740	  &	    r  &  2  &  0.20 $\pm$ 0.01  &  0.36 $\pm$ 0.01 & 0.64 -- 0.08 \\
 J2124$-$3358	  	&	    m  &  1  &  0.86 $\pm$ 0.02  &  ... 	    & 0.92 -- 0.58 \\
 J2229$+$6114	  &	    r  &  1  &  0.49 $\pm$ 0.01  &  ... 	    & 0.64 -- 0.14 \\
 J2238$+$59	  	&	    g  &  2  &  ...  &  0.50 $\pm$ 0.01 & 0.60 -- 0.92 \\
\enddata

\tablenotetext{a}{Types are r=radio-selected, g=gamma-ray-selected, m=millisecond}
\tablenotetext{b}{For some pulse profiles the current dataset does not allow clear discrimination between
a single, broad pulse and two unresolved pulses. See the discussion in \citet{LAT6pulsars}
regarding PSRs J1420$-$6048 and J1509$-$5850.}

\tablecomments{~Light curve shape parameters evaluated from the full energy range light curve (see 
Section 2.1.3). These include the peak multiplicity (3rd column), the lag $\delta$ of the first gamma 
peak from the main radio peak for the radio-detected pulsars (4th column), and the phase difference 
$\Delta$ between the main gamma-ray peaks (5th column). Column 6 lists the off-pulse phase range used in the 
spectral analysis. The boldface entries for PSR J1124-5916 are the corrected values as per an Erratum sent
to the ApJ in December 2010.}

\end{deluxetable}
\normalsize
\clearpage
\noindent

% \table{tbl-spec}
%
% Incorporated comments from F. Camilo - 22 Aug 2009
% 
% 
%%%%%%%%%%%%%%%%%%%%%%%%%%%%inizio variazioni
\clearpage
\begin{landscape}
\begin{deluxetable}{llrrrrrrcc}
\tabletypesize{\scriptsize}
\tablecaption{Spectral fitting results for LAT-detected pulsars
\label{tbl-spec}}
\tablewidth{0pt}

\tablehead{
\colhead{PSR} & \colhead{Type$^a$} & \colhead{Photon Flux ($F_{100}$)} & \colhead{Energy Flux ($G_{100}$)} & \colhead{$\Gamma$}  & \colhead{$E_{\rm cutoff}$}  & \colhead{TS}  & \colhead{TS$_{\rm cutoff}$} & \colhead{Luminosity} & \colhead{Efficiency$^b$}\\
\colhead{} &\colhead{} & \colhead{($\rm 10^{-8} \,ph \ cm^{-2} \,s^{-1}$)} & \colhead{($\rm 10^{-11} \,erg \ cm^{-2}\, s^{-1}$)} & \colhead{}
 & \colhead{($\rm GeV$)}  & \colhead{}  & \colhead{}  & \colhead{($\rm 10^{33}\, erg \,s^{-1}$)}  & \colhead{$(f_\Omega = 1)$}
}

\startdata
J0007$+$7303		&	g	&	30.7	$\pm$	1.3\phn	&	38.2	$\pm$	1.3\phn	&	1.38	$\pm$	0.05	&	4.6	$\pm$	0.4\phn	&	7384		&	274.7	&	89	$\pm$	38		&	0.20	$\pm$ 0.08		\\
J0030$+$0451		&	m	&	5.83	$\pm$	0.78		&	5.26	$\pm$	0.42		&	1.22	$\pm$	0.19	&	1.8	$\pm$	0.4\phn	&	960		&	59.2		&	0.57	$\pm$	0.35		&	0.17	$\pm$ 0.10		\\
J0205$+$6449		&	r	&	13.2	$\pm$	2.0\phn	&	6.64 $\pm$	0.65		&	2.09	$\pm$	0.17	&	3.5	$\pm$	1.4\phn	&	346		&	12.5		&	54	--	81			& 	0.002 --	0.003		\\
J0218$+$4232		&	m	&	6.2	$\pm$	1.7\phn	&	3.62	$\pm$	0.64		&	2.02	$\pm$	0.28	&	5.1	$\pm$	4.2\phn	&	119		&	4.7		&	27	--	69			&	0.11	--	0.29			\\
J0248$+$6021		&	r	&	3.7	$\pm$	1.8\phn	&	3.07	$\pm$	0.70		&	1.15	$\pm$	0.59	&	1.4	$\pm$	0.6\phn	&	103		&	18.5		&	15	--	300			&	0.07	--	1.4			\\
J0357$+$32		&	g	&	10.4	$\pm$	1.2\phn	&	6.38	$\pm$	0.44		&	1.29	$\pm$	0.22	&	0.9	$\pm$	0.2\phn	&	949		&	71.6		&	\nodata				&	\nodata				\\
J0437$-$4715		&	m	&	3.65	$\pm$	0.84		&	1.86	$\pm$	0.26		&	1.74	$\pm$	0.38	&	1.3	$\pm$	0.7\phn	&	172		&	9.9		&	0.054 $\pm$ 0.008		&	0.02	$\pm$ 0.003		\\
J0534$+$2200$^c$	&	r	&	209	$\pm$	4\phd\phn &	130.6 $\pm$	3.4\phn	&	1.97	$\pm$	0.02	&	5.8	$\pm$	0.5\phn	&	21507	&	80.2		&	620	$\pm$ 	310		&	0.001 $\pm$ 0.001		\\
J0613$-$0200		&	m	&	3.38	$\pm$	0.85		&	3.23	$\pm$	0.42		&	1.38	$\pm$	0.29	&	2.7	$\pm$	1.0\phn	&	285		&	18.5		&	0.89	$_{-0.42}^{+0.71}$	&	0.07	$_{-0.03}^{+0.06}$	\\
J0631$+$1036		&	r	&	2.8	$\pm$	1.2\phn	&	3.04	$\pm$	0.61		&	1.38	$\pm$	0.42	&	3.6	$\pm$	1.8\phn	&	86		&	10.0		&	2.0	--	48			&	0.01	--	0.27			\\
J0633$+$0632		&	g	&	8.4	$\pm$	1.4\phn	&	8.0	$\pm$	0.77		&	1.29	$\pm$	0.22	&	2.2	$\pm$	0.6\phn	&	370		&	50.8		&	\nodata				&	\nodata				\\
J0633$+$1746		&	g	&	305.3 $\pm$	3.5\phn	&	338.1 $\pm$	3.5\phn	&	1.08	$\pm$	0.02	&	1.9	$\pm$	0.05		&	62307	&	5120.4	&	25$_{-12}^{+24}$		&	0.78	$_{-0.38}^{+0.74}$	\\
J0659$+$1414		&	r	&	10	$\pm$	1.4\phn	&	3.17	$\pm$	0.36		&	2.37	$\pm$	0.50	&	0.7	$\pm$	0.5\phn	&	206		&	6.9		&	0.31	$\pm$	0.08		&	0.01	$\pm$ 0.002		\\
J0742$-$2822		&	r	&	3.18	$\pm$	1.2\phn	&	1.82	$\pm$	0.42		&	1.76	$\pm$	0.48	&	2.0	$\pm$	1.4\phn	&	47		&	4.2		&	9.0	$_{-9}^{+12}$		&	0.07	$_{-0.07}^{+0.09}$	\\
J0751$+$1807		&	m	&	1.35	$\pm$	0.66		&	1.09	$\pm$	0.38		&	1.56	$\pm$	0.70	&	3.0	$\pm$	4.3\phn	&	37		&	3.8		&	0.47	$_{-0.35}^{+1}$	&	0.08	$_{-0.06}^{+0.17}$	\\
J0835$-$4510		&	r	&	1061	$\pm$	7.0\phn	&	879.4 $\pm$	5.4\phn	&
1.57	$\pm$	0.01	&	3.2	$\pm$	0.1		&	219585	&	5971.0	&	87	$\pm$	12		&	0.01	$\pm$	0.002	\\
J1028$-$5819		&	r	&	19.6	$\pm$	3.1\phn	&	17.7	$\pm$	1.4\phn	&	1.25	$\pm$	0.20	&	1.9	$\pm$	0.5\phn	&	620		&	75.1		&	120	$\pm$	73		&	0.14	$\pm$	0.09		\\
J1048$-$5832		&	r	&	19.7	$\pm$	3.0\phn	&	17.2	$\pm$	1.3\phn	&	1.31	$\pm$	0.18	&	2.0	$\pm$	0.4\phn	&	881		&	81.8		&	150	$\pm$	90		&	0.08	$\pm$	0.05		\\
J1057$-$5226		&	r	&	30.45 $\pm$	1.7\phn	&	27.2	$\pm$	0.98		&	1.06	$\pm$	0.10	&	1.3	$\pm$	0.1\phn	&	4961		&	366.3	&	17	$\pm$	9		&	0.56	$\pm$	0.31		\\
J1124$-$5916		&	r	&	5.2	$\pm$	1.8\phn	&	3.79	$\pm$	0.70		&	1.43	$\pm$	0.40	&	1.7	$\pm$	0.7\phn	&	111		&	16.7		&	100	$_{-53}^{+34}$		&	0.01	$_{-0.004}^{+0.003}$	\\
J1418$-$6058		&	g	&	27.7	$\pm$	8.3\phn	&	23.5	$\pm$	3.8\phn	&	1.32	$\pm$	0.24	&	1.9	$\pm$	0.4\phn	&	162		&	54.1		&	110	--	700			&	0.02	--	0.14			\\
J1420$-$6048		&	r	&	24.2	$\pm$	7.9\phn	&	15.8	$\pm$	3.5\phn	&	1.73	$\pm$	0.24	&	2.7	$\pm$	1.0\phn	&	63		&	21.4		&	590	$\pm$	380		&	0.06	$\pm$	0.04		\\
J1459$-$60		&	g	&	17.8	$\pm$	3.4\phn	&	10.56 $\pm$	1.2\phn	&	1.83	$\pm$	0.24	&	2.7	$\pm$	1.1\phn	&	337		&	21.1		&	\nodata				&	\nodata				\\
J1509$-$5850		&	r	&	8.7	$\pm$	1.4\phn	&	9.7	$\pm$	1.2\phn	&	1.36	$\pm$	0.28	&	3.5	$\pm$	1.1\phn	&	262		&	26.3		&	78	$\pm$	49		&	0.15	$\pm$	0.10		\\
J1614$-$2230		&	m	&	2.89	$\pm$	1.2\phn	&	2.74	$\pm$	0.50		&	1.34	$\pm$	0.43	&	2.4	$\pm$	1.0\phn	&	149		&	13.3		&	5.3	$\pm$3.4			&	1.0	$\pm$	0.7		\\
J1709$-$4429		&	r	&	149.8 $\pm$	4.1\phn	&	124 $\pm$	2.6\phn	&	1.70	$\pm$	0.04	&	4.9	$\pm$	0.4\phn	&	16009	&	373.6	&	290	--	1900			&	0.09	--	0.57			\\
J1718$-$3825		&	r	&	9.1	$\pm$	5.8\phn	&	6.7	$\pm$	1.9\phn	&	1.26	$\pm$	0.74	&	1.3	$\pm$	0.6\phn	&	105		&	19.7		&	120	$\pm$	80		&	0.09	$\pm$	0.06		\\
J1732$-$31		&	g	&	25.3	$\pm$	3.0\phn	&	24.2	$\pm$	1.4\phn	&	1.27	$\pm$	0.14	&	2.2	$\pm$	0.3\phn	&	1002		&	131.2	&	\nodata				&	\nodata				\\
J1741$-$2054		&	g	&	20.3	$\pm$	2.0\phn	&	12.8	$\pm$	0.8\phn	&	1.39	$\pm$	0.17	&	1.2	$\pm$	0.2\phn	&	935		&	92.6		&	2.2	$\pm$	1.3		&	0.24	$\pm$	0.14		\\
J1744$-$1134		&	m	&	4.3	$\pm$	1.6\phn	&	2.8	$\pm$	0.6\phn	&	1.02	$\pm$	0.71	&	0.7	$\pm$	0.4\phn	&	78		&	20.0		&	0.43	$\pm$	0.13		&	0.1	$\pm$	0.03		\\
J1747$-$2958		&	r	&	18.2	$\pm$	4.2\phn	&	13.1	$\pm$	1.7\phn	&	1.11	$\pm$	0.34	&	1.0	$\pm$	0.2\phn	&	213		&	59.3		&	63	--	390			&	0.02	--	0.16			\\
J1809$-$2332		&	g	&	49.5	$\pm$	3.0\phn	&	41.3	$\pm$	1.6\phn	&	1.52	$\pm$	0.07	&	2.9	$\pm$	0.3\phn	&	3451		&	201.9	&	140	$\pm$	140		&	0.33	$\pm$	0.33		\\
J1813$-$1246		&	g	&	28.1	$\pm$	3.5\phn	&	16.9	$\pm$	1.3\phn	&	1.83	$\pm$	0.14	&	2.9	$\pm$	0.8\phn	&	482		&	39.7		&	\nodata				&	\nodata				\\
J1826$-$1256		&	g	&	41.8	$\pm$	4.1\phn	&	33.4	$\pm$	1.8\phn	&	1.49	$\pm$	0.11	&	2.4	$\pm$	0.3\phn	&	1152		&	138		&	\nodata				&	\nodata				\\
J1833$-$1034		&	r	&	20.5	$\pm$	4.6\phn	&	10.1	$\pm$	1.4\phn	&	2.24	$\pm$	0.18	&	7.7	$\pm$	4.8\phn	&	110		&	4.9		&	270	$\pm$	60		&	0.01	$\pm$	0.002	\\
J1836$+$5925$^d$	&	g	&	65.6	$\pm$	1.8\phn	&	59.9	$\pm$	1.3\phn	&	1.35	$\pm$	0.04	&	2.3	$\pm$	0.1\phn	&	20982	&	674.6	&	$<$46				&	$<$4.0				\\
J1907$+$06		&	g	&	40.25 $\pm$	3.8\phn	&	27.5	$\pm$	1.6\phn	&	1.84	$\pm$	0.10	&	4.6	$\pm$	1.0\phn	&	1209		&	59.3		&	\nodata				&	\nodata				\\
J1952$+$3252		&	r	&	17.6	$\pm$	1.9\phn	&	13.4	$\pm$	0.9\phn	&	1.75	$\pm$	0.12	&	4.5	$\pm$	1.2\phn	&	1008		&	36.4		&	64	$\pm$	32		&	0.02	$\pm$	0.01		\\
J1958$+$2846		&	g	&	7.65	$\pm$	1.6\phn	&	8.45	$\pm$	0.83		&	0.77	$\pm$	0.31	&	1.2	$\pm$	0.2\phn	&	491		&	89.2		&	\nodata				&	\nodata				\\
J2021$+$3651		&	r	&	67.35 $\pm$	4.4\phn	&	47.0	$\pm$	1.8\phn	&	1.65	$\pm$	0.07	&	2.6	$\pm$	0.3\phn	&	3138		&	223.5	&	250	$_{-240}^{+500}$	&	0.07	$_{-0.07}^{+0.15}$	\\
J2021$+$4026$^d$	&	g	&	152.6 $\pm$	4.9\phn	&	97.6	$\pm$	2.0\phn	&	1.79	$\pm$	0.04	&	3.0	$\pm$	0.2\phn	&	10180	&	331.4	&	260	$\pm$	150		&	2.2	$\pm$	1.3		\\
J2032$+$4127		&	g	&	6	$\pm$	2.3\phn	&	11.1	$\pm$	1.4\phn	&	0.68	$\pm$	0.46	&	2.1	$\pm$	0.6\phn	&	487		&	56.3		&	34	--	170			&	0.13	--	0.64			\\
J2043$+$2740		&	r	&	2.41	$\pm$	0.90		&	1.55	$\pm$	0.32		&	1.07	$\pm$	0.66	&	0.8	$\pm$	0.3\phn	&	79		&	15.1		&	6.0	$\pm$	3.8		&	0.09	$\pm$	0.06		\\
J2124$-$3358		&	m	&	1.95	$\pm$	0.49		&	2.75	$\pm$	0.42		&	1.05	$\pm$	0.34	&	2.7	$\pm$	1.0\phn	&	226		&	22.9		&	0.21	$_{-0.14}^{+0.42}$	&	0.05	$_{-0.04}^{+0.11}$	\\
J2229$+$6114		&	r	&	32.6	$\pm$	2.2\phn	&	22.0	$\pm$	1.0\phn	&	1.74	$\pm$	0.08 &	3.0	$\pm$	0.5\phn	&	1929		&	96.0		&	17	--	1100			&	0.001	--	0.05		\\
J2238$+$59		&	g	&	6.8	$\pm$	1.8\phn	&	5.44	$\pm$	0.71		&	1.00	$\pm$	0.43	&	1.0	$\pm$	0.3\phn	&	219		&	37.2		&	\nodata				&	\nodata				\\				
\enddata

\tablenotetext{\textit{a}}{Types are r=radio-selected, g=gamma-ray-selected, m=millisecond.}
\tablenotetext{\textit{b}}{Here, $f_\Omega$ is assumed to be 1, which can result in an efficiency $>$ 1.}
\tablenotetext{\textit{c}}{For the Crab the spectral parameters come from \citet{FermiCrab}.}
\tablenotetext{\textit{d}}{For J1836+5925 and J2021+4026 the spectral parameters come from the phase-averaged analysis (see Section 2.2).}

\tablecomments{Results of the unbinned maximum likelihood spectral fits  
for the LAT gamma-ray pulsars (see Section 2.2). Columns 3 and 4 list the on-pulse photon flux $F_{\rm 100}$ and 
on-pulse energy flux $G_{\rm 100}$ respectively. 
The fits used an exponentially cutoff power-law model (see Eq. 4) 
with photon index $\Gamma$ and cutoff energy $E_{\rm cutoff}$ given in columns 5 and 6. 
The systematic uncertainties on $F_{\rm 100}$, $G_{\rm 100}$, and $\Gamma$ due to uncertainties in the 
Galactic diffuse emission model have been added in quadrature with the statistical errors. 
Uncertainties in the instrument response induce additional biases of $\delta F_{100} = (+30\%,\,-10\%)$,
$\delta G_{100} = (+20\%,\,-10\%)$, $\delta\Gamma = (+0.3,\,-0.1)$, and $\delta E_{\rm cutoff} = (+20\%,\,-10\%)$.
The test statistic ($TS$) for the source significance is provided in column 7. The significance of an exponential 
cutoff (as compared to a simple power-law) is indicated by $TS_{\rm cutoff}$ in column 8, where a value 
$< 10$ indicates that the two models are comparable. The total gamma-ray luminosity $L_\gamma$ and 
the resulting calculated gamma-ray conversion efficiency $\eta_\gamma\equiv L_\gamma/\dot{E}$ 
(where $f_\Omega=1$ as described in Section 3.2) are listed in columns 9 and 10, respectively. The uncertainties
in $L_\gamma$ and $\eta$ include the flux and distance uncertainties. Nevertheless, the strong dependence
of these variables on the measured distance (see Table \ref{tab:dist}) and beaming factor means that they
should be considered with care.}

\end{deluxetable}
\normalsize
\noindent

\clearpage
\end{landscape}

%\end{document}

% \table{tab:dist}
%
% Teresa M. \& Peter D. - 28 July 2009 
% PeterH - 31 July 2009
% ECF - 1 Aug 2009
% DAS - 2 Aug 2009: break Table out into a separate file
% Incorporated comments from F. Camilo - 22 Aug 2009
% ECF - 16 Sep 2009 (Updates to address Fernando2)
%
% 
%%%%%%%%%%%%%%%%%%%%%%%%%%%%inizio variazioni

\tabletypesize{\scriptsize}
\begin{deluxetable}{lccc}
\tablecaption{Pulsar distance estimates}
\tablewidth{0pt}

\tablehead{
\colhead{Pulsar Name}      & 
\colhead{Distance (kpc)}   & 
\colhead{Method$^a$ }      &
\colhead{ (Ref$^b$)}}
\startdata
J0007+7303     &    1.4$\pm$0.3 	    & K     & (30)  \\% Pineault, S., et al., 1993, AJ, 105, 1060 \\	     
J0030+0451     &    0.300$\pm$0.090	    & P     & (23)  \\% Lommen, A.N., et al., 2006, ApJ, 642, 1012 \\	      
J0205+6449     &    2.6--3.2		    & K     & (14,32)  \\% Green & Gull 1982; Roberts,D.A., et al., 1993, A\&A, 274, 438 (a) \\    
J0218+4232     &    2.5--4		    & O     & (1) \\%Bassa et al 2003	   
J0248+6021     &    2--9		    & O    & (6)   \\%Cognard private communication	     \\ 	 
J0437$-$4715   &    0.1563$\pm$0.0013	    & P     & (9)  \\% Deller et al. 2008 \\	   
J0534+2200     &    2.0$\pm$0.5 	    & O     & (35) \\% Trimble 1973 \\  	       
J0613$-$0200   &    0.48$^{+0.19}_{-0.11}$    & P     & (18)  \\% Hotan, A. W, et al., 2006, MNRAS, 369, 1502  \\      
J0631+1036     &    0.75--3.62  	     & O    & (39)  \\%Zepka, A., et al., 1996, ApJ, 456, 305 (b)  \\	     
J0633+1746     &    0.250$^{+0.120}_{-0.062}$ & P     & (11) \\%Faherty et al. 2007 \\  	
J0659+1414     &    0.288$^{ +0.033}_{-0.027}$ & P    & (2)   \\% Brisken, W. F., et al., 2003, ApJ, 593, 89 \\      
J0742$-$2822   &    2.07$^{+1.38}_{-1.07}$    & DM    & (33)  \\%Taylor, Manchester & Lyne 1993, ApJS 88, 529   \\   
J0751+1807     &    0.6$^{ +0.6}_{-0.2}$      & P     & (28) \\%Nice et al. 2005\\
J0835$-$4510   &    0.287$^{+0.019}_{-0.017}$ & P     & (10)  \\%Dodson, R. et al., 2003, ApJ, 596, 1137 \\	     
J1028$-$5819   &    2.33$\pm$0.70	    & DM    & (19)  \\%Keith, M.J., et al. 2008, MNRAS, 389, 1881 (d) \\     
J1048$-$5832   &    2.71$\pm$0.81	    & DM    & (21)  \\%Johnston s., et al., 1996, MNRAS, 279, 661 (b) \\     
J1057$-$5226   &    0.72 $\pm$0.2	    & DM    & (37) \\%Weltevrede \& Wright 2009 \\    
J1124$-$5916   &    4.8$^{+0.7}_{-1.2 }$      & O     & (13)  \\%Gonzalez, M., Safi-Harb, S., 2003, ApJ, 583, 91 \\  
J1418$-$6058   &    2--5		    & O     & (27,38) \\%Ng et al 2005;  Yadigaroglu & Romani 1997	  
J1420$-$6048   &    5.6$\pm$1.7 	    & DM    & (8)  \\%D'amico et al 2001 ApJ...552L..45D
J1509$-$5850   &    2.6$\pm$0.8 	    & DM    & (24)  \\% Manchester et al. 2005 in AJ 129, 1993; 25% error \\	 
J1614$-$2230   &    1.27$\pm$0.39	    & DM    & (7)   \\%Crawford, F., et al., 2006, ApJ, 652, 1499 (d) \\     
J1709$-$4429   &    1.4--3.6		    & O    & (26,33)  \\%McGowan et al 2004;Taylor, Manchester & Lyne 1993, ApJS 88, 529 \\	  
J1718$-$3825   &    3.82$\pm$1.15	   & DM    & (25)  \\%Manchester, R.N., et al., 2001, MNRAS, 328, 17 (b) \\ 
J1741$-$2054   &    0.38$\pm$0.11	    & DM    & (3)   \\%Camilo et al 2009 \\	
J1744$-$1134   &    0.357$^{+0.043}_{-0.035}$ & P    & (34)   \\%Toscano, M., et al., 1999, ApJ, 523, 171 \\	     
J1747$-$2958   &    2--5	 	    & O    & (5,12)   \\%Camilo, F., et al., 2002a, ApJ, 579, 25 (d); Gaensler2004 \\	      
J1809$-$2332   &    1.7$\pm$1.0 	    & K     & (29) \\%Oka et al 1999
J1833$-$1034   &    4.7$\pm$0.4 	    & K     & (4)   \\%Camilo, F., et al., 2006, ApJ, 637, 456 \\	     
J1836+5925     &    $<$0.8		    & O     & (16) \\%Halpern, Camilo and Gotthelf 2007 ApJ 668 1154		     
J1952+3252     &    2.0$\pm$0.5 	    & K     & (15)  \\%Greidanus, H., Strom, R.G., 1990, A\&A, 240, 376 \\    
J2021+3651     &    2.1$^{+2.1}_{-1.0}$       & O     & (36)  \\%Van Etten, A., et al., 2008, ApJ, 680, 1417 \\ 	 
J2021+4026     &    1.5$\pm$0.45	    & K     & (22) \\%Landecker et al 1980, AASS 39 133 	     
J2032+4127     &    1.6--3.6	    & O	   & (3)   \\%Camilo 2009 \\		       
J2043+2740     &    1.80$\pm$0.54	    & DM    &(31)   \\%Ray, W., et al., 1996 apj 470 1103 \\	     
J2124$-$3358   &    0.25$^{+0.25}_{-0.08}$    & P   & (18)  \\%Hotan et al 2006    \\  
J2229+6114     &    0.8--6.5		    & O     & (17,20)  \\%Kothes et al 2001; Halpern, J.P.,  et al., 2001b, ApJ, 547, 323 (e) \\   
\enddata

\tablenotetext{\textit{a}}{K distance evaluation from kinematic model; P from parallax; DM from dispersion measure
using the \citet{Cordes2002} model; O from other measurements. For DM measurements, we assume a minimum distance uncertainty of 30\%,
as discussed in Section 3.1.}
\tablenotetext{\textit{b}}{For DM, the reference gives the DM measurement.}

\tablecomments{The best known distances of 37 pulsars detected by {\it Fermi}. Nine of the pulsars in the catalog have no distance estimate and are not included in this table.}

\tablerefs{
(1)  \citet{Bassa2003};
(2)  \citet{Brisken2003};
(3)  \citet{Camilo2009};
(4)  \citet{Camilo2006}; 
(5)  \citet{Camilo2002a};          
(6)  \citet{Cognard2009};
(7)  \citet{Crawford2006};
(8)  \citet{DAmico2001};
(9)  \citet{Deller2008};
(10) \citet{Dodson2003};
(11) \citet{Faherty2007};
(12) \citet{Gaensler2004};
(13) \citet{Gonzalez2003};
(14) \citet{Green1982};
(15) \citet{Greidanus1990};
(16) \citet{Halpern2007};   
(17) \citet{Halpern2001a};                                                      
(18) \citet{Hotan2006};
%(19) \citet{Hui2007};
(19) \citet{Keith2008};
% (20) \citet{Koribalski1995};
(20) \citet{Kothes2001};
(21) \citet{Johnston1996};
(22) \citet{Landecker1980};
(23) \citet{Lommen2006}; 
%(24) \citet{Manchester2005};
(24) \citet{ATNFcatalog};
(25) \citet{Manchester2001};
(26) \citet{McGowan2004};
(27) \citet{Ng2005};  
(28) \citet{Nice2005};
(29) \citet{Oka1999};
(30) \citet{Pineault1993};  
(31) \citet{Ray1996};       
(32) \citet{Roberts1993}; 
(33) \citet{TaylorDM1993};
(34) \citet{Toscano1999};
(35) \citet{Trimble1973};
(36) \citet{VanEtten2008};
(37) \citet{Weltevrede2009};
(38) \citet{Yadigaroglu1997};
(39) \citet{Zepka1996}
  }              
\label{tab:dist}
\end{deluxetable}
\normalsize

%%%%%%%%%%%%%%%%%%%%%%%%%%%%%%%%%%%fine variazioni

% \table{tbl-assoc}
%
% Incorporated comments from F. Camilo - 22 Aug 2009
% Comments from FC not yet addressed are:
% 
% In positional associations, J2032+4127 should also have HEGRA TeV J2032+4130
% - or just TeV J2032+4130 against it.  I'm going to guess that this column
% is incomplete. 
% 
% 
%%%%%%%%%%%%%%%%%%%%%%%%%%%%inizio variazioni

\begin{deluxetable}{lclll}
\tabletypesize{\scriptsize}
\tablecaption{Positional associations with known GeV and TeV sources for LAT-detected pulsars
\label{tbl-assoc}}
\tablewidth{0pt}

\tablehead{
\colhead{PSR} & 
\colhead{Alt. name} & 
\colhead{LAT BSL association$^a$} & 
\colhead{GeV associations$^b$} & 
\colhead{Other associations}
}

\startdata
J0007$+$7303 	& \nodata 			& 0FGL J0007.4$+$7303 & 3EG J0010+7309 		& SNR CTA 1 $^{2}$ \\
   				&   				&   					& EGR J0008$+$7308 	& PWN G119.5$+$10.2 $^{1}$ \\
   				&   				&   					& GEV J0008$+$7304 	&     \\
   				&   				&   					& 1AGL J0006$+$7311 	&     \\
J0030$+$0451 	& \nodata 			& 0FGL J0030.3$+$0450 & EGR J0028$+$0457 	& \nodata \\
J0205$+$6449 	& \nodata 			& \nodata 				& \nodata 				& SNR/PWN 3C 58 $^{2}$ \\
   				&   				&   					&   					& PWN G119.5$+$10.2 $^{1}$ \\
J0218$+$4232 	& \nodata 			& \nodata 				& \nodata 				& \nodata \\
J0248$+$6021 	& \nodata 			& \nodata 				& \nodata 				& \nodata \\
J0357$+$32 		& \nodata 			& 0FGL J0357.5$+$3205 & \nodata 				& \nodata \\
J0437$-$4715 		& \nodata 			& \nodata 				& \nodata 				& PWN G253.4$-$42.0 $^{1}$ \\
J0534$+$2200 	& Crab 			& 0FGL J0534.6$+$2201 & 3EG J0534$+$2200 	& SNR/PWN G184.6$-$5.8 $^{1,2}$ \\
   				& PSR B0531$+$21 &   					& EGR J0534$+$2159 	& HESS J0534$+$220 $^{4}$ \\
   				&   				&   					& GEV J0534$+$2159 	&     \\
   				&   				&   					& 1AGL J0535$+$2205 	&     \\
J0613$-$0200 		& \nodata 			& 0FGL J0613.9$-$0202 	& \nodata 				& \nodata \\
J0631$+$1036 	& \nodata 			& 0FGL J0631.8$+$1034 & \nodata 				& \nodata \\
J0633$+$0632 	& \nodata 			& 0FGL J0633.5$+$0634 & 3EG J0631$+$0642 	& \nodata \\
   				&   				&   					& EGR J0633$+$0646 	&   \\
   				&   				&   					& GEV J0633$+$0645 	&   \\
J0633$+$1746 	& Geminga 		& 0FGL J0634.0$+$1745 & 3EG J0633$+$1751 	& PWN G195.1$+$4.3 $^{1}$ \\
   				&   				&   					& EGR J0633$+$1750 	& MGRO J0632$+$17 $^{3,11}$ \\
   				&   				&   					& GEV J0634$+$1746 	&     \\
   				&   				&   					& 1AGL J0634$+$1748 	&     \\
J0659$+$1414 	& PSR B0656$+$14 & \nodata 				& \nodata 				& SNR 203.0$+$12.0 \\
J0742$-$2822 		& PSR B0740$-$28 	& \nodata 				& \nodata 				& \nodata \\
J0751$+$1807 	& \nodata 			& \nodata 				& \nodata 				& \nodata \\
J0835$-$4510 		& Vela 			& 0FGL J0835.4$-$4510 	& 3EG J0834$-$4511 	& SNR/PWN G263.9$-$3.3 $^{1,2}$ \\
   				& PSR B0833$-$45 &   					& EGR J0834$-$4512	& HESS J0835$-$455 $^{5}$ \\
   				&   				&   					& GEV J0835$-$4512 	&     \\
   				&   				&   					& 1AGL J0835$-$4509 	&     \\
J1028$-$5819 		& \nodata 			& 0FGL J1028.6$-$5817 	& 3EG J1027$-$5817 	& \nodata \\
   				&   				&   					& GEV J1025$-$5809 	&     \\
J1048$-$5832 		& PSR B1046$-$58 & 0FGL J1047.6$-$5834 	& 3EG J1048$-$5840 	& PWN G287.4$+$0.58 $^{1}$ \\
   				&   				&   					& EGR J1048$-$5839	&   \\
   				&   				&   					& GEV J1047$-$5840	&   \\
J1057$-$5226 		& PSR B1055$-$52 & 0FGL J1058.1$-$5225 	& 3EG J1058$-$5234 	& \nodata \\
   				&   				&   					& EGR J1058$-$5221 	&   \\
   				&   				&   					& GEV J1059$-$5218 	&     \\
   				&   				&   					& 1AGL J1058$-$5239 	&     \\
J1124$-$5916 		& \nodata 			& \nodata 				& \nodata 				& MSH 11$-$54 \\
   				&   				&   					&   					& SNR/PWN G292.0$+$1.8 $^{1,2}$ \\
J1418$-$6058 		& \nodata 			& 0FGL J1418.8$-$6058 	& 3EG J1420$-$6038 	& PWN G313.3$+$0.1 $^{1}$ \\
   				&   				&   					& GEV J1417$-$6100	& HESS J1418$-$609 $^{6}$ \\
   				&   				&   					& 1AGL J1419$-$6055	&   \\
J1420$-$6048 		& \nodata 			& \nodata 				& 3EG J1420$-$6038 	& PWN G313.6$+$0.3 $^{1}$ \\
   				&   				&   					& EGR J1418$-$6040 	& HESS J1420$-$607 $^{6}$ \\
   				&   				&   					& GEV J1417$-$6100	&   \\
   				&   				&   					& 1AGL J1419$-$6055 	&     \\
J1459$-$60 		& \nodata 			& 0FGL J1459.4$-$6056 	& \nodata 				& \nodata \\
J1509$-$5850$^c$ 	& \nodata 			& 0FGL J1509.5$-$5848 	& 1AGL J1506$-$5859	& PWN G319.97$-$0.62 $^{1}$ \\
J1614$-$2230 		& \nodata 			& \nodata 				& 3EG J1616$-$2221 	& \nodata \\
J1709$-$4429 		& PSR B1706$-$44 & 0FGL J1709.7$-$4428 	& 3EG J1710$-$4439 	& SNR/PWN G343.1$-$2.3 $^{1,2}$ \\
   				&   				&   					& EGR J1710$-$4435	& HESS J1708$-$443 $^{7}$ \\
   				&   				&   					& GEV J1709$-$4430 	&     \\
   				&   				&   					& 1AGL J1709$-$4428 	&     \\
J1718$-$3825 		& \nodata 			& \nodata 				& \nodata 				& HESS J1718$-$385 $^{8}$ \\
J1732$-$31 		& \nodata 			& 0FGL J1732.8$-$3135 	& 3EG J1734$-$3232 	& \nodata \\
   				&   				&   					& EGR J1732$-$3126	&   \\
   				&   				&   					& GEV J1732$-$3130	&   \\
J1741$-$2054 		& \nodata 			& 0FGL J1742.1$-$2054 	& 3EG J1741$-$2050 	& \nodata \\
J1744$-$1134 		& \nodata  		& \nodata 				& \nodata 				& \nodata \\
J1747$-$2958 		& \nodata 			& \nodata 				& 1AGL J1746$-$3017	& PWN G359.23$-$0.82 $^{1}$ \\
J1809$-$2332 		& \nodata 			& 0FGL J1809.5$-$2331 	& 3EG J1809$-$2328 	& PWN G7.4$-$2.0 $^{1}$ \\
   				&   				&   					& GEV J1809$-$2327 	&     \\
   				&   				&   					& 1AGL J1809$-$2333 	&     \\
J1813$-$1246 		& \nodata 			& 0FGL J1813.5$-$1248 	& GEV J1814$-$1228	& \nodata \\
J1826$-$1256 		& \nodata 			& 0FGL J1825.9$-$1256 	& 3EG J1826$-$1302 	& PWN G18.5$-$0.4 $^{1}$ \\
   				&   				&   					& GEV J1825$-$1310 	&     \\
   				&   				&   					& 1AGL J1827$-$1277 	&     \\
J1833$-$1034 		& \nodata 			& \nodata 				& \nodata 				& SNR/PWN G21.5$-$0.9 $^{1,2}$ \\
   				&   				&   					&   					& HESS J1833$-$105 $^{9}$ \\
J1836$+$5925 	& \nodata 			& 0FGL J1836.2$+$5924 	& 3EG J1835$+$5918 	& \nodata \\
   				&   				&   					& GEV J1835$+$5921 	&     \\
   				&   				&   					& 1AGL J1836$+$5923 	&     \\
J1907$+$06 		& \nodata 			& 0FGL J1907.5$+$0602 	& GEV J1907$+$0557	& MGRO J1908$+$063 $^{3}$ \\
   				&   				&   					& 1AGL J1908$+$0613	& HESS J1908$+$06 $^{10}$ \\
J1952$+$3252$^d$ & PSR B1951$+$32 & 0FGL J1953.2$+$3249 & \nodata 			& SNR CTB 80 $^2$ \\
   				&   				&   					&   					& PWN G69.0$+$2.7 $^{1}$ \\
J1958$+$2846 	& \nodata 			& 0FGL J1958.1$+$2848 & 3EG J1958$+$2909 	& \nodata \\
   				&   				&   					& GEV J1957$+$2859 	&     \\
J2021$+$3651 	& \nodata 			& 0FGL J2020.8$+$3649 & GEV J2020$+$3658	& PWN G75.2$+$0.1 $^{1}$ \\
   				&   				&   					& 1AGL J2021$+$3652	& MGRO J2019$+$37 $^{3}$ \\
J2021$+$4026 	& \nodata 			& 0FGL J2021.5$+$4026 & 3EG J2020$+$4017 	& SNR $\gamma$ Cygni $^2$ \\
   				&   				&   					& 1AGL J2021$+$3652 	& SNR G78.2$+$2.1 $^{2}$ \\
J2032$+$4127 	& \nodata 			& 0FGL J2032.2$+$4122 & 3EG J2033$+$4118 	& MGRO J2031$+$41 $^{3}$ \\
   				&   				&   					& EGR J2033$+$4117 	&     \\
   				&   				&   					& 1AGL J2032$+$4102 	&     \\
J2043$+$2740 	& \nodata 			& \nodata 				& \nodata 				& \nodata \\
J2124$-$3358 		& \nodata 			& 0FGL J2124.7$-$3358 	& \nodata 				& PWN G10.9$-$45.4 $^{1}$ \\
J2229$+$6114 	& \nodata 			& 0FGL J2229.0$+$6114 & 3EG J2227$+$6122 	& PWN G106.6$+$2.9 $^{1}$ \\
   				&   				&   					& EGR J2227$+$6114 	& MGRO J2228$+$61  $^{3,11}$ \\
   				&   				&   					& GEV J2227$+$6101 	&     \\
   				&   				&   					& 1AGL J2231$+$6109 	&     \\
J2238$+$59 		& \nodata 			& \nodata 				& \nodata 				& \nodata \\
\enddata

\tablenotetext{\textit{a}}{Source designator from the LAT Bright Source List \citep{BSL}.}
\tablenotetext{\textit{b}}{Source designator(s) from the 3rd EGRET \citep[3EG: ][]{3rdCat},  
Revised EGRET \citep[EGR: ][]{EGRcatalog}, High-energy EGRET \citep[GEV][]{Lamb1997} 
and/or the first AGILE \citep[1AGL:][]{AGILEcatalog} catalogs.}
\tablenotetext{\textit{c}}{PSR J1509$-$5850 should not be confused with PSR B1509$-$58 observed by \textit{CGRO} \citep{Kuiper1999}.}
\tablenotetext{\textit{d}}{While pulsations from PSR J1952$+$3252 were detected in EGRET data, it was never cataloged as a point source.}

\tablecomments{Alternate names for the pulsars in this catalog are given in column 2.
%while
%the LAT Bright Source List designation is provided in column 3. Column 4 shows 
%positional associations with cataloged EGRET sources. 
Positional associations with SNRs, 
PWNe and selected TeV sources are provided in column 5.}

\tablerefs{{1. \citet{Roberts2005}}, {2. \citet{Green2009}}, {3. \citet{Milagro2009}}, {4. \citet{HESSCrab2006}}, {5. \citet{HESSVela2006}}, {6.
\citet{HESSRabbit2006}}, {7. \citet{HESS1709_2009}}, {8. \citet{HESS1718_2007}}, {9. \citet{HESS1833_2007}}, {10. \citet{HESS1908_2009}}, {11. \citet{ATEL2172} }. }

\end{deluxetable}

% \documentclass[a4paper,10pt]{article}
% 
% 
% %opening
% \title{}
% \author{}
% 
% \begin{document}

%--------------------------------------------------------------------------------------------------------------%

%--------------------------------------------------------------------------------------------------------------%

%--------------------------------------------------------------------------------------------------------------%

%--------------------------------------------------------------------------------------------------------------%

%--------------------------------------------------------------------------------------------------------------%

%--------------------------------------------------------------------------------------------------------------%

%--------------------------------------------------------------------------------------------------------------%

%--------------------------------------------------------------------------------------------------------------%

% \begin{figure}[!ht]
% \includegraphics[width=0.5\textwidth]{Figures/Hindex.eps}
% \caption{ Spectral index distribution of the $\gamma$-ray pulsars detected by the \textit{Fermi} LAT.
% \label{Hindex}}
% \end{figure}
% %--------------------------------------------------------------------------------------------------------------%
% 
% 
% 
% \begin{figure}[!ht]
% \includegraphics[width=0.5\textwidth]{Figures/Hcutoff.eps}
% \caption{ Cutoff energy distribution of the $\gamma$-ray pulsars detected by the \textit{Fermi} LAT.
% \textit{Plot needs to be updated, with Crab value of $5.8$ replacing the $13.5$ GeV value shown here.}
% In fact, both this and the proceeding figure are likely to be dropped.
% \label{Hcut-off}}
% \end{figure}
% %--------------------------------------------------------------------------------------------------------------%

%--------------------------------------------------------------------------------------------------------------%

%--------------------------------------------------------------------------------------------------------------%

% \begin{figure}[!ht]
% \includegraphics[width=0.5\textwidth]{Figures/ZvsTS.eps}
% \caption{ Confidence level of the Z test expressed in sigmas versus the square root of TS. The black line is a linear fit passing from the origin of the axes.
% \label{ZvsTS}}
% \end{figure}
% %--------------------------------------------------------------------------------------------------------------%
% 
% 
% 
% \begin{figure}[!ht]
% \includegraphics[width=0.5\textwidth]{Figures/FluxvsTS_2.eps}
% \caption{ Pulsar fluxes versus square root of TS. The black dashed line is a linear fit. The purple line is the previous fit shifted toward lower fluxes. It represent our flux limit as function of the TS.
% \label{FluxvsTS}}
% \end{figure}
% %--------------------------------------------------------------------------------------------------------------%
% 
% 
% 
% \begin{figure}[!ht]
% \includegraphics[width=0.5\textwidth]{Figures/ZvsFlux.eps}
% \caption{ Confidence level of the Z test expressed in sigmas versus pulsar flux.
% \label{ZvsFlux}}
% \end{figure}
% %--------------------------------------------------------------------------------------------------------------%

%--------------------------------------------------------------------------------------------------------------%

% \begin{figure}[!ht]
% \includegraphics[width=0.5\textwidth]{Figures/CutoffvsBsurf_Labels.eps}
% \caption{ High energy cutoff vs magnetic field at neutron star surface. Milli-second pulsar are not included in this plot.
% \label{EcvsBsurf}}
% \end{figure}
%--------------------------------------------------------------------------------------------------------------%
%\clearpage

% \begin{figure}[!ht]
% \includegraphics[width=0.5\textwidth]{Figures/PeakSepvsLsd.eps}
% \caption{ Peak separation versus Luminosity. Red diamonds: Radio-selected pulsars including milliseconds. 
% Blue triangles: Gamma-selected pulsars. Purple square: EGRET pulsars.
% \label{DeltavsEdot}}
% \end{figure}
%--------------------------------------------------------------------------------------------------------------%

%--------------------------------------------------------------------------------------------------------------%

%--------------------------------------------------------------------------------------------------------------%

% \end{document}

\clearpage
\appendix
\renewcommand{\thefigure}{A-\arabic{figure}}
\setcounter{figure}{0}

\section*{Appendix: Gamma-ray Pulsar Light Curves}
%%%%%%%
%%% Dave Smith changed "CAPTION FOR THE FIGURE" to "Light curves for PSR J...",
%%% 30 July 2009.
%%%%%%%
%----start of new page----
\begin{sidewaysfigure}
\centering
%%----start of figure----
\begin{minipage}[t]{0.45\linewidth}
\centering
\includegraphics[width=\linewidth]{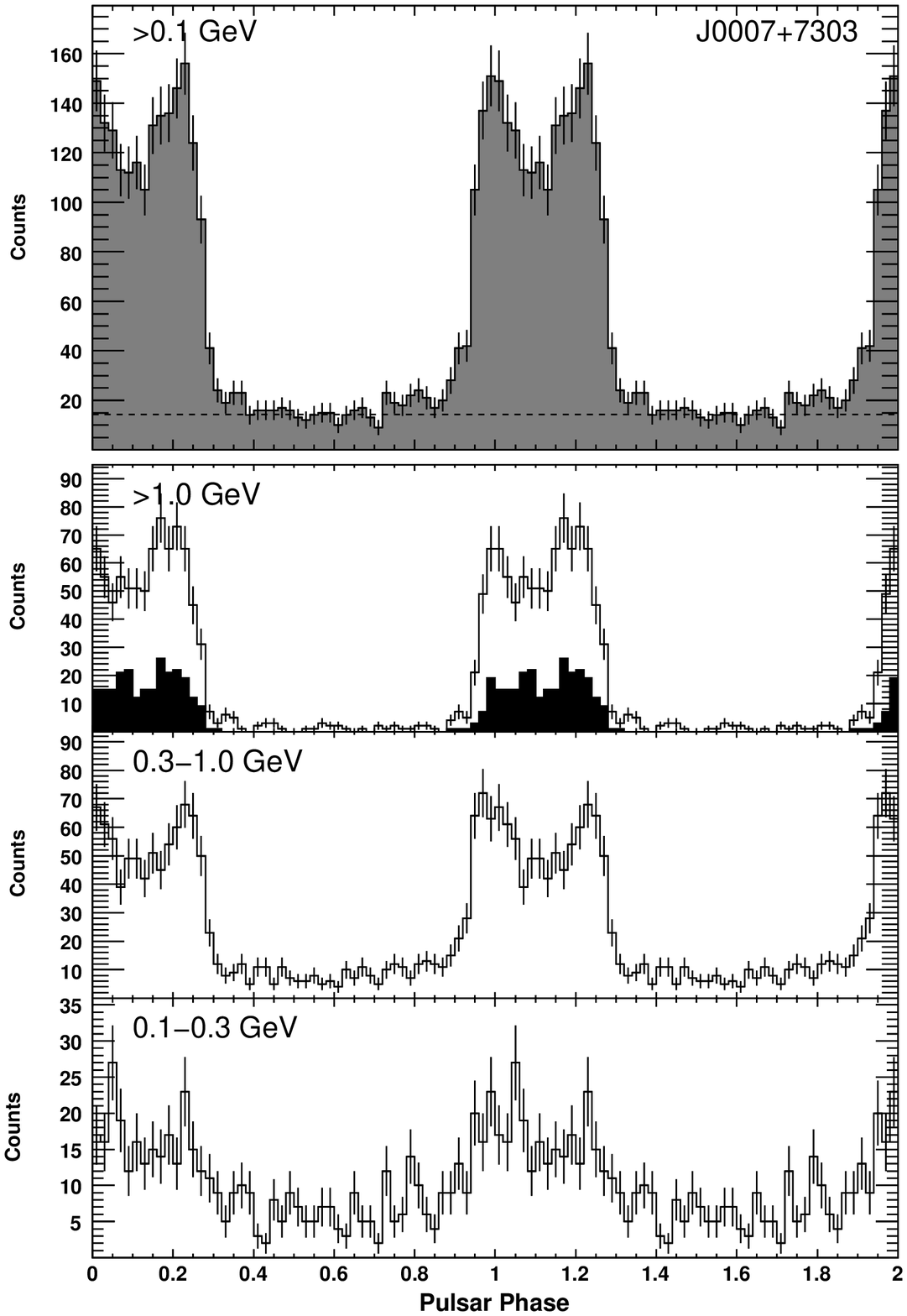}
\caption{Light curves for PSR J0007+7303 ($P=316$\,ms). 
\label{fig:J0007p7303_lightcurve}}
\end{minipage}%
\hspace{1cm}%
%%----start of figure----
\begin{minipage}[t]{0.45\linewidth}
\centering
\includegraphics[width=\linewidth]{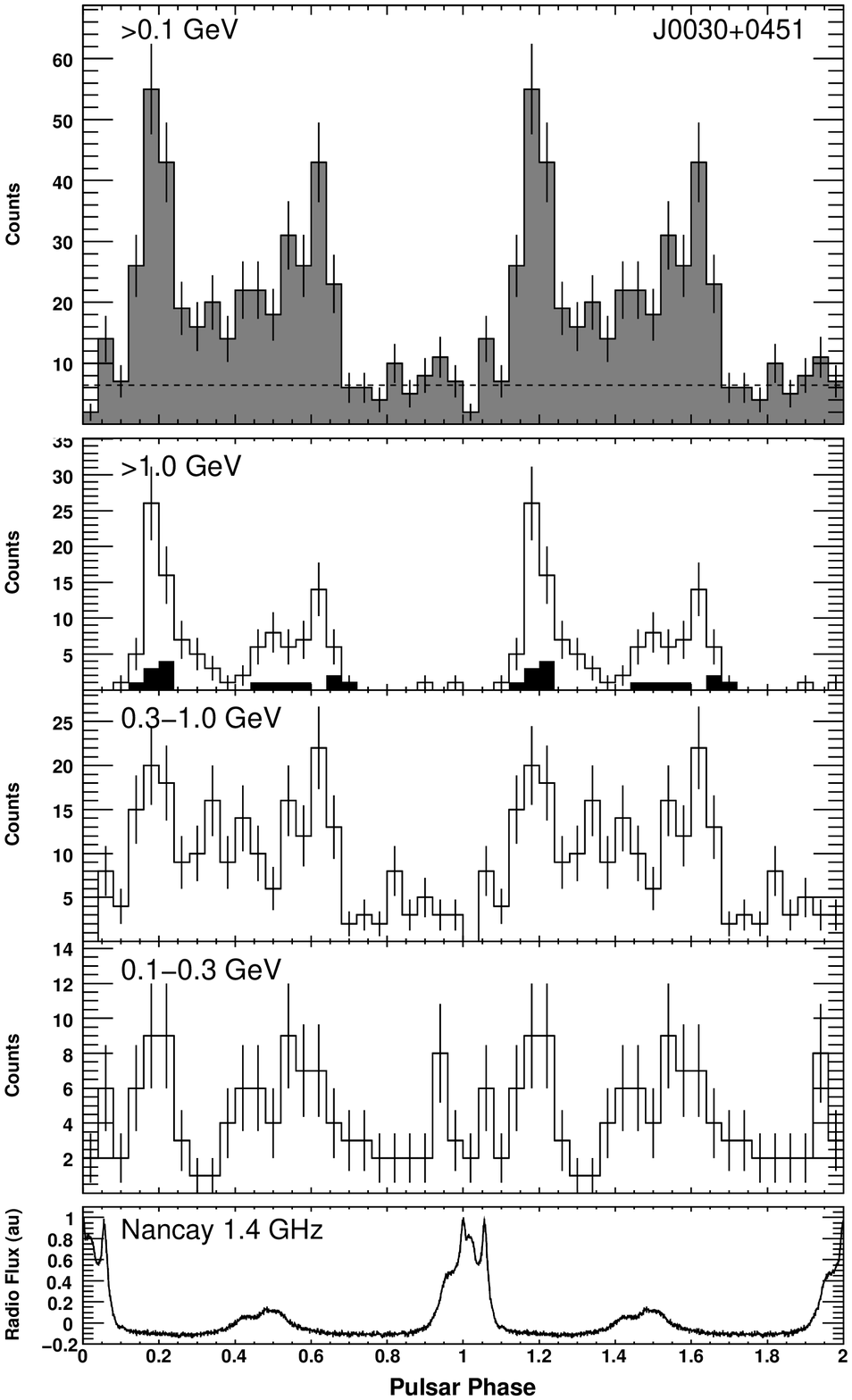}
\caption{Light curves for PSR J0030+0451 ($P=4.87$\,ms).
\label{fig:J0030p0451_lightcurve}}
\end{minipage}\\
\end{sidewaysfigure}
\clearpage

%----start of new page----
\begin{sidewaysfigure}
\centering
%%----start of figure----
\begin{minipage}[t]{0.45\linewidth}
\centering
\includegraphics[width=\linewidth]{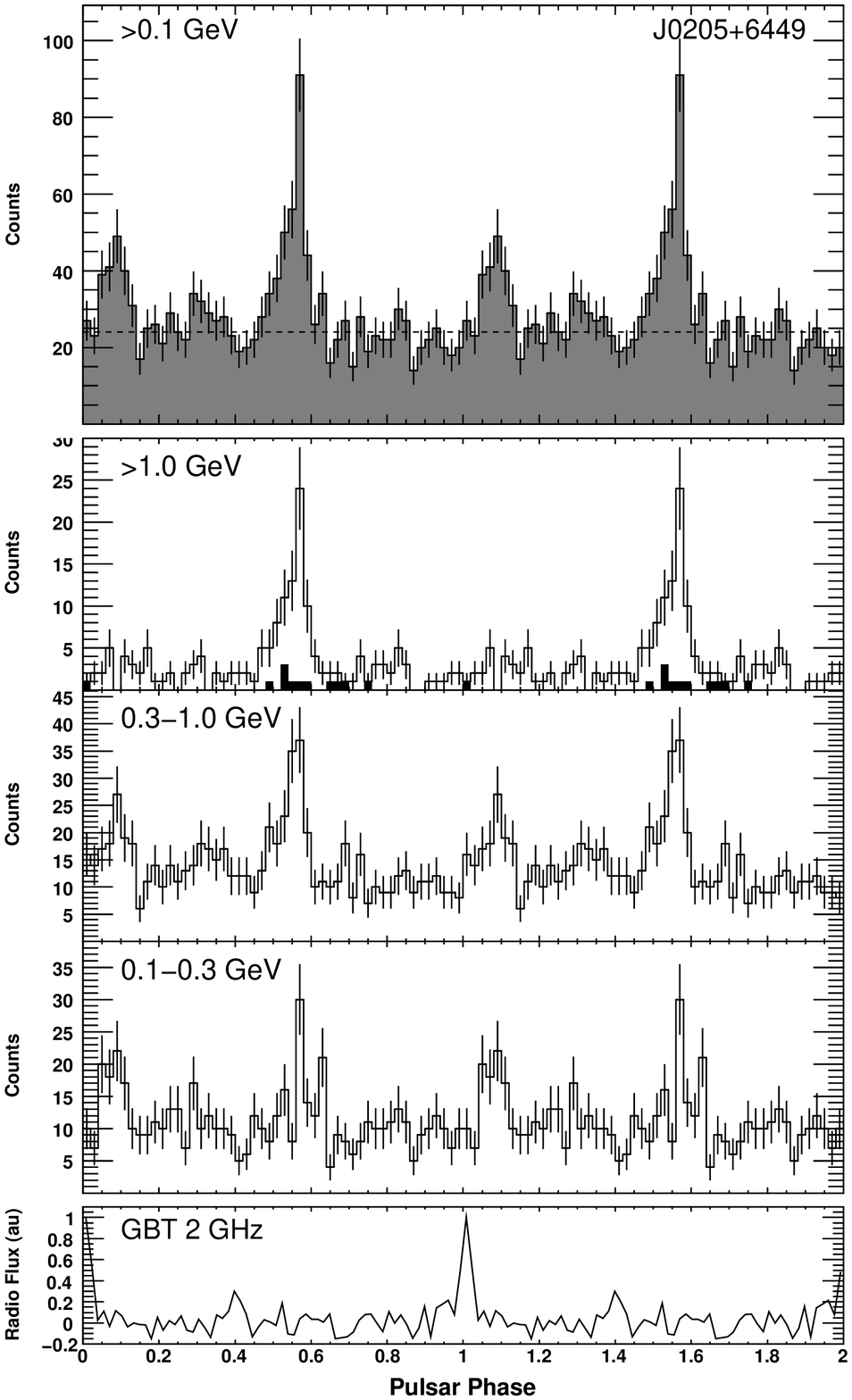}
\caption{Light curves for PSR J0205+6449 ($P=65.7$\,ms).
\label{fig:J0205p6449_lightcurve}}
\end{minipage}%
\hspace{1cm}%
%%----start of figure----
\begin{minipage}[t]{0.45\linewidth}
\centering
\includegraphics[width=\linewidth]{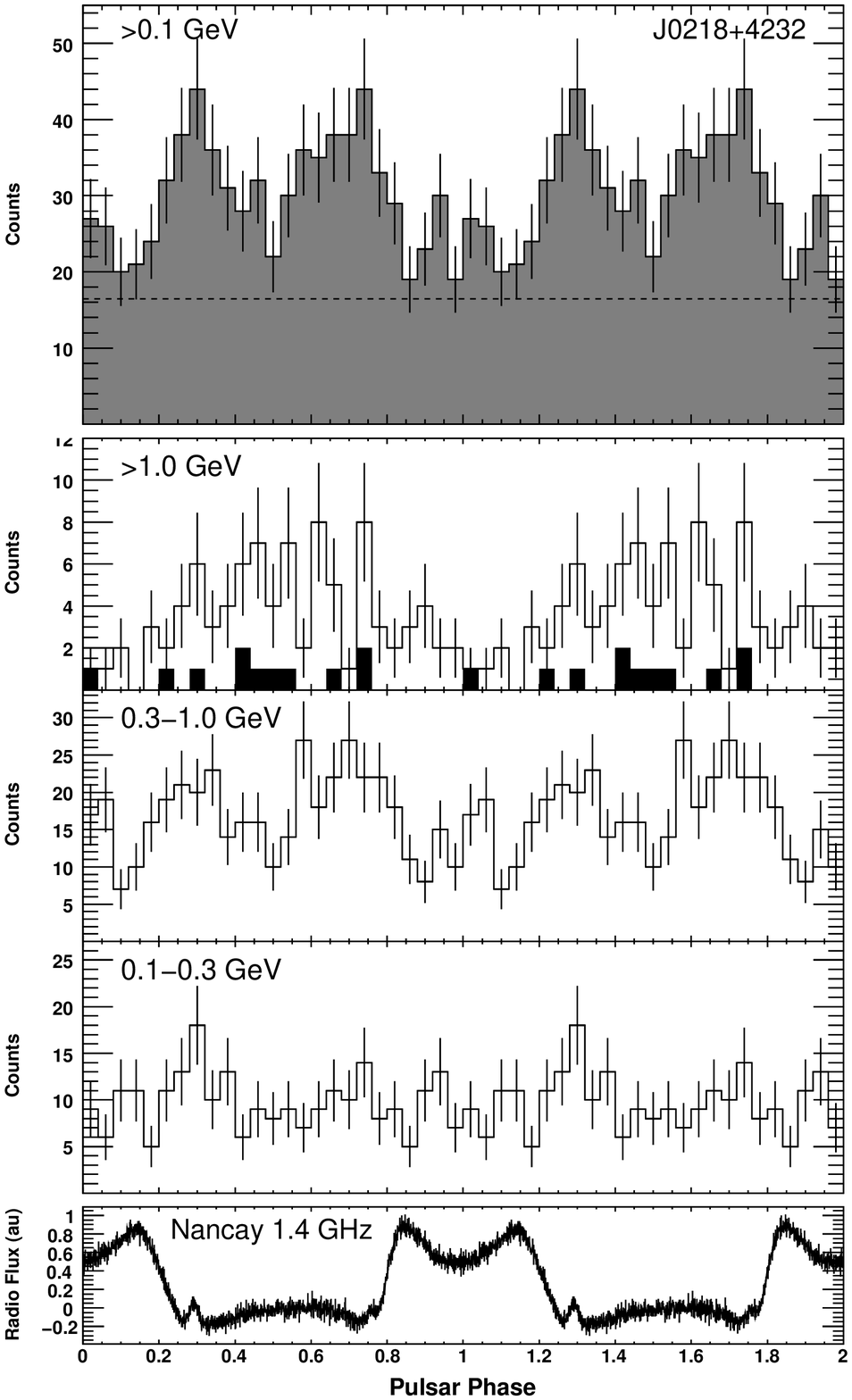}
\caption{Light curves for PSR J0218+4232 ($P=2.32$\,ms).
\label{fig:J0218p4232_lightcurve}}
\end{minipage}
\end{sidewaysfigure}
\clearpage

%----start of new page----
\begin{sidewaysfigure}
\centering
%%----start of figure----
\begin{minipage}[t]{0.45\linewidth}
\centering
\includegraphics[width=\linewidth]{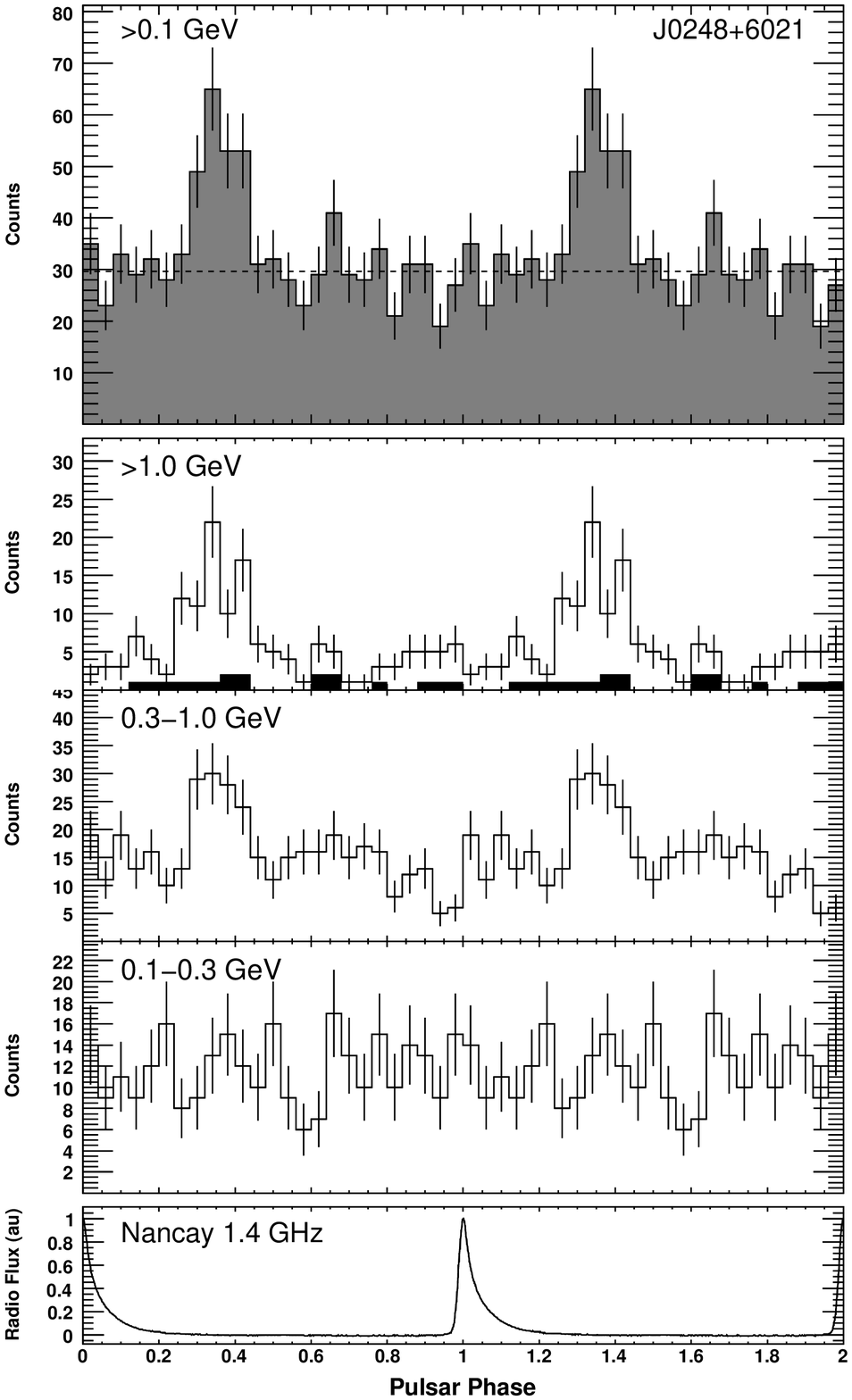}
\caption{Light curves for PSR J0248+6021 ($P=217$\,ms).
\label{fig:J0248p6021_lightcurve}}
\end{minipage}%
\hspace{1cm}%
%%----start of figure----
\begin{minipage}[t]{0.45\linewidth}
\centering
\includegraphics[width=\linewidth]{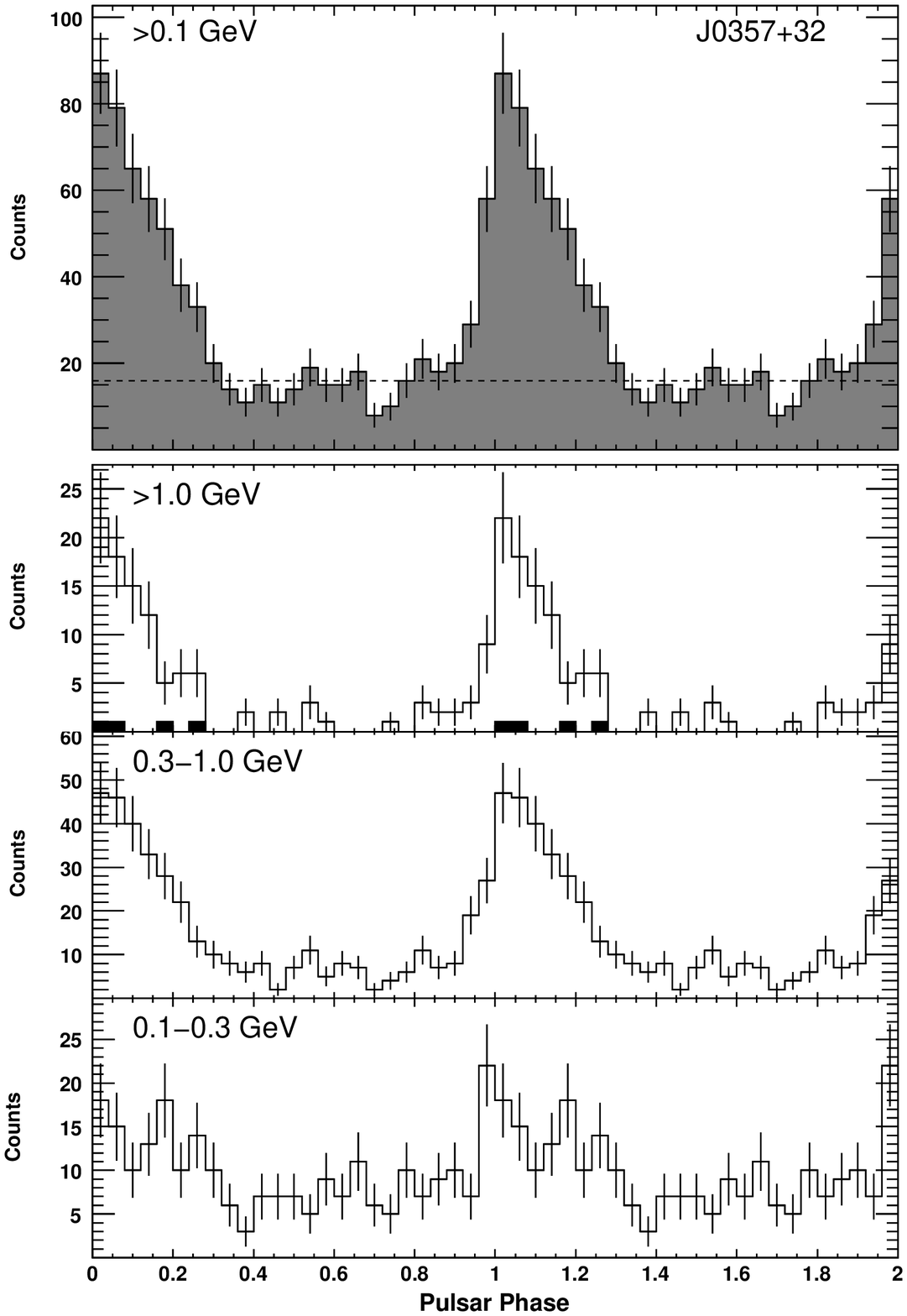}
\caption{Light curves for PSR J0357+32 ($P=444$\,ms).
\label{fig:J0357p32_lightcurve}}
\end{minipage}\\
\end{sidewaysfigure}
\clearpage

%----start of new page----
\begin{sidewaysfigure}
\centering
%%----start of figure----
\begin{minipage}[t]{0.45\linewidth}
\centering
\includegraphics[width=\linewidth]{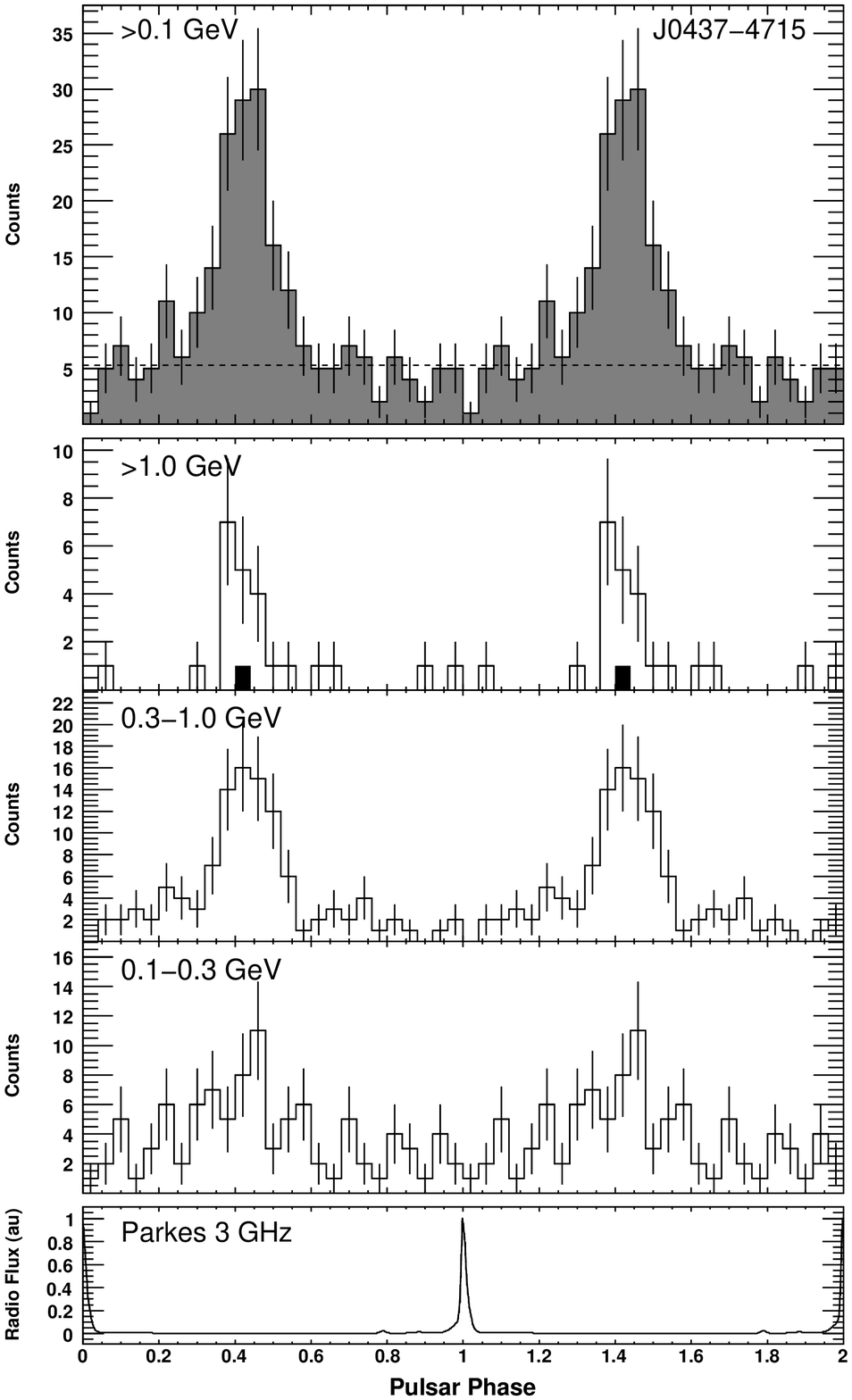}
\caption{Light curves for PSR J0437$-$4715 ($P=5.76$\,ms). 
\label{fig:J0437m4715_lightcurve}}
\end{minipage}%
\hspace{1cm}%
%%----start of figure----
\begin{minipage}[t]{0.45\linewidth}
\centering
\includegraphics[width=\linewidth]{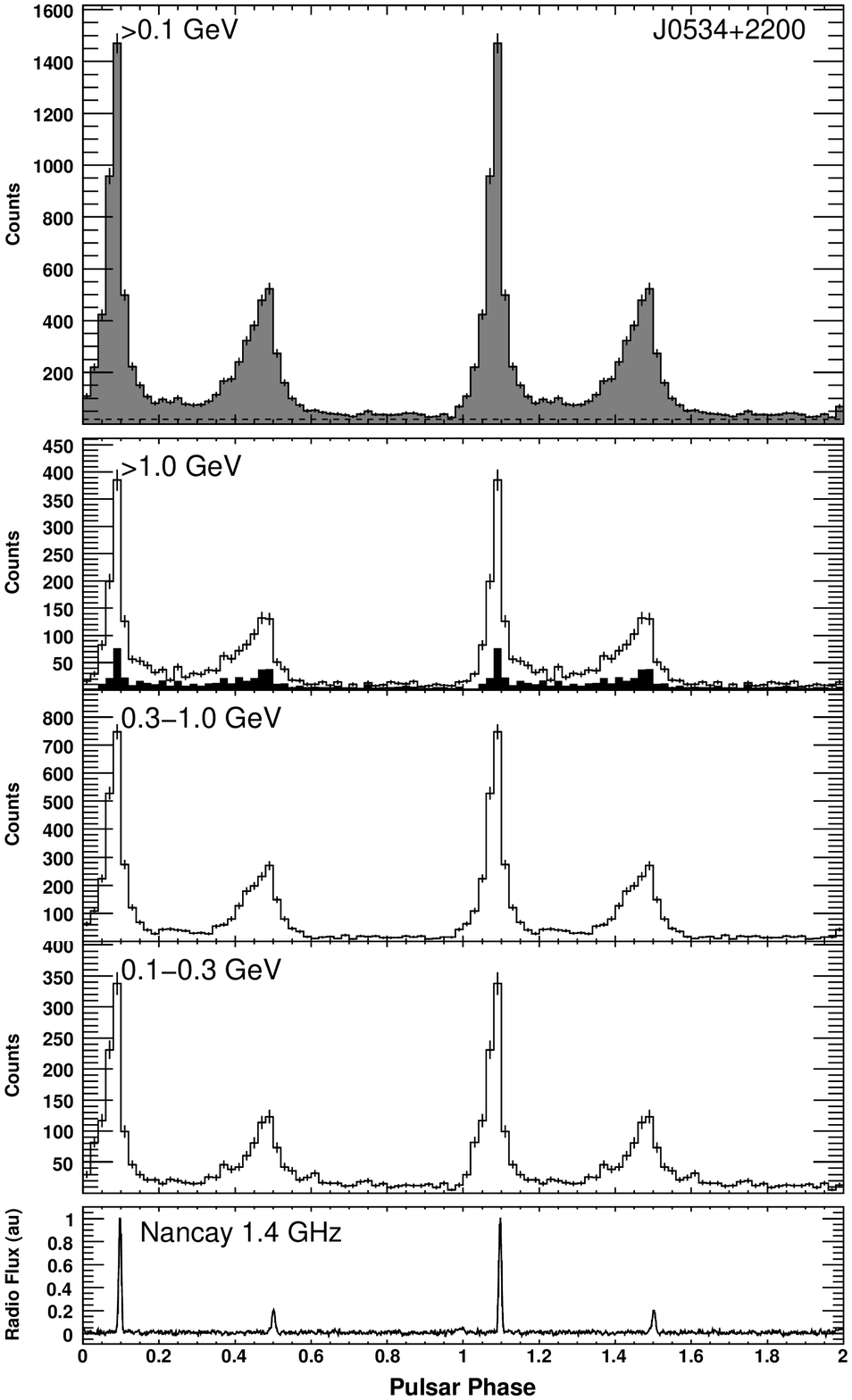}
\caption{Light curves for PSR J0534+2200 ($P=33.1$\,ms, Crab pulsar). The zero of phase
is set to the radio precursor.
\label{fig:J0534p2200_lightcurve}}
\end{minipage}\\
\end{sidewaysfigure}
\clearpage

%----start of new page----
\begin{sidewaysfigure}
\centering
%%----start of figure----
\begin{minipage}[t]{0.45\linewidth}
\centering
\includegraphics[width=\linewidth]{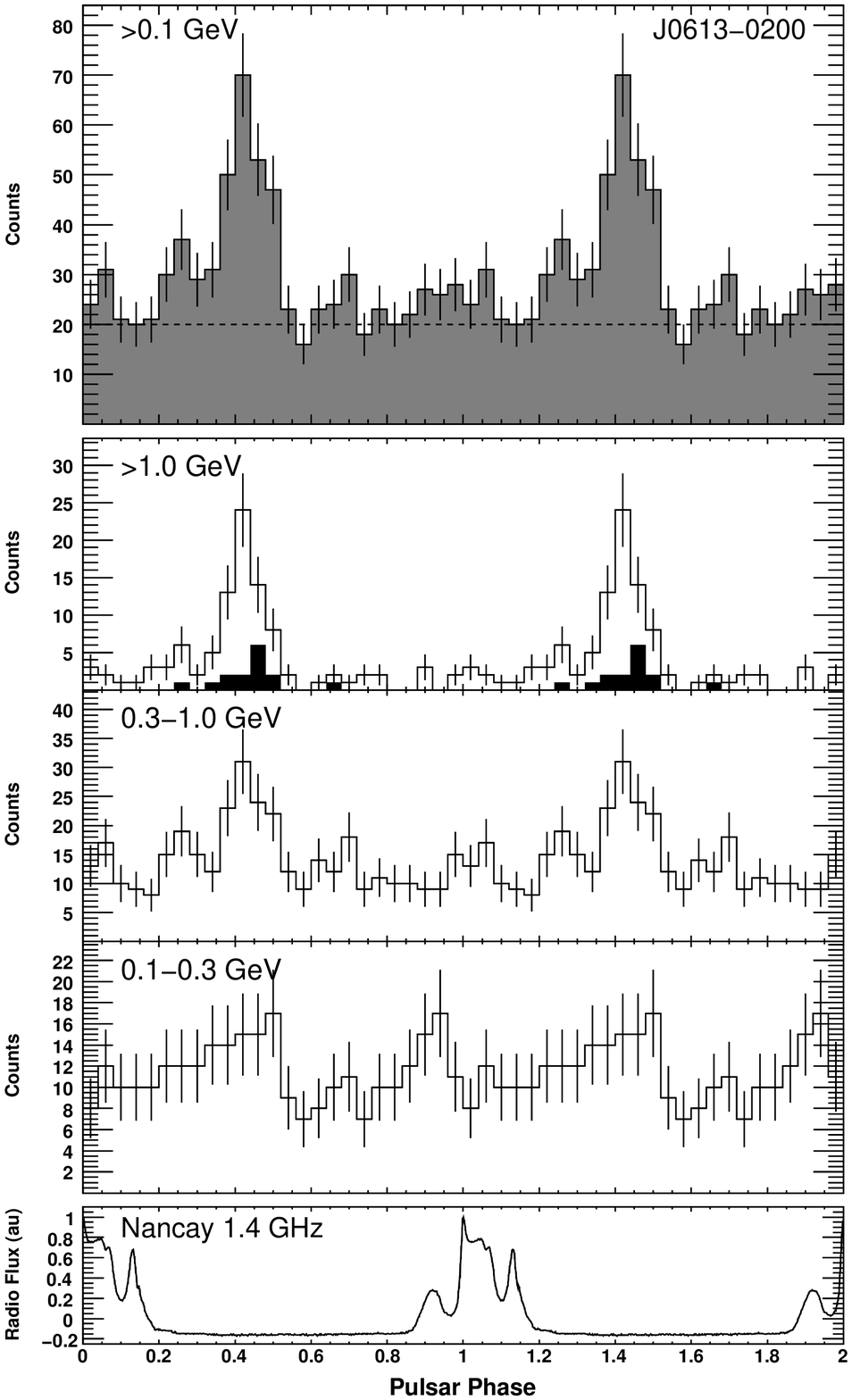}
\caption{Light curves for PSR J0613$-$0200 ($P=3.06$\,ms).
\label{fig:J0613m0200_lightcurve}}
\end{minipage}%
\hspace{1cm}%
%%----start of figure----
\begin{minipage}[t]{0.45\linewidth}
\centering
\includegraphics[width=\linewidth]{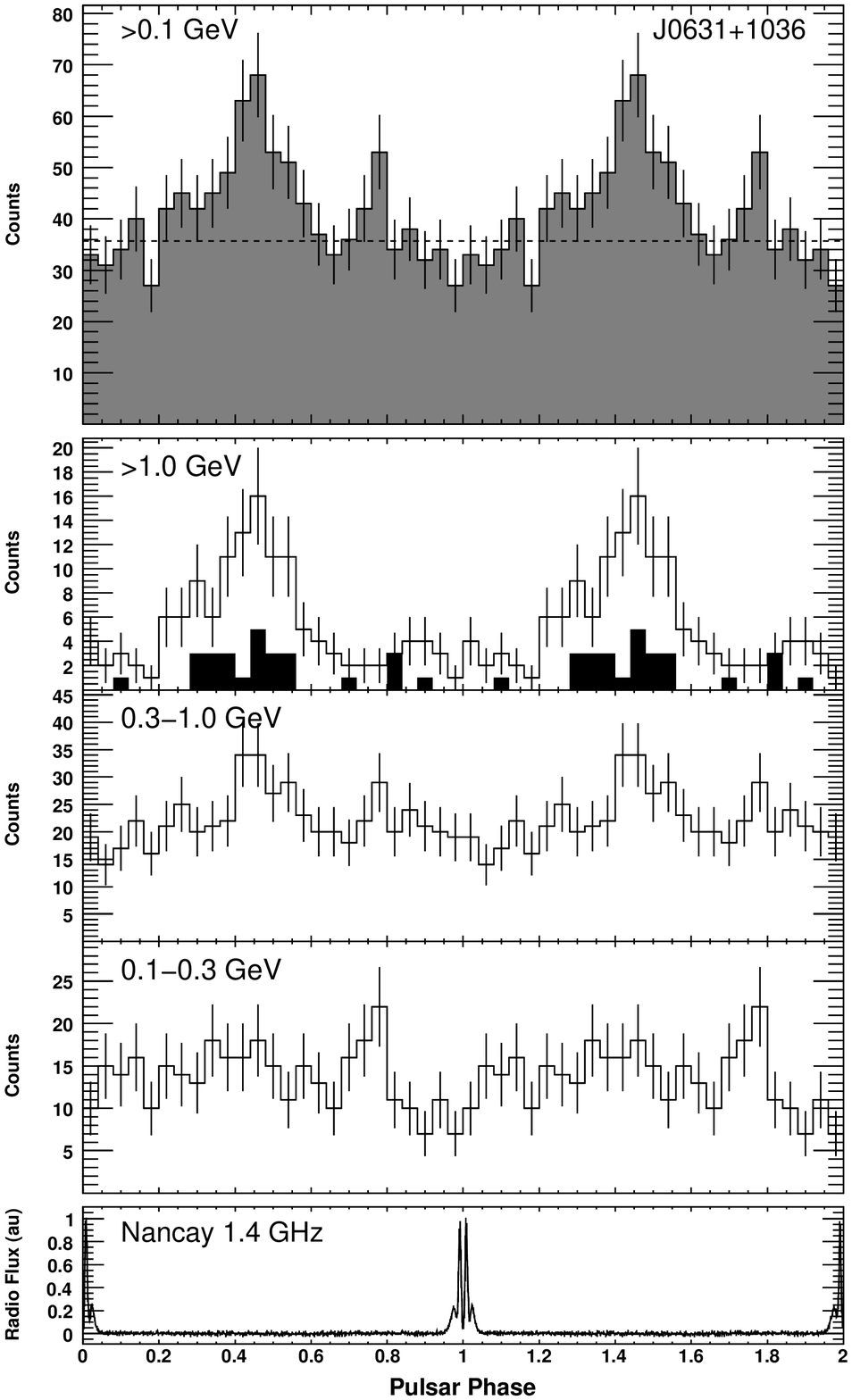}
\caption{Light curves for PSR J0631+1036 ($P=288$\,ms).
\label{fig:J0631p1036_lightcurve}}
\end{minipage}\\
\end{sidewaysfigure}
\clearpage

%----start of new page----
\begin{sidewaysfigure}
\centering
%%----start of figure----
\begin{minipage}[t]{0.45\linewidth}
\centering
\includegraphics[width=\linewidth]{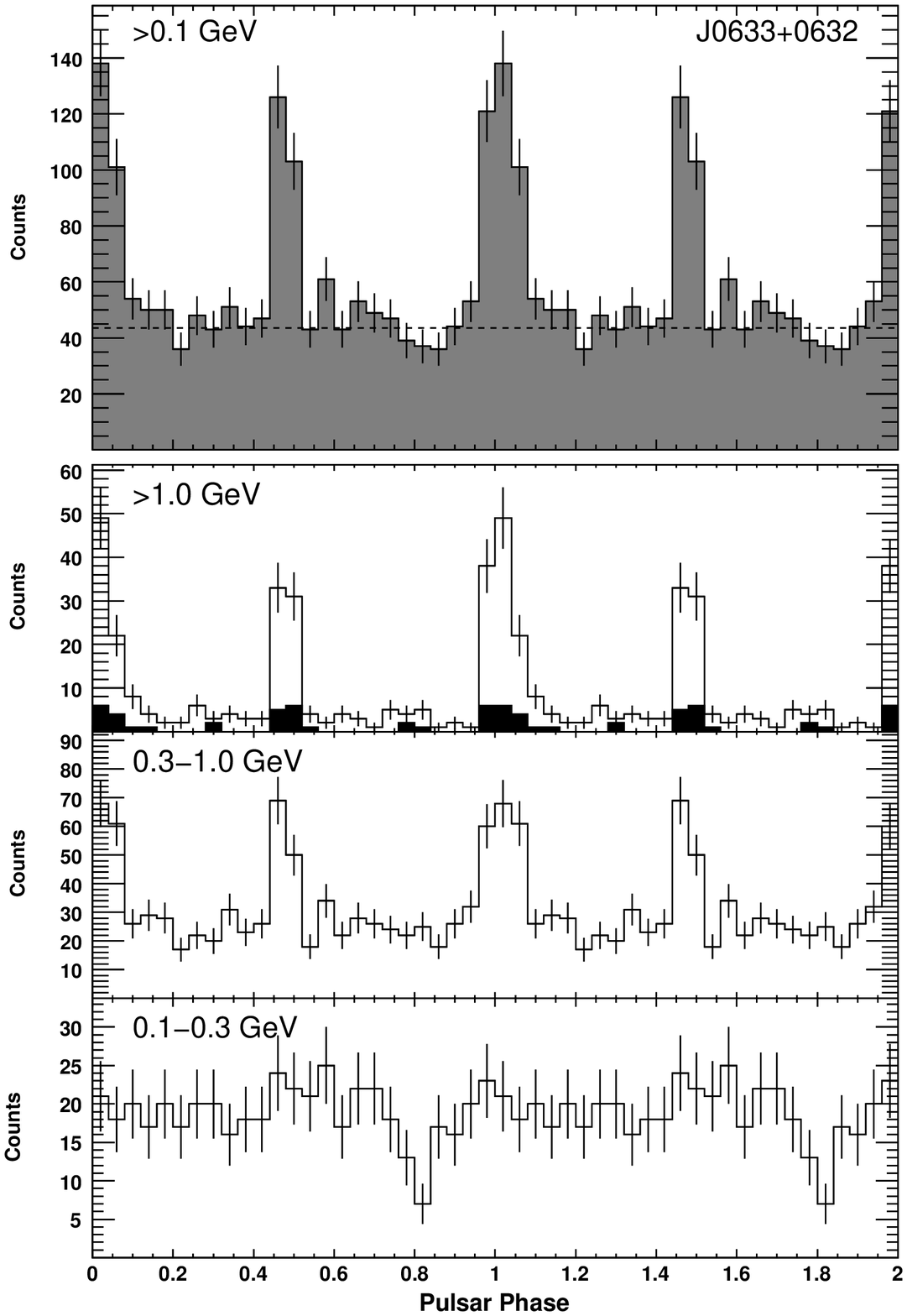}
\caption{Light curves for PSR J0633+0632 ($P=297$\,ms).
\label{fig:J0633p0632_lightcurve}}
\end{minipage}%
\hspace{1cm}%
%%----start of figure----
\begin{minipage}[t]{0.45\linewidth}
\centering
\includegraphics[width=\linewidth]{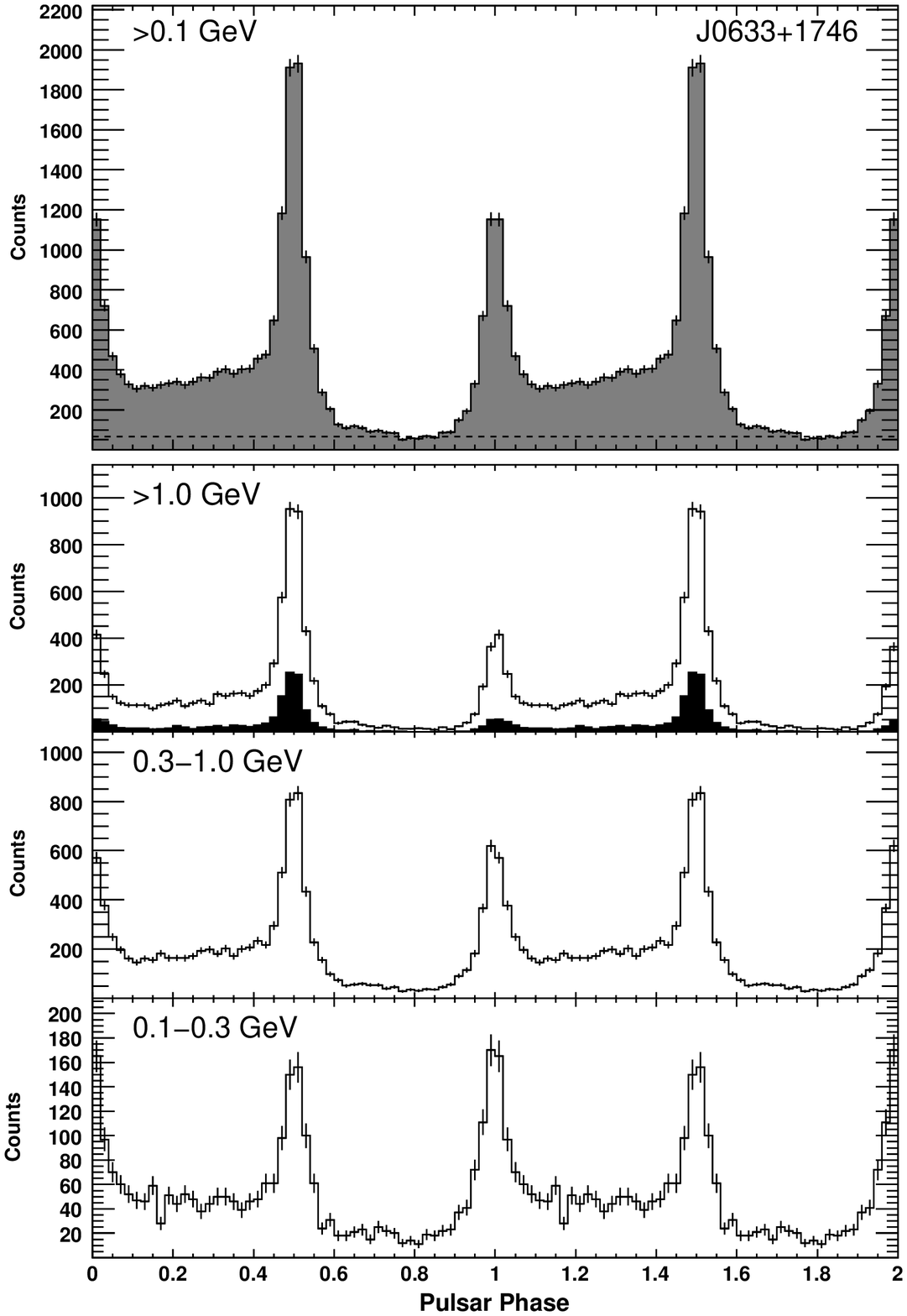}
\caption{Light curves for PSR J0633+1746 ($P=237$\,ms, Geminga pulsar).
\label{fig:J0633p1746_lightcurve}}
\end{minipage}\\
\end{sidewaysfigure}
\clearpage

%----start of new page----
\begin{sidewaysfigure}
\centering
%%----start of figure----
\begin{minipage}[t]{0.45\linewidth}
\centering
\includegraphics[width=\linewidth]{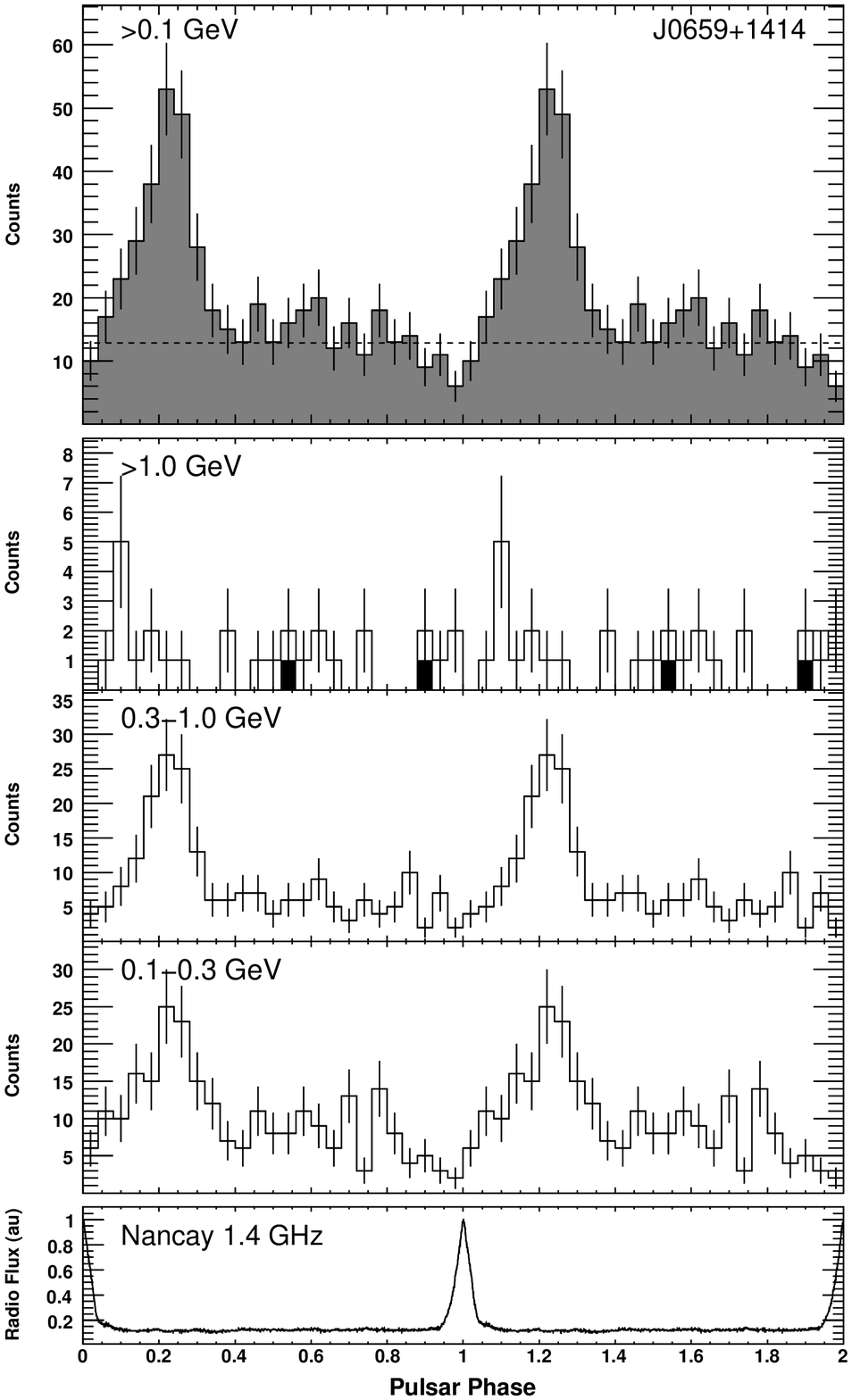}
\caption{Light curves for PSR J0659+1414 ($P=385$\,ms).
\label{fig:J0659p1414_lightcurve}}
\end{minipage}%
\hspace{1cm}%
%%----start of figure----
\begin{minipage}[t]{0.45\linewidth}
\centering
\includegraphics[width=\linewidth]{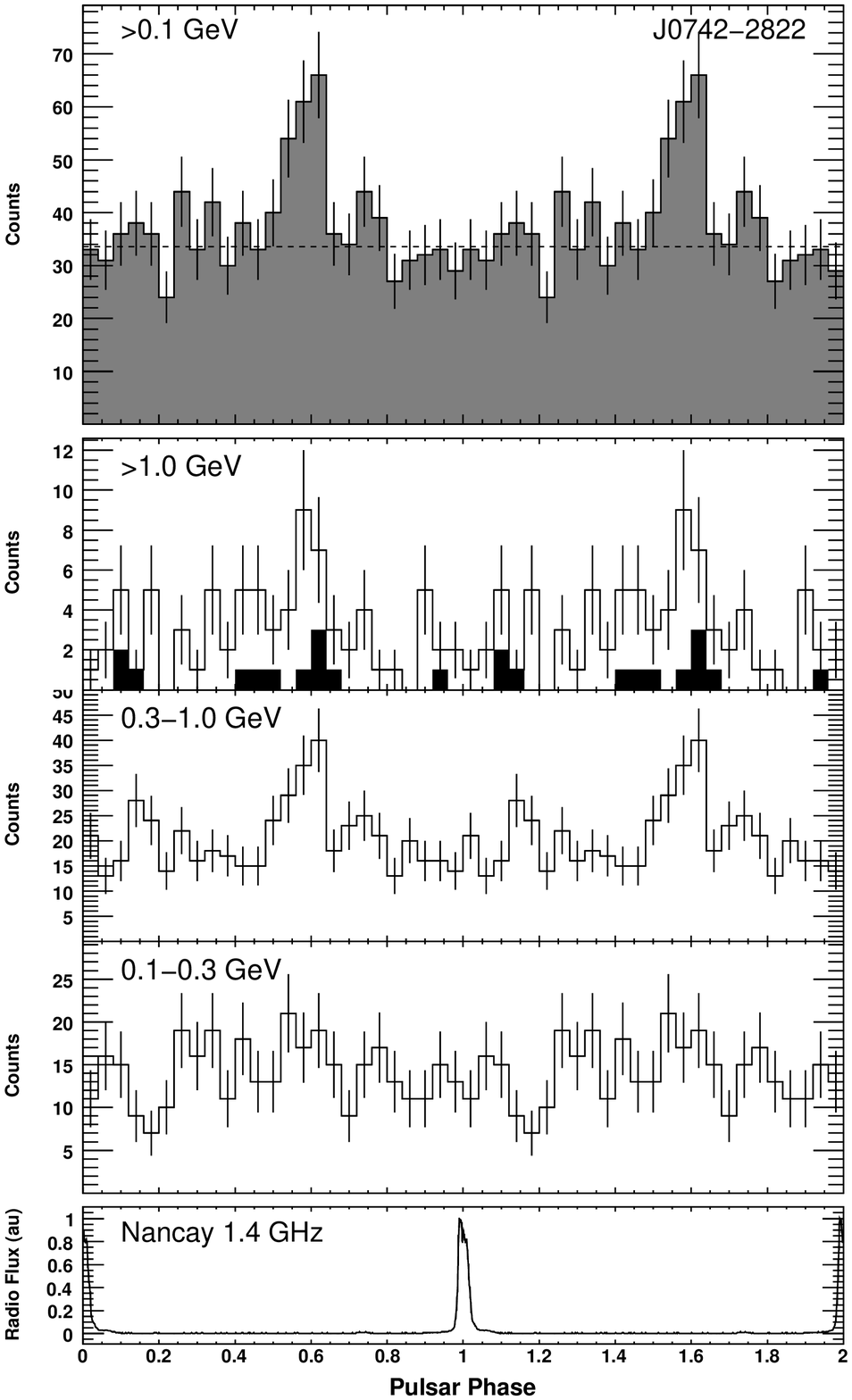}
\caption{Light curves for PSR J0742$-$2822 ($P=167$\,ms).
\label{fig:J0742m2822_lightcurve}}
\end{minipage}\\
\end{sidewaysfigure}
\clearpage

%----start of new page----
\begin{sidewaysfigure}
\centering
%%----start of figure----
\begin{minipage}[t]{0.45\linewidth}
\centering
\includegraphics[width=\linewidth]{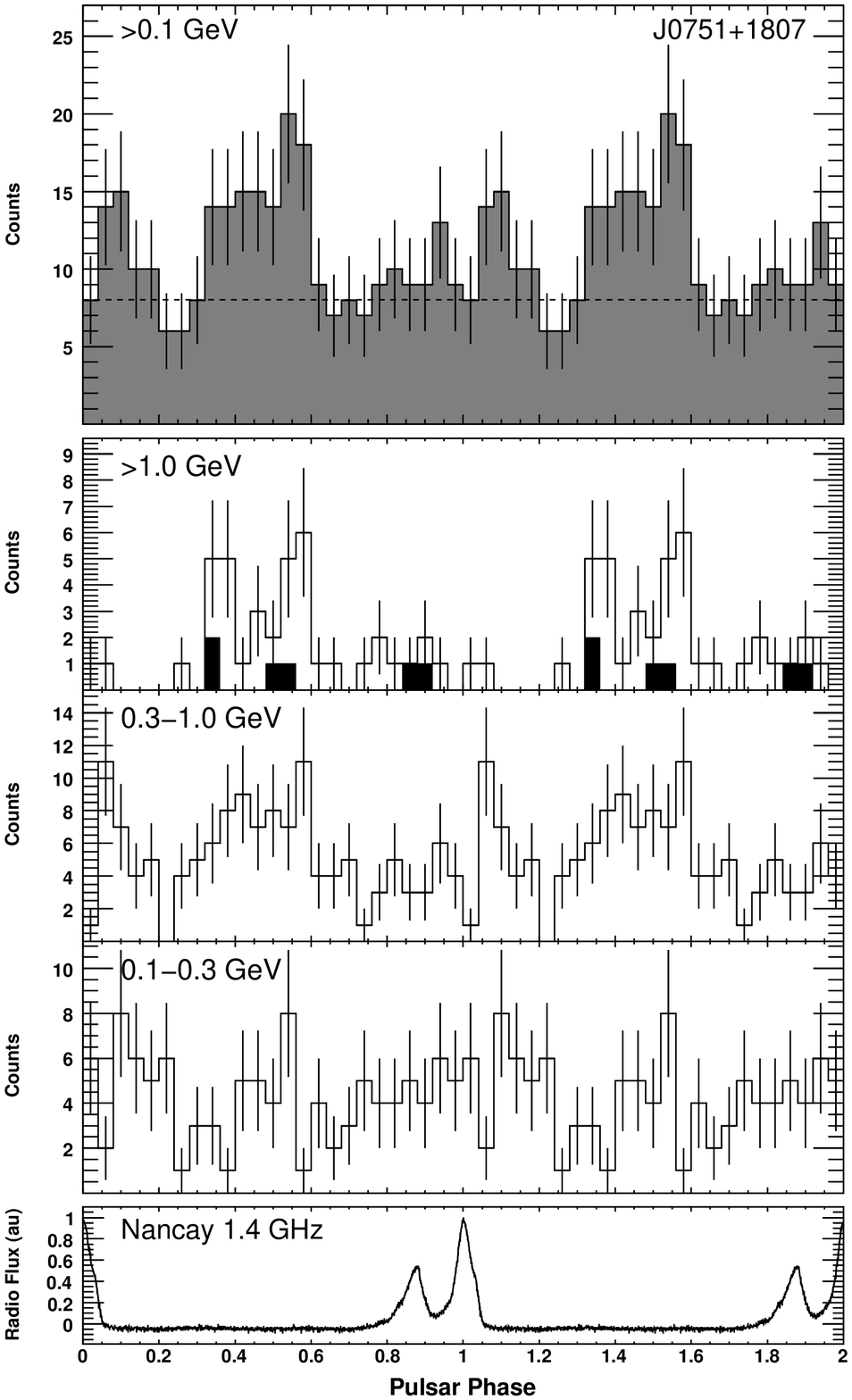}
\caption{Light curves for PSR J0751+1807 ($P=3.48$\,ms).
\label{fig:J0751p1807_lightcurve}}
\end{minipage}%
\hspace{1cm}%
%%----start of figure----
\begin{minipage}[t]{0.45\linewidth}
\centering
\includegraphics[width=\linewidth]{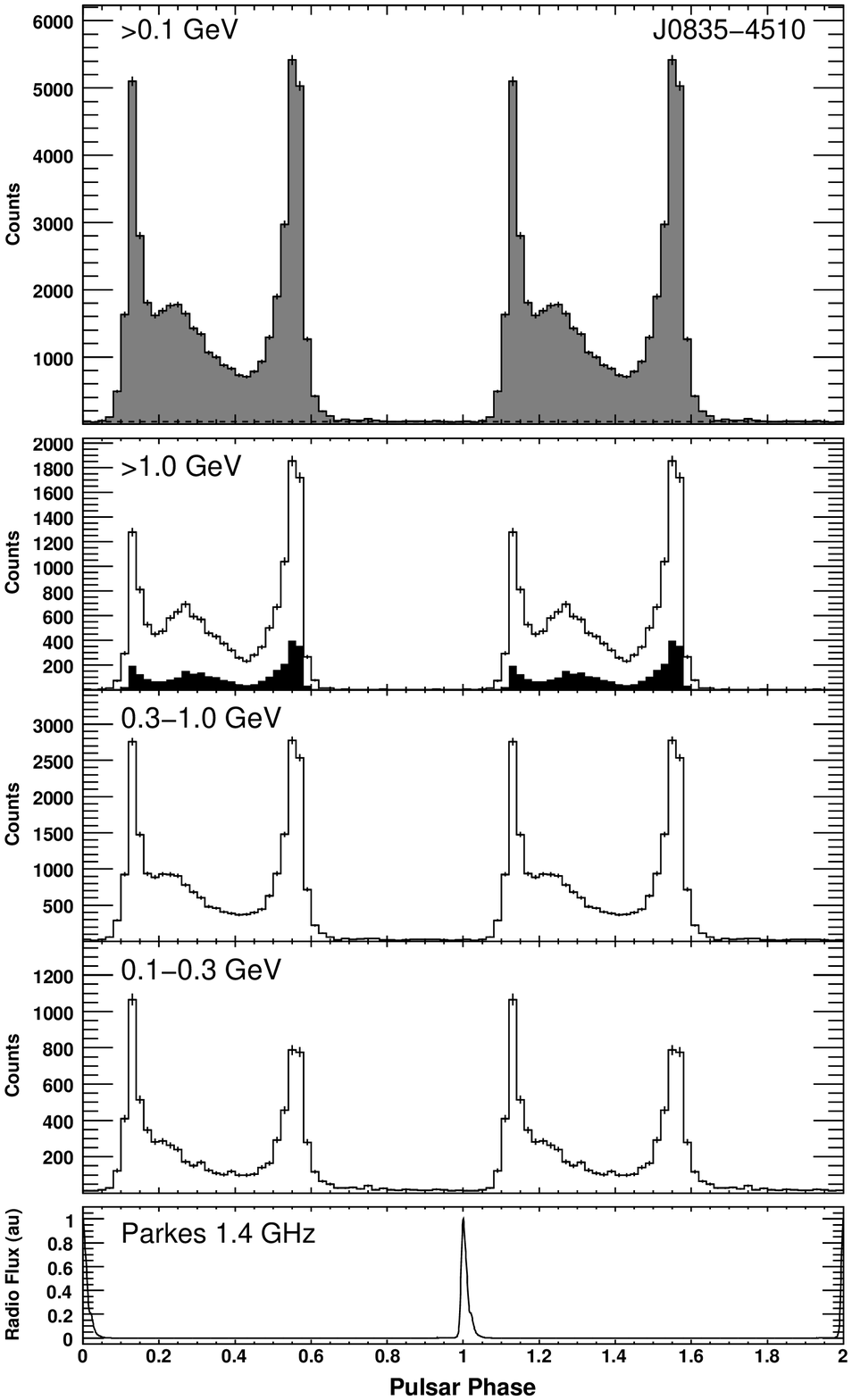}
\caption{Light curves for PSR J0835$-$4510 ($P=89.3$\,ms, Vela pulsar).
\label{fig:J0835m4510_lightcurve}}
\end{minipage}\\
\end{sidewaysfigure}
\clearpage

%----start of new page----
\begin{sidewaysfigure}
\centering
%%----start of figure----
\begin{minipage}[t]{0.45\linewidth}
\centering
\includegraphics[width=\linewidth]{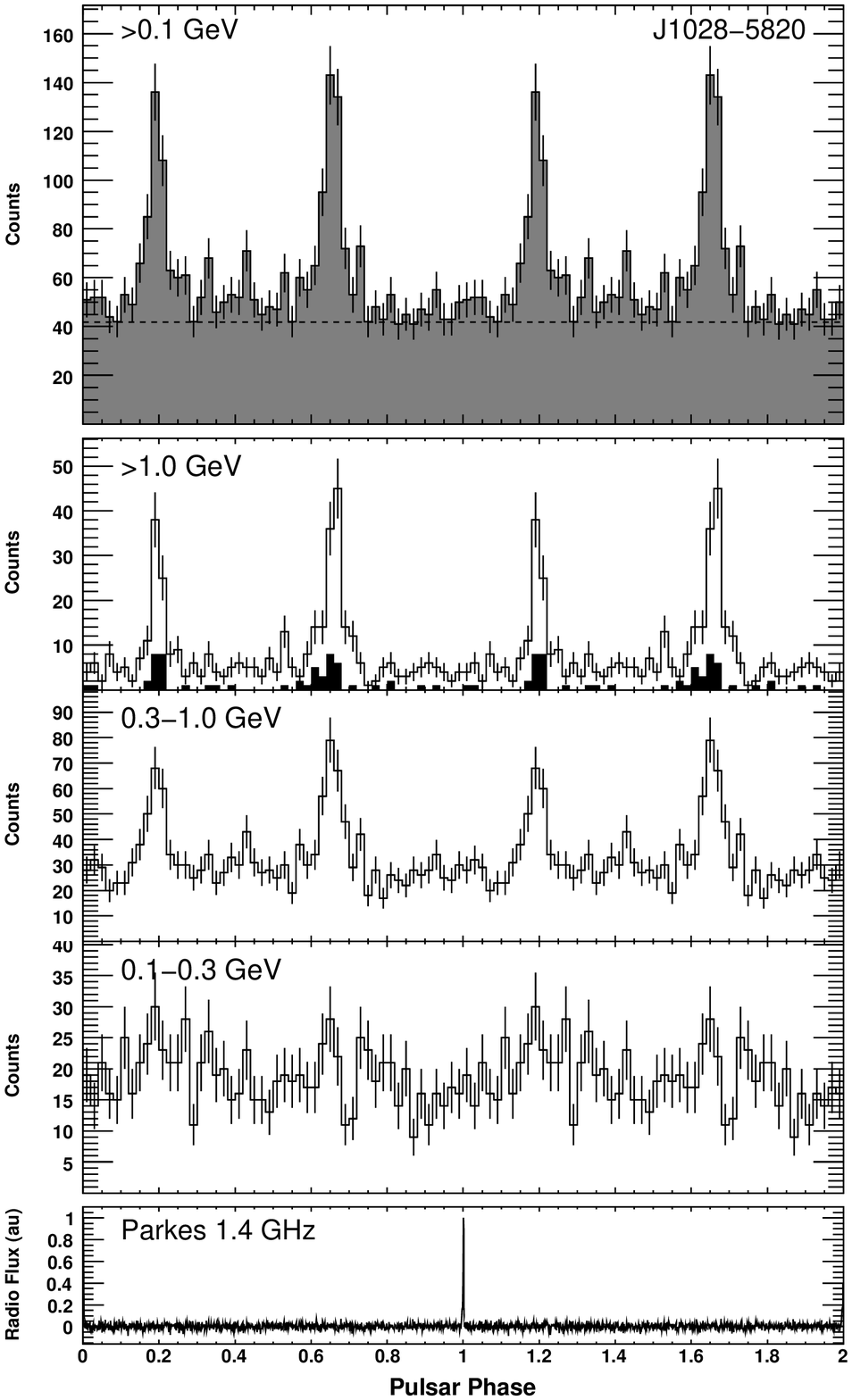}
\caption{Light curves for PSR J1028$-$5819 ($P=91.4$\,ms).
\label{fig:J1028m5820_lightcurve}}
\end{minipage}%
\hspace{1cm}%
%%----start of figure----
\begin{minipage}[t]{0.45\linewidth}
\centering
\includegraphics[width=\linewidth]{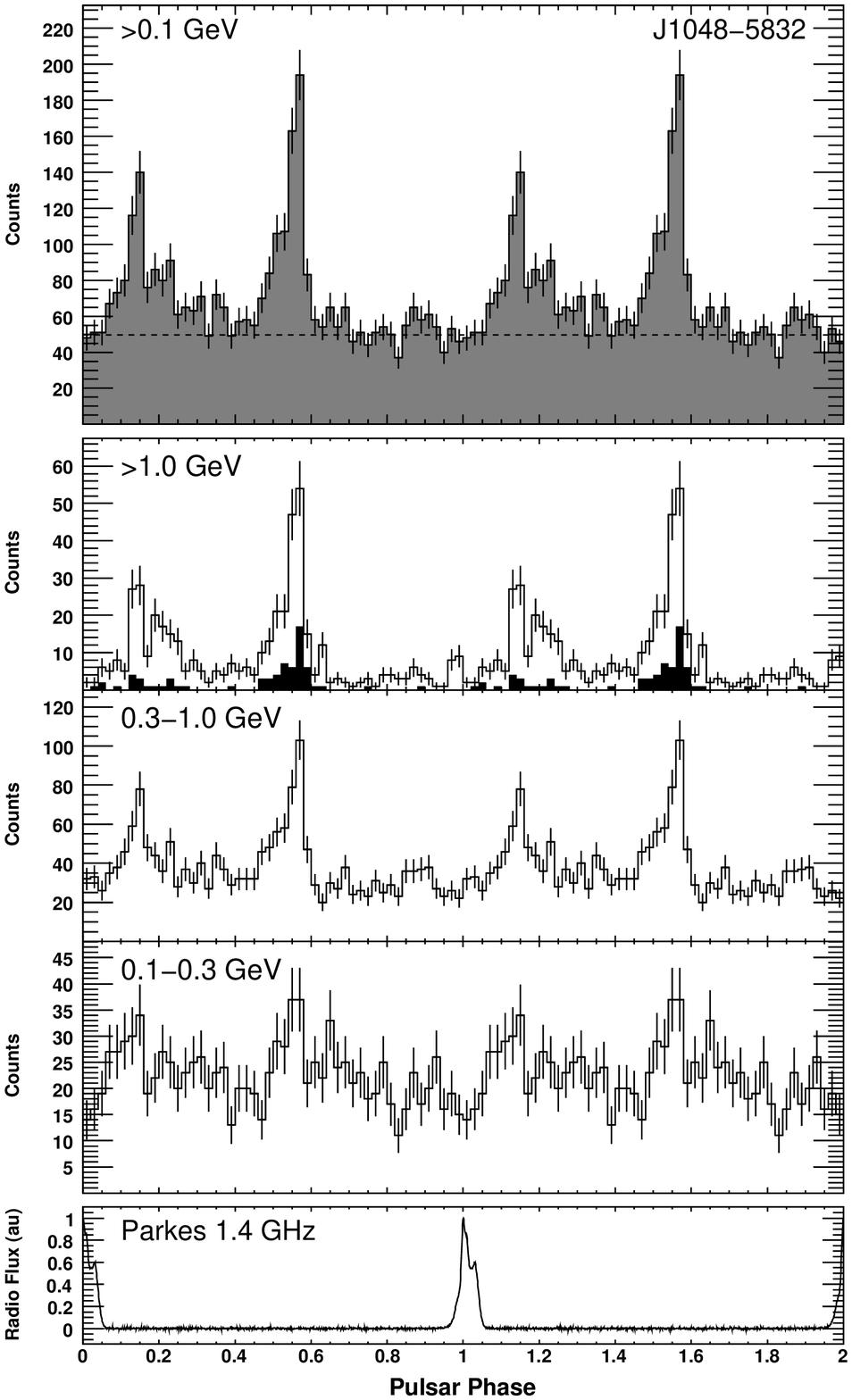}
\caption{Light curves for PSR J1048$-$5832 ($P=124$\,ms).
\label{fig:J1048m5832_lightcurve}}
\end{minipage}\\
\end{sidewaysfigure}
\clearpage

%----start of new page----
\begin{sidewaysfigure}
\centering
%%----start of figure----
\begin{minipage}[t]{0.45\linewidth}
\centering
\includegraphics[width=\linewidth]{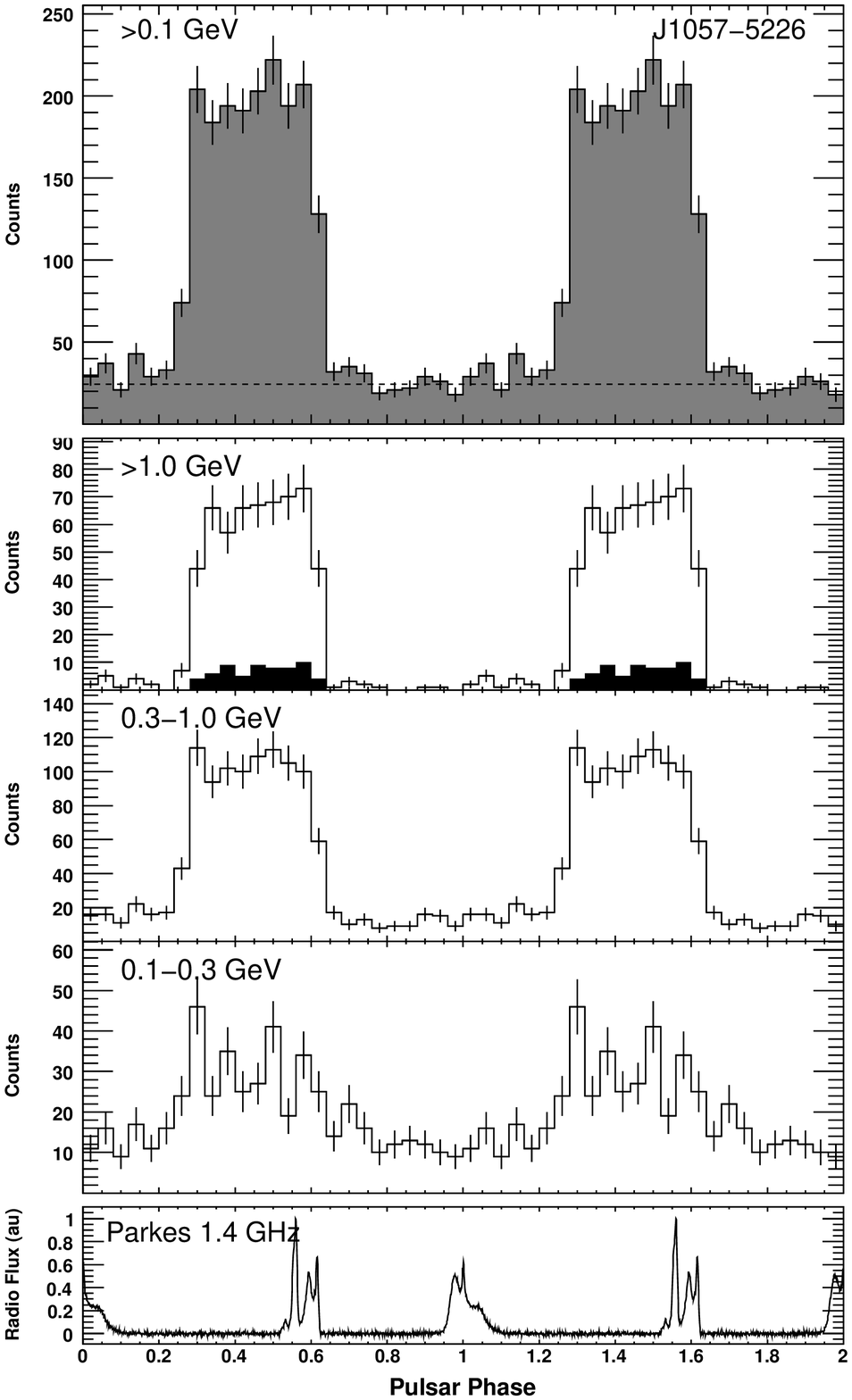}
\caption{Light curves for PSR J1057$-$5226 ($P=197$\,ms, PSR B1055$-$52).
\label{fig:J1057m5226_lightcurve}}
\end{minipage}%
\hspace{1cm}%
%%----start of figure----
\begin{minipage}[t]{0.45\linewidth}
\centering
\includegraphics[width=\linewidth]{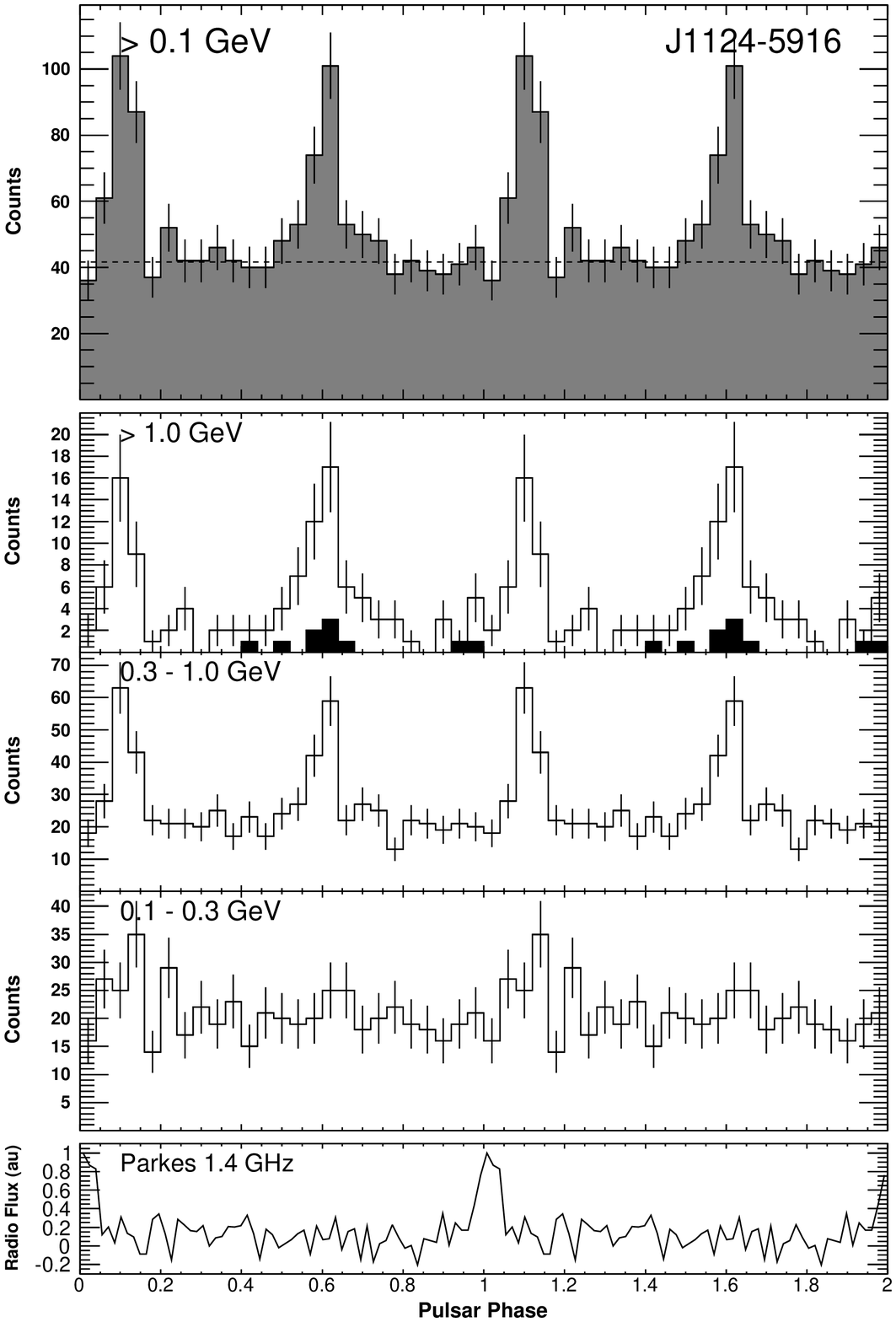}
\caption{Light curves for PSR J1124$-$5916 ($P=135$\,ms). The radio phase alignment has
been corrected, as per an Erratum sent to the ApJ (December 2010).
\label{fig:J1124m5916_lightcurve}}
\end{minipage}\\
\end{sidewaysfigure}
\clearpage

%----start of new page----
\begin{sidewaysfigure}
\centering
%%----start of figure----
\begin{minipage}[t]{0.45\linewidth}
\centering
\includegraphics[width=\linewidth]{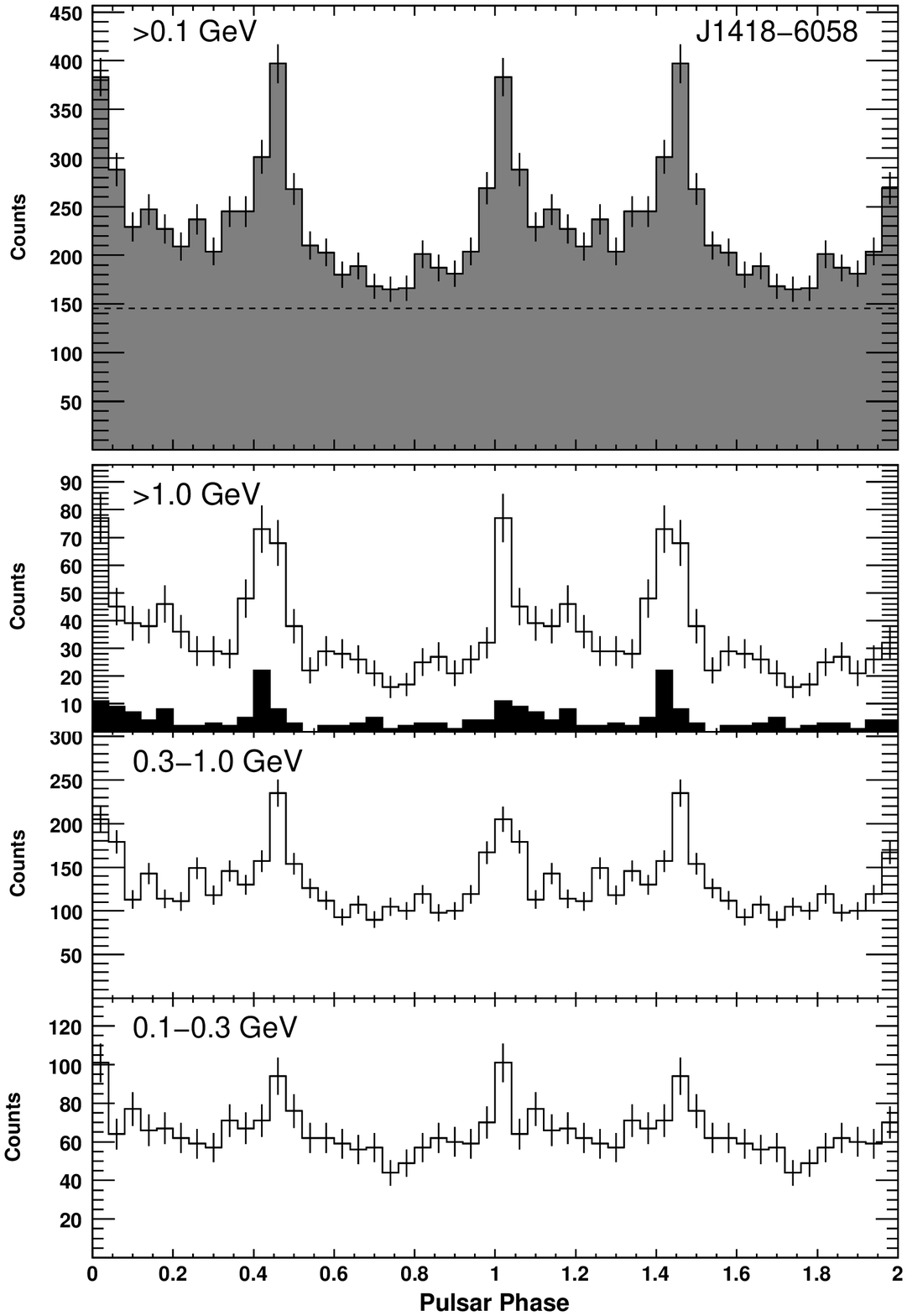}
\caption{Light curves for PSR J1418$-$6058 ($P=111$\,ms).
\label{fig:J1418m6058_lightcurve}}
\end{minipage}%
\hspace{1cm}%
%%----start of figure----
\begin{minipage}[t]{0.45\linewidth}
\centering
\includegraphics[width=\linewidth]{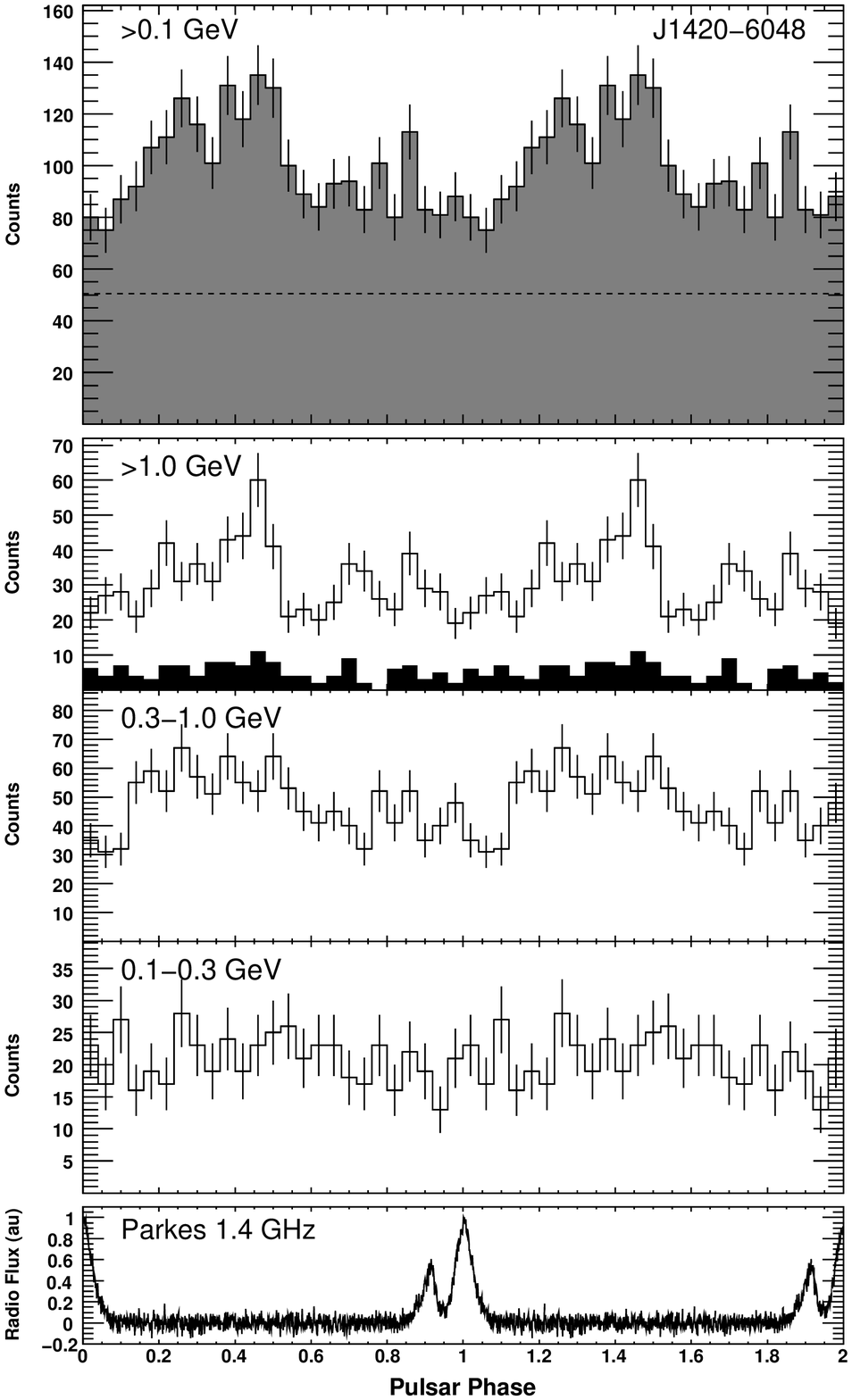}
\caption{Light curves for PSR J1420$-$6048 ($P=68.2$\,ms).
\label{fig:J1420m6048_lightcurve}}
\end{minipage}\\
\end{sidewaysfigure}
\clearpage

%%----start of figure----
% \begin{minipage}[t]{0.45\linewidth}
% \centering
% \includegraphics[width=\linewidth]{LightCurves/J1418m6058_catalog_lightcurve.eps}
% \caption{Light curves for PSR J1418--6058.
% \label{fig:J1418m6058_lightcurve}}
% \end{minipage}%
% \hspace{1cm}%

%----start of new page----
\begin{sidewaysfigure}
\centering
%%----start of figure----
\begin{minipage}[t]{0.45\linewidth}
\centering
\includegraphics[width=\linewidth]{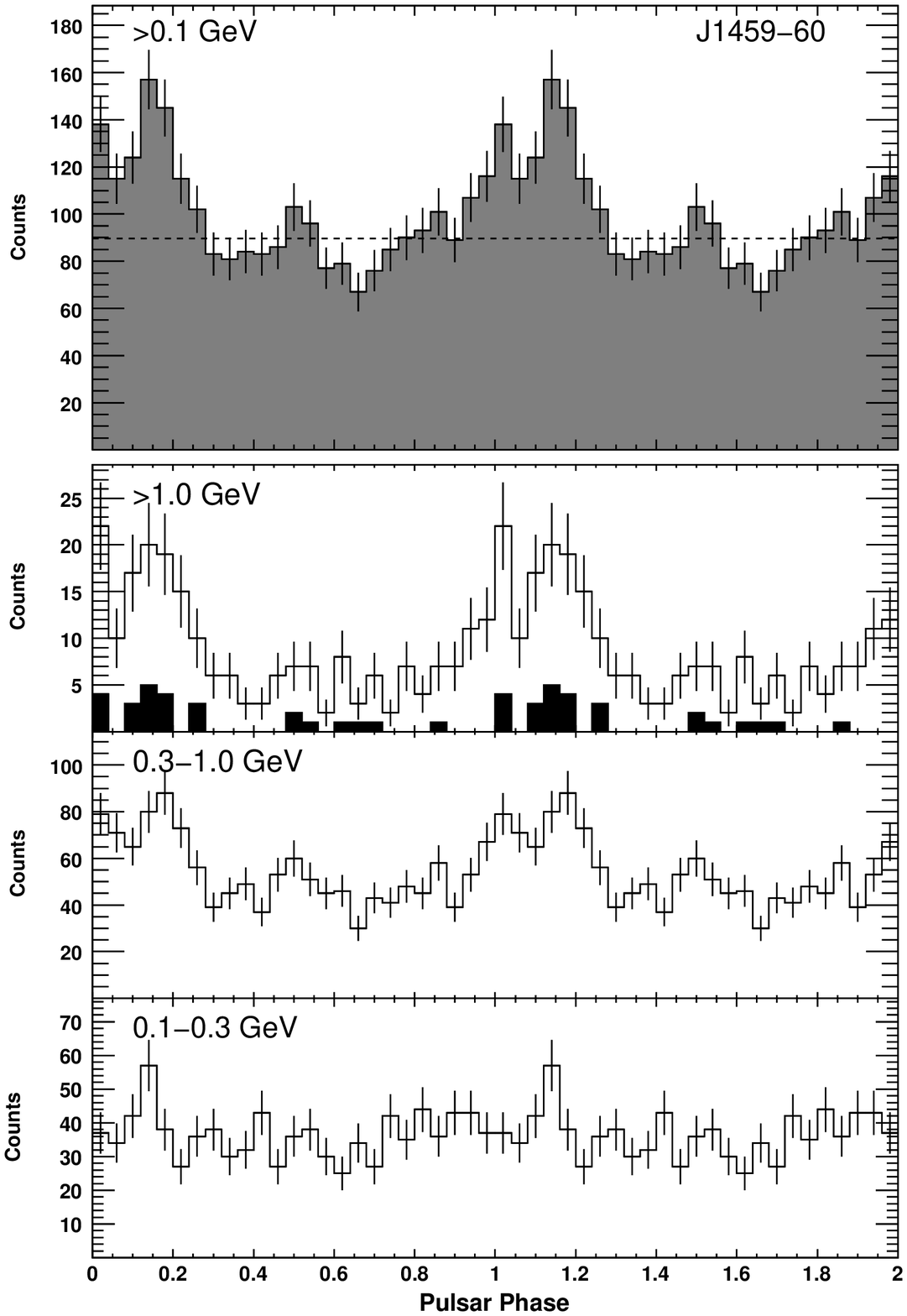}
\caption{Light curves for PSR J1459$-$60 ($P=103$\,ms).
\label{fig:J1459m60_lightcurve}}
\end{minipage}%
\hspace{1cm}%
%%----start of figure----
\begin{minipage}[t]{0.45\linewidth}
\centering
\includegraphics[width=\linewidth]{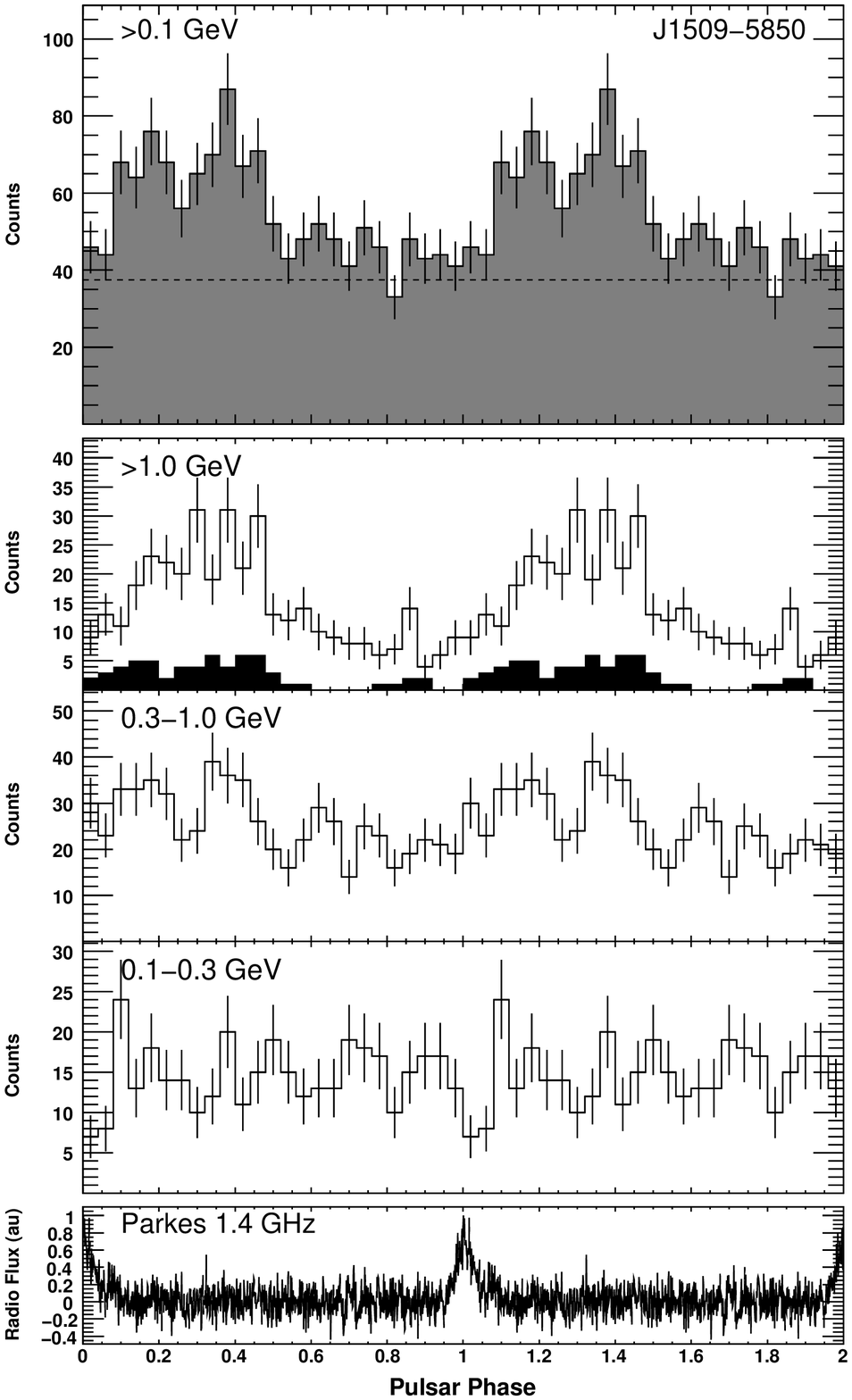}
\caption{Light curves for PSR J1509$-$5850 ($P=88.9$\,ms).
\label{fig:J1509m5850_lightcurve}}
\end{minipage}\\
\end{sidewaysfigure}
\clearpage

%----start of new page----
\begin{sidewaysfigure}
\centering
%%----start of figure----
\begin{minipage}[t]{0.45\linewidth}
\centering
\includegraphics[width=\linewidth]{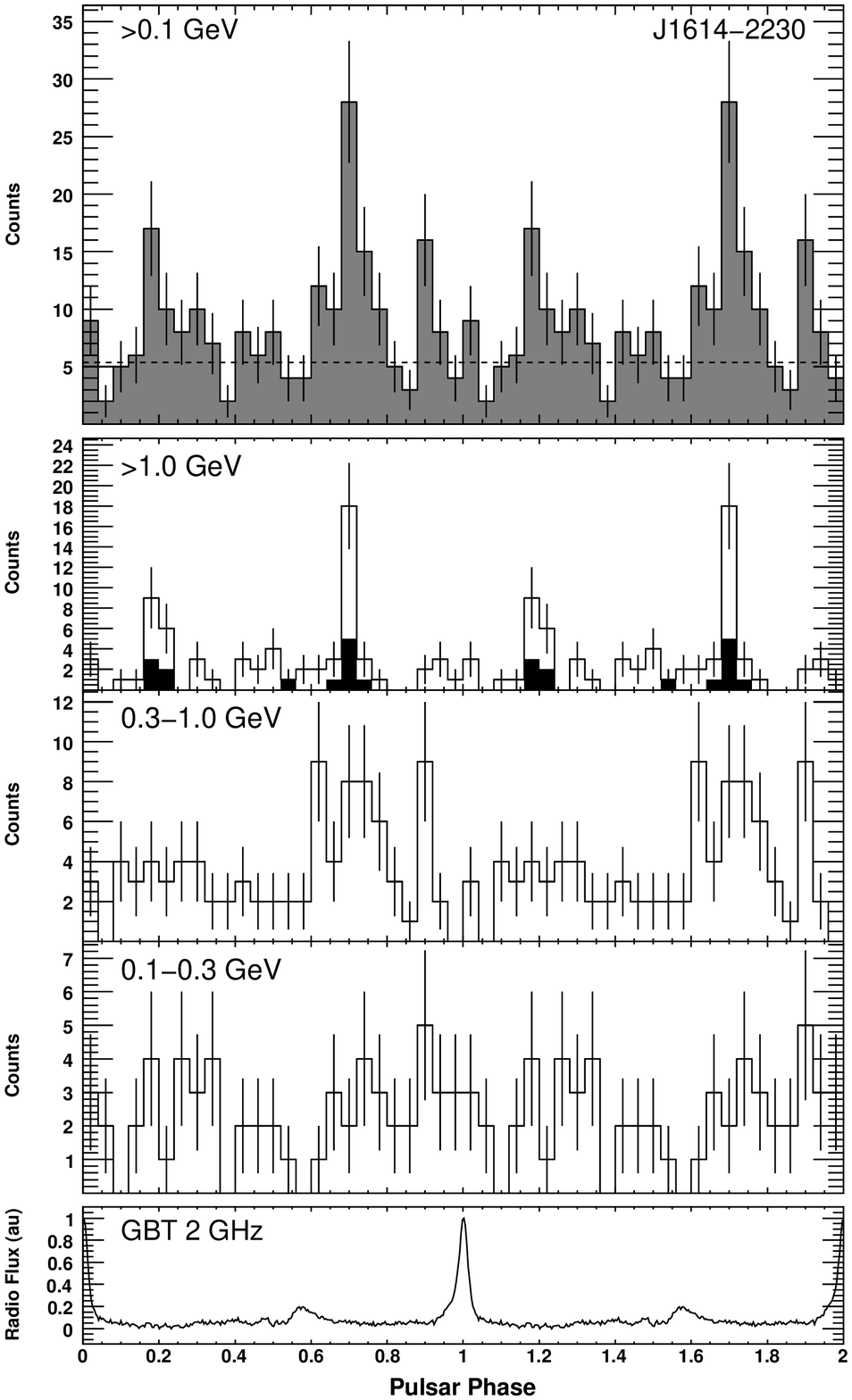}
\caption{Light curves for PSR J1614$-$2230 ($P=3.15$\,ms).
\label{fig:J1614m2230_lightcurve}}
\end{minipage}%
\hspace{1cm}%
%%----start of figure----
\begin{minipage}[t]{0.45\linewidth}
\centering
\includegraphics[width=\linewidth]{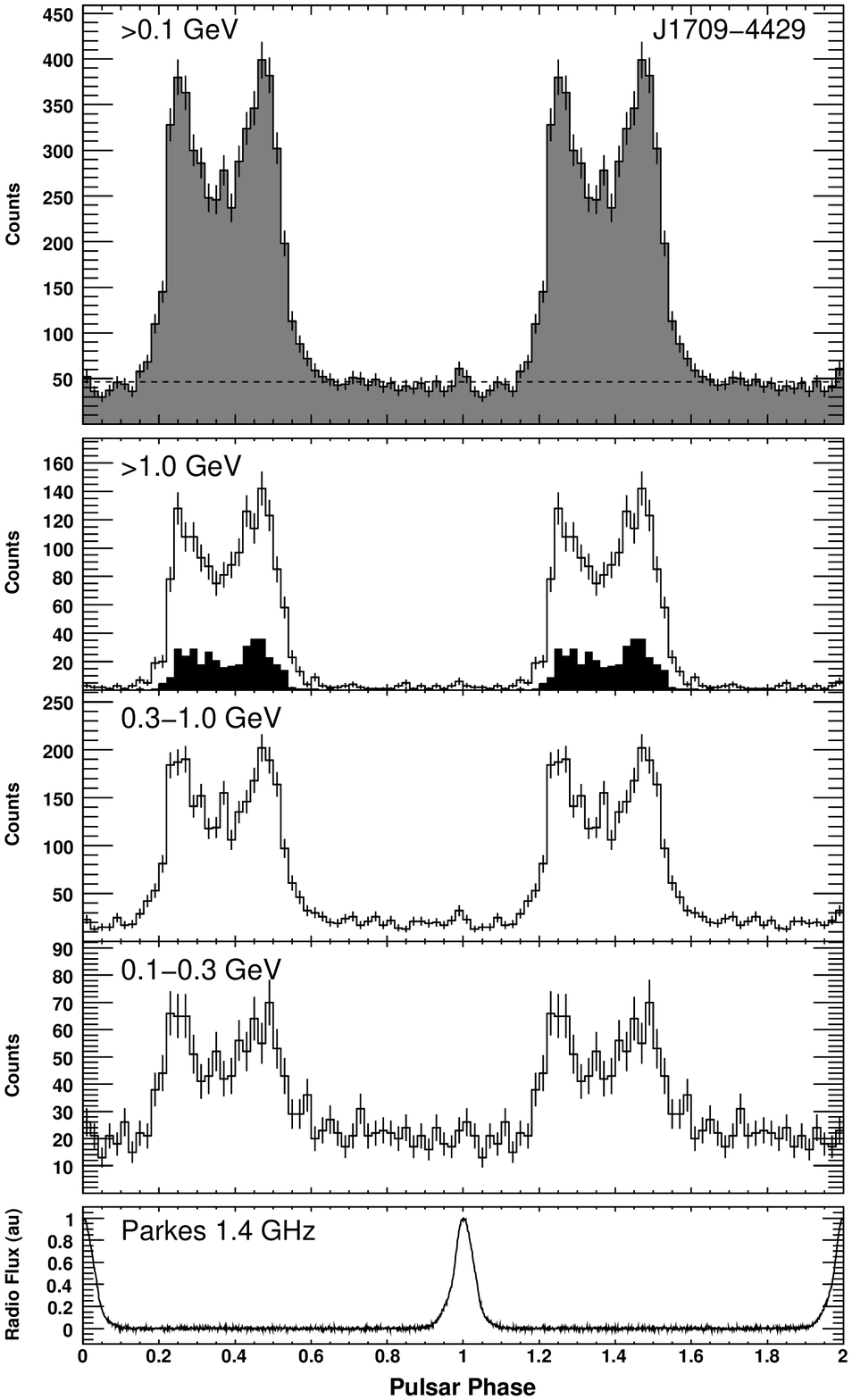}
\caption{Light curves for PSR J1709$-$4429 ($P=102$\,ms, PSR B1706$-$44).
\label{fig:J1709m4429_lightcurve}}
\end{minipage}\\
\end{sidewaysfigure}
\clearpage

%----start of new page----
\begin{sidewaysfigure}
\centering
%%----start of figure----
\begin{minipage}[t]{0.45\linewidth}
\centering
\includegraphics[width=\linewidth]{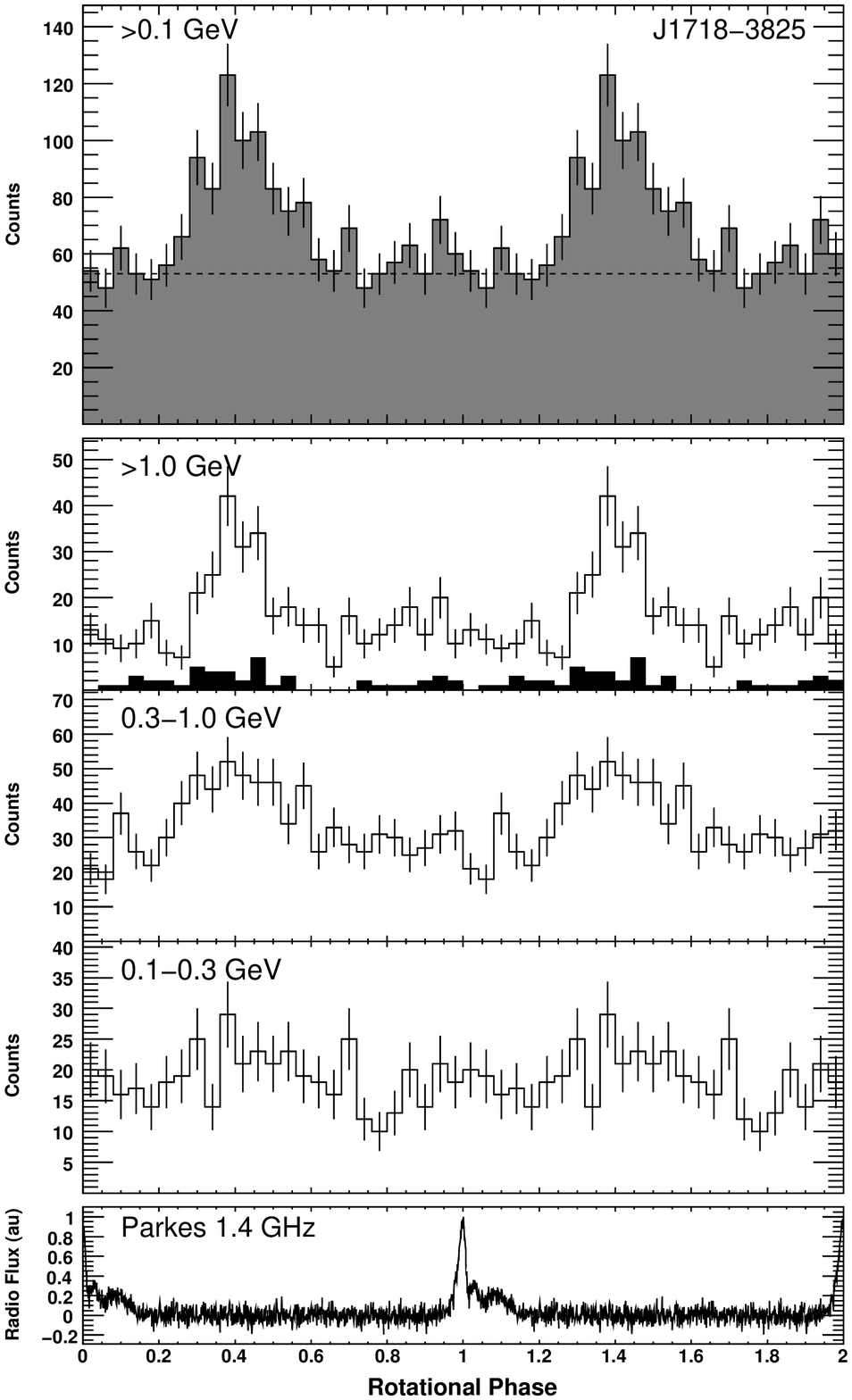}
\caption{Light curves for PSR J1718$-$3825 ($P=74.7$\,ms).
\label{fig:J1718m3825_lightcurve}}
\end{minipage}%
\hspace{1cm}%
%%----start of figure----
\begin{minipage}[t]{0.45\linewidth}
\centering
\includegraphics[width=\linewidth]{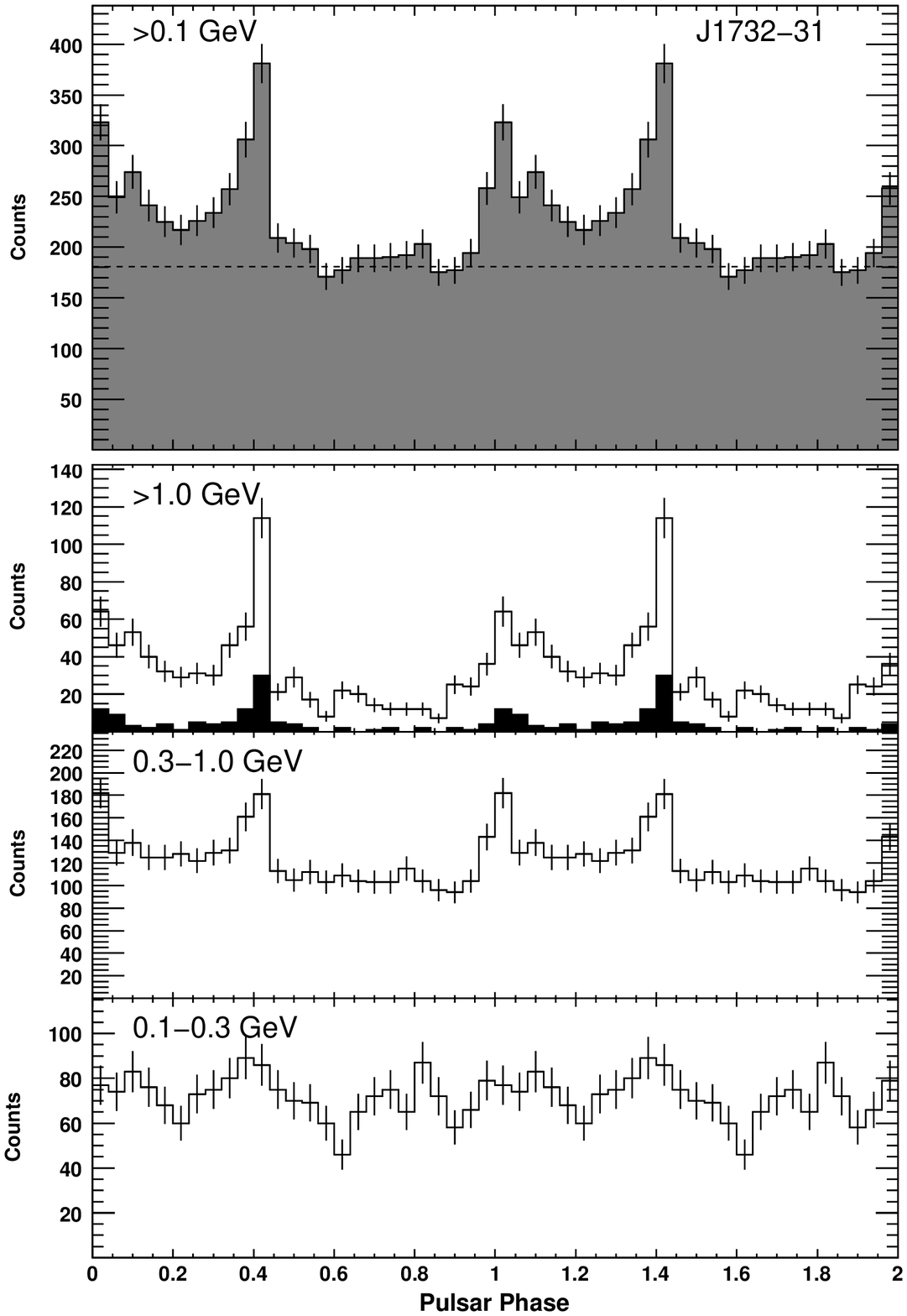}
\caption{Light curves for PSR J1732$-$31 ($P=197$\,ms).
\label{fig:J1732m31_lightcurve}}
\end{minipage}\\
\end{sidewaysfigure}
\clearpage

%----start of new page----
\begin{sidewaysfigure}
\centering
%%----start of figure----
\begin{minipage}[t]{0.45\linewidth}
\centering
\includegraphics[width=\linewidth]{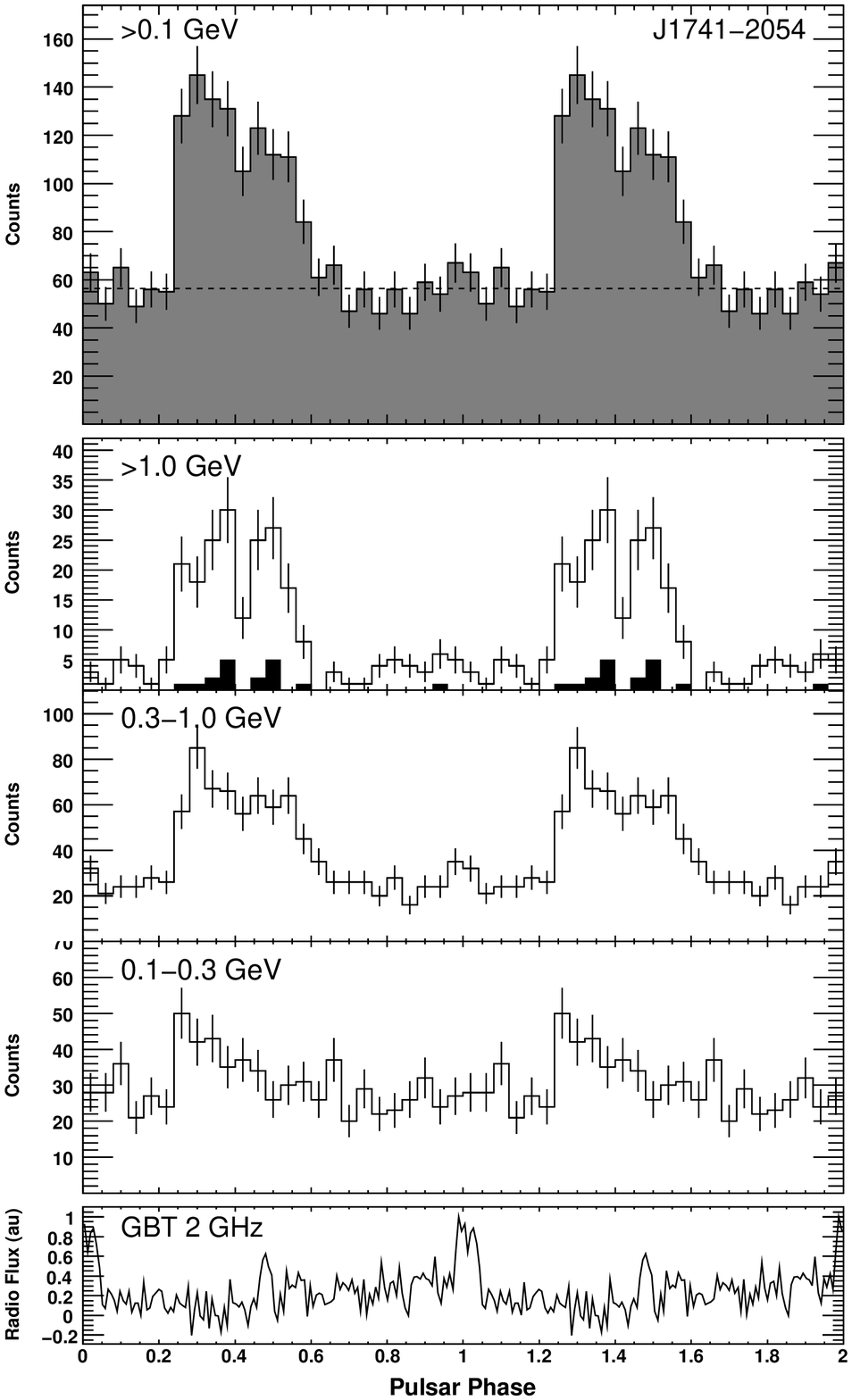}
\caption{Light curves for PSR J1741$-$2054 ($P=414$\,ms). While this pulsar is detected in the radio \citep{Camilo2009}, it was discovered by the LAT and is considered a gamma-ray-selected pulsar.
\label{fig:J1741m2054_lightcurve}}
\end{minipage}%
\hspace{1cm}%
%%----start of figure----
\begin{minipage}[t]{0.45\linewidth}
\centering
\includegraphics[width=\linewidth]{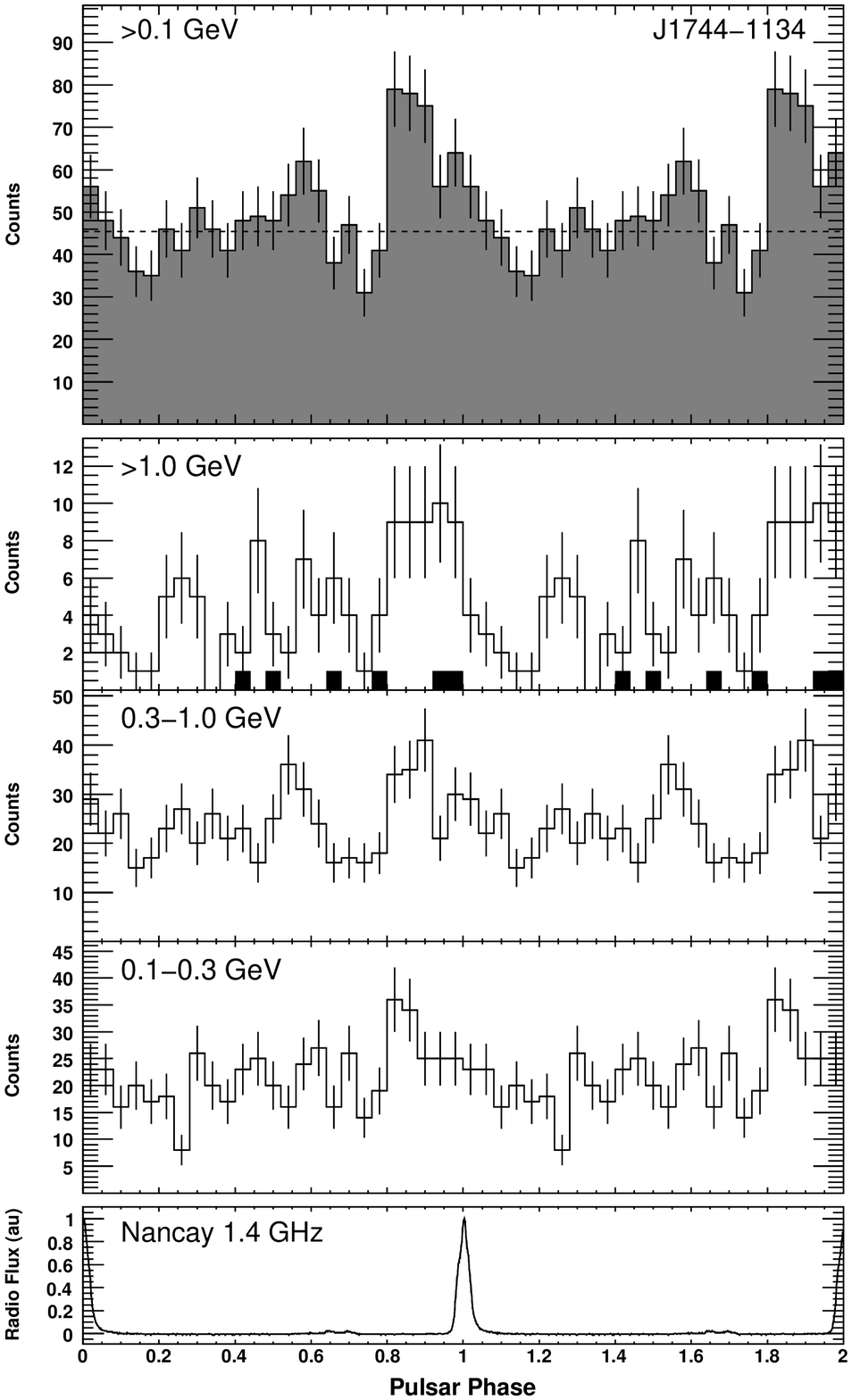}
\caption{Light curves for PSR J1744$-$1134 ($P=4.08$\,ms).
\label{fig:J1744m1134_lightcurve}}
\end{minipage}\\
\end{sidewaysfigure}
\clearpage

%----start of new page----
\begin{sidewaysfigure}
\centering
%%----start of figure----
\begin{minipage}[t]{0.45\linewidth}
\centering
\includegraphics[width=\linewidth]{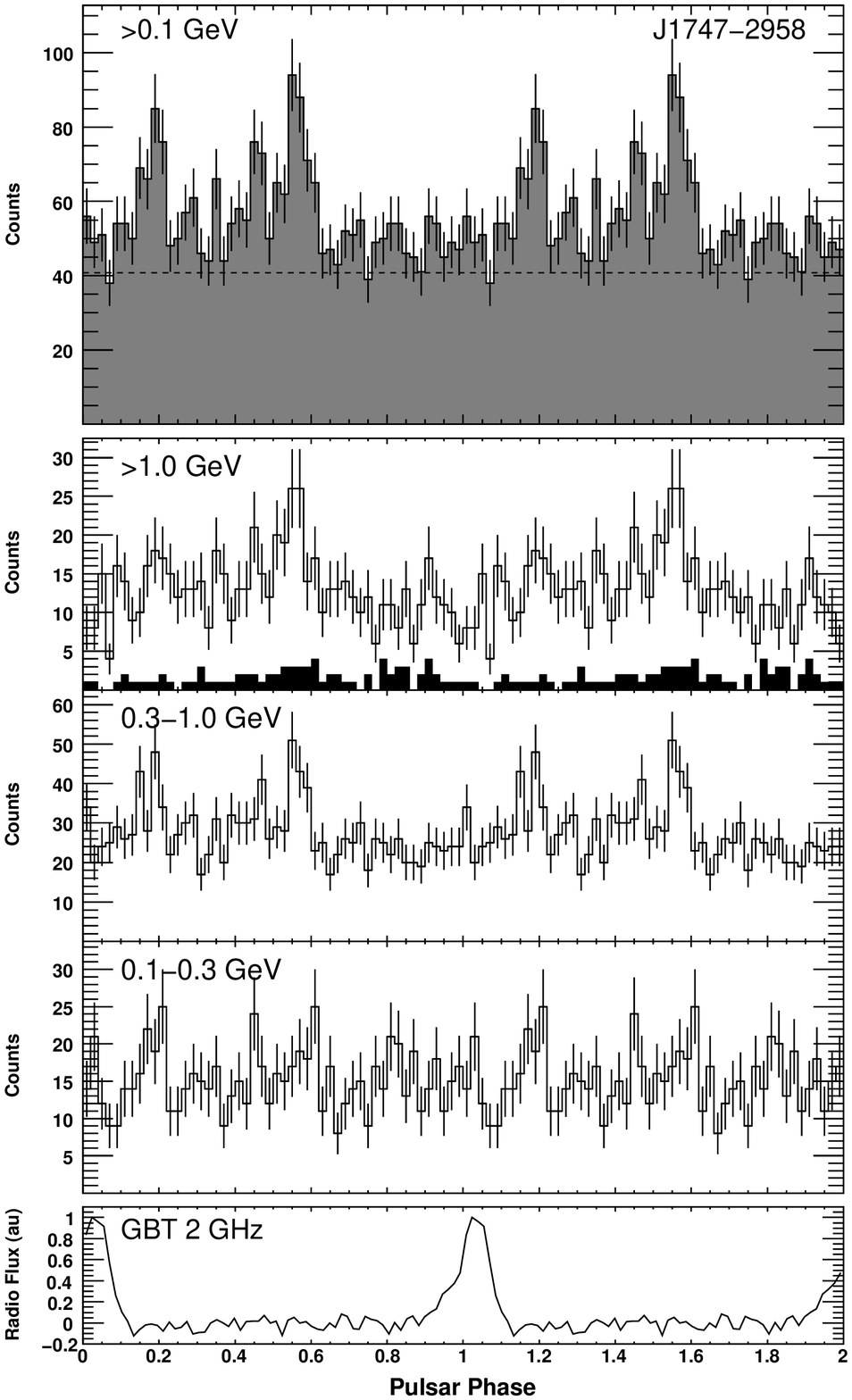}
\caption{Light curves for PSR J1747$-$2958 ($P=98.8$\,ms).
\label{fig:J1747m2958_lightcurve}}
\end{minipage}%
\hspace{1cm}%
%%----start of figure----
\begin{minipage}[t]{0.45\linewidth}
\centering
\includegraphics[width=\linewidth]{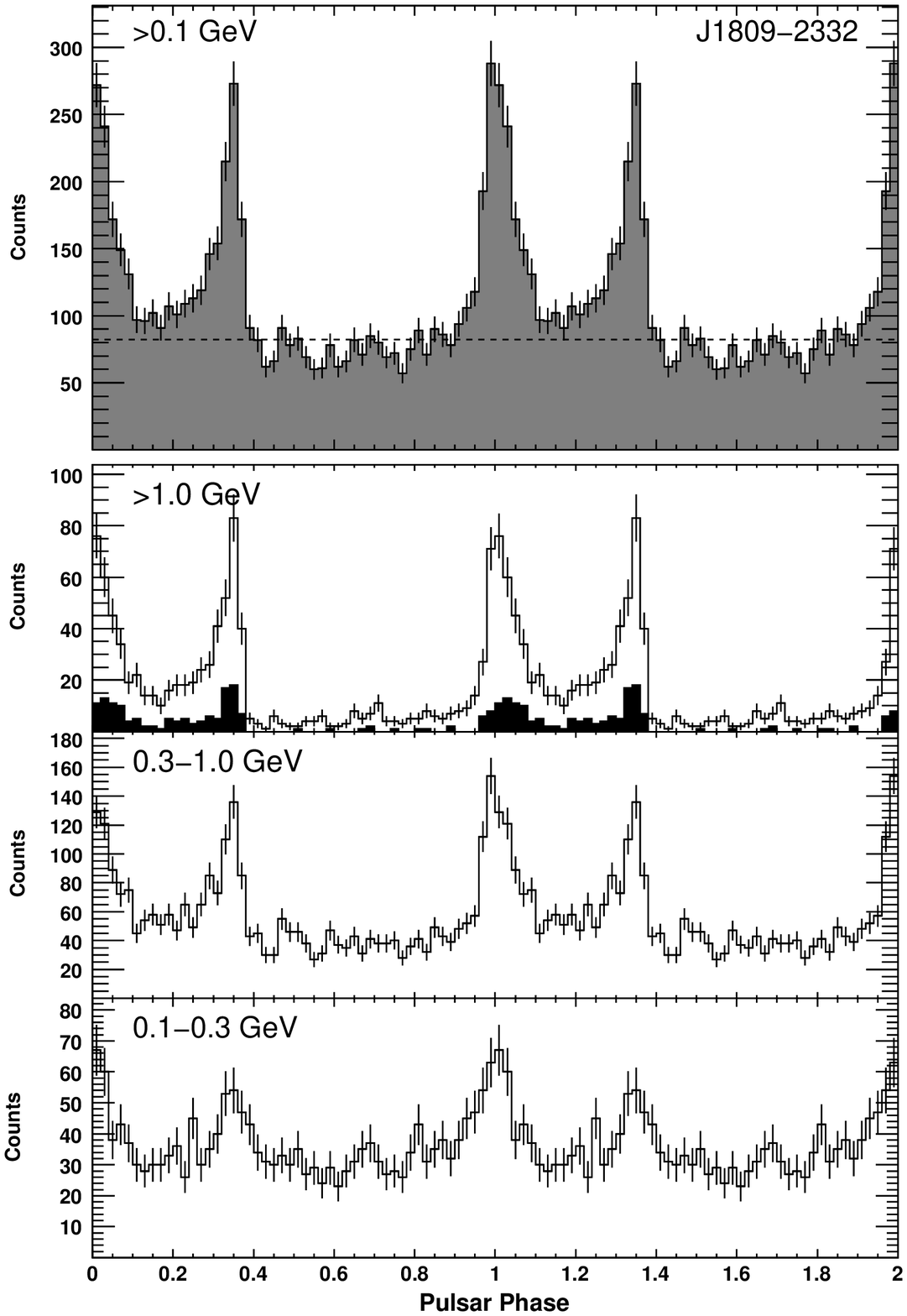}
\caption{Light curves for PSR J1809$-$2332 ($P=147$\,ms).
\label{fig:J1809m2332_lightcurve}}
\end{minipage}\\
\end{sidewaysfigure}
\clearpage

%----start of new page----
\begin{sidewaysfigure}
\centering
%%----start of figure----
\begin{minipage}[t]{0.45\linewidth}
\centering
\includegraphics[width=\linewidth]{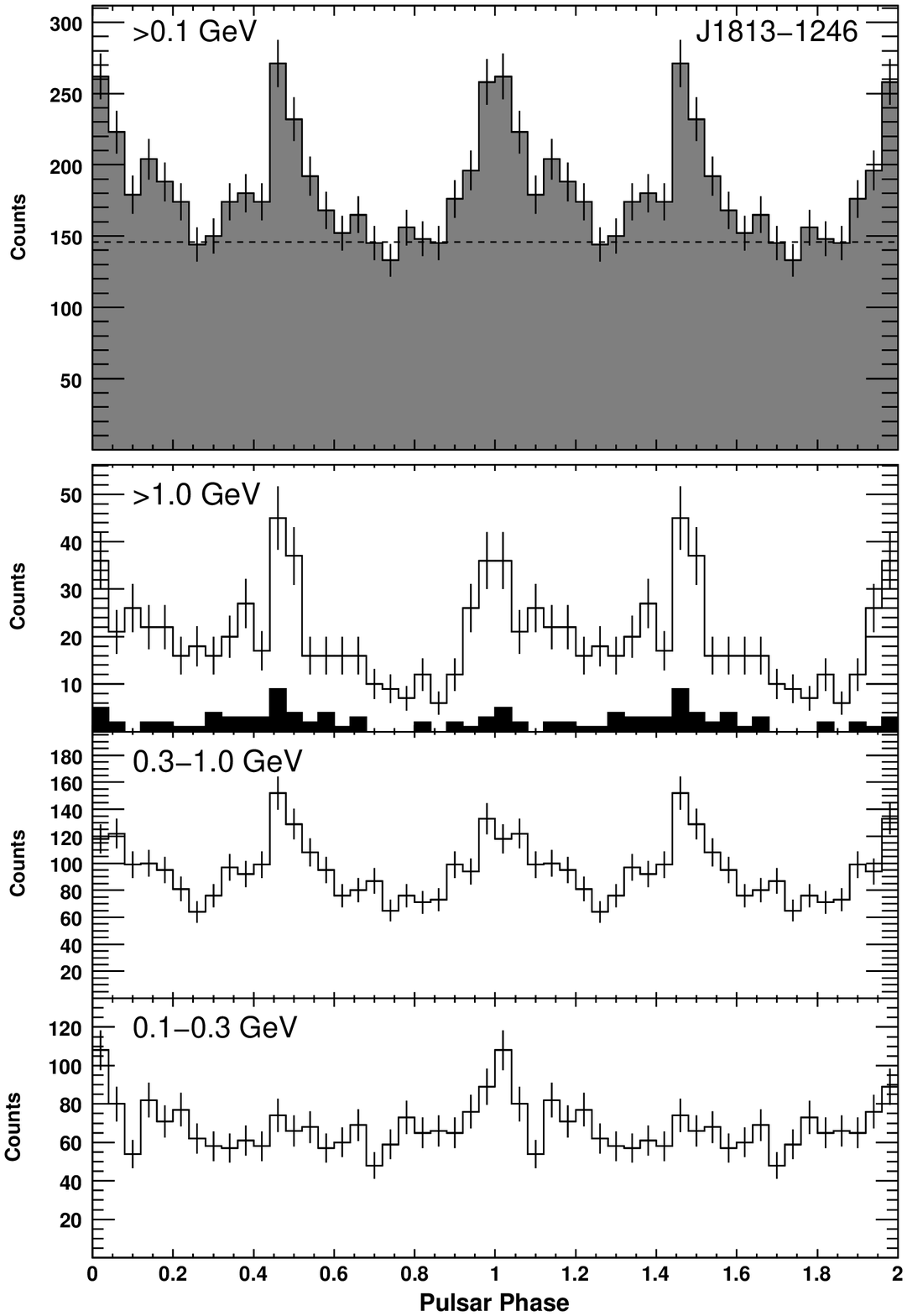}
\caption{Light curves for PSR J1813$-$1246 ($P=48.1$\,ms).
\label{fig:J1813m1246_lightcurve}}
\end{minipage}%
\hspace{1cm}%
%%----start of figure----
\begin{minipage}[t]{0.45\linewidth}
\centering
\includegraphics[width=\linewidth]{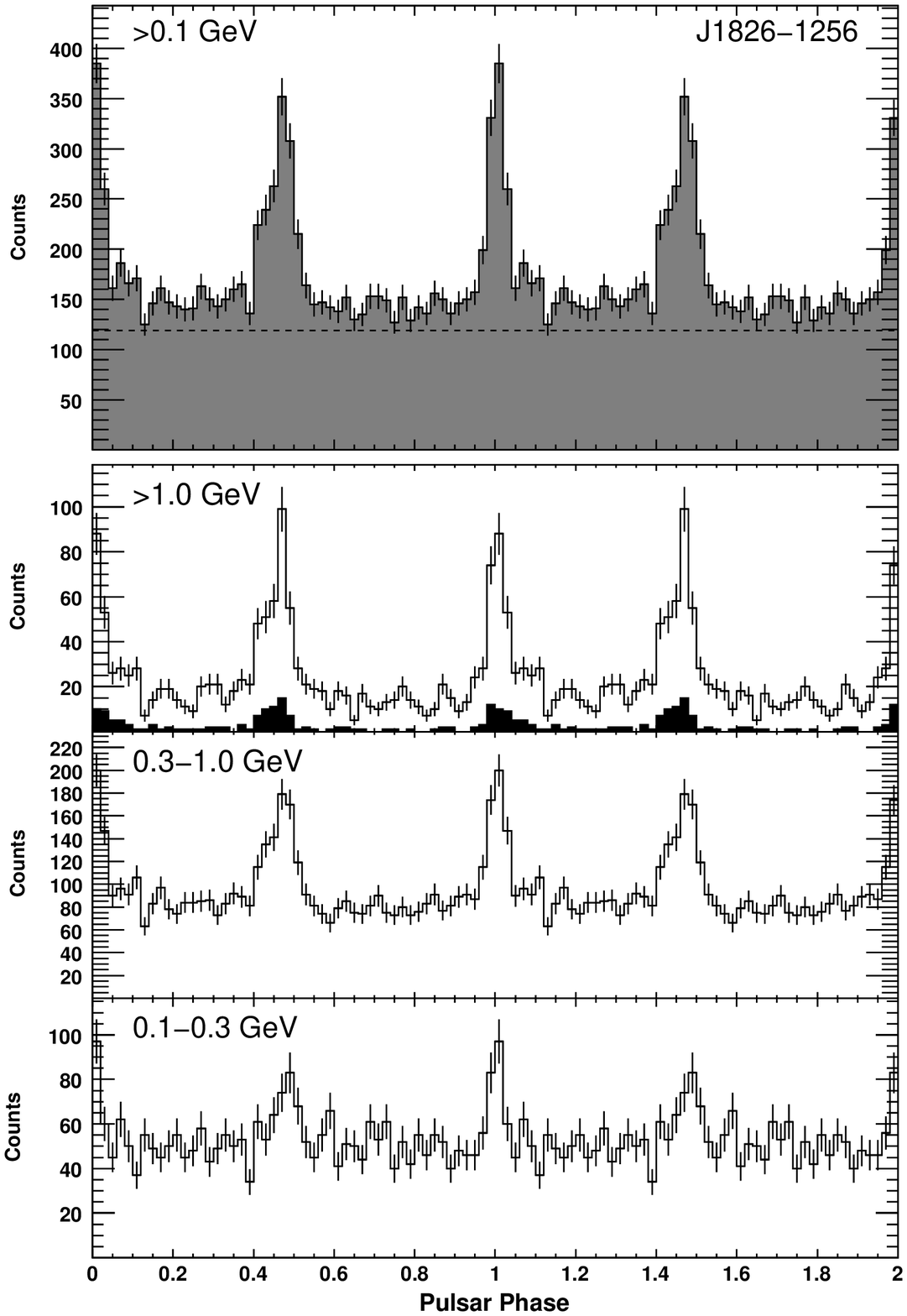}
\caption{Light curves for PSR J1826$-$1256 ($P=110$\,ms).
\label{fig:J1826m1256_lightcurve}}
\end{minipage}\\
\end{sidewaysfigure}
\clearpage

%----start of new page----
\begin{sidewaysfigure}
\centering
%%----start of figure----
\begin{minipage}[t]{0.45\linewidth}
\centering
\includegraphics[width=\linewidth]{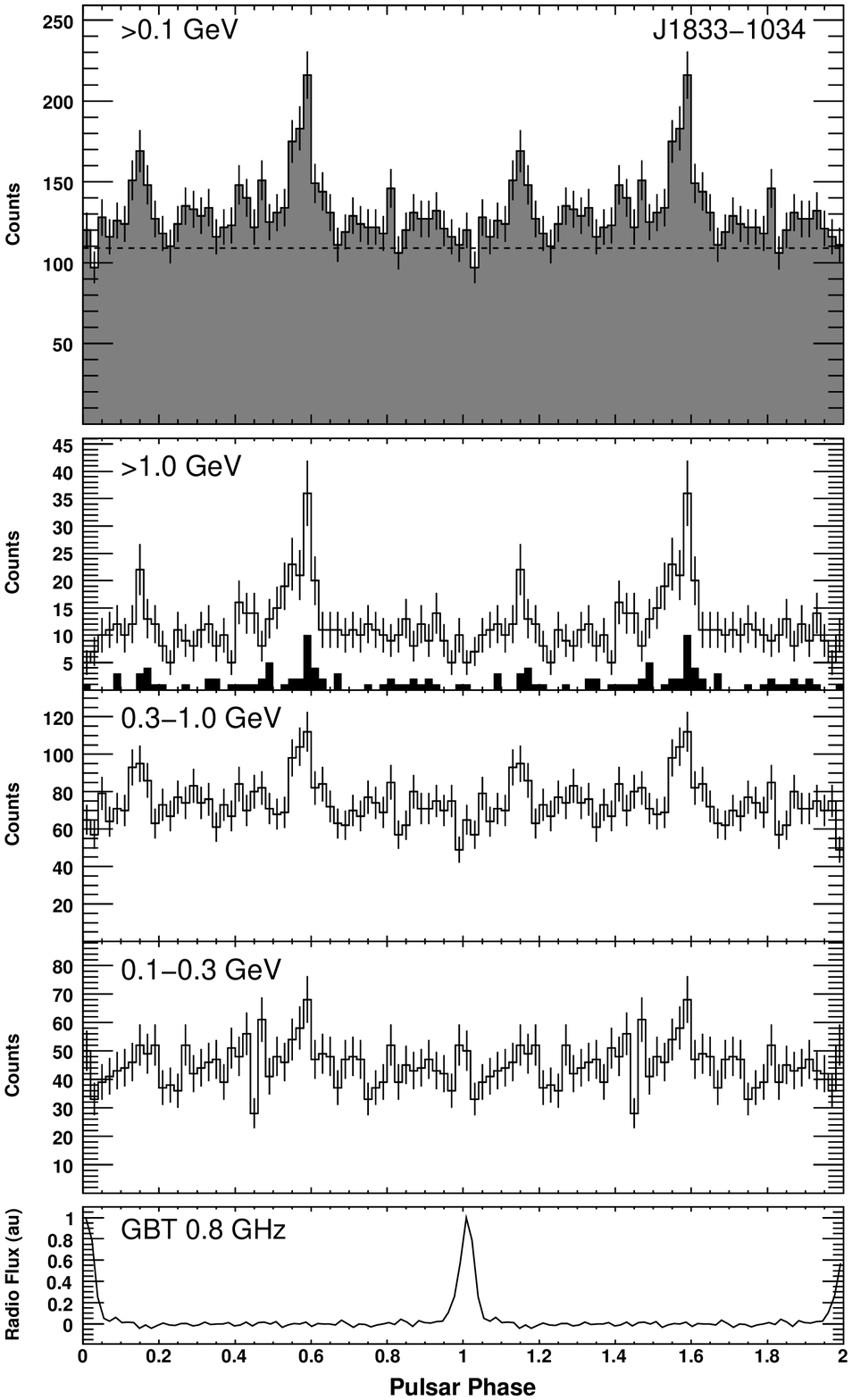}
\caption{Light curves for PSR J1833$-$1034 ($P=61.9$\,ms).
\label{fig:J1833m1034_lightcurve}}
\end{minipage}%
\hspace{1cm}%
%%----start of figure----
\begin{minipage}[t]{0.45\linewidth}
\centering
\includegraphics[width=\linewidth]{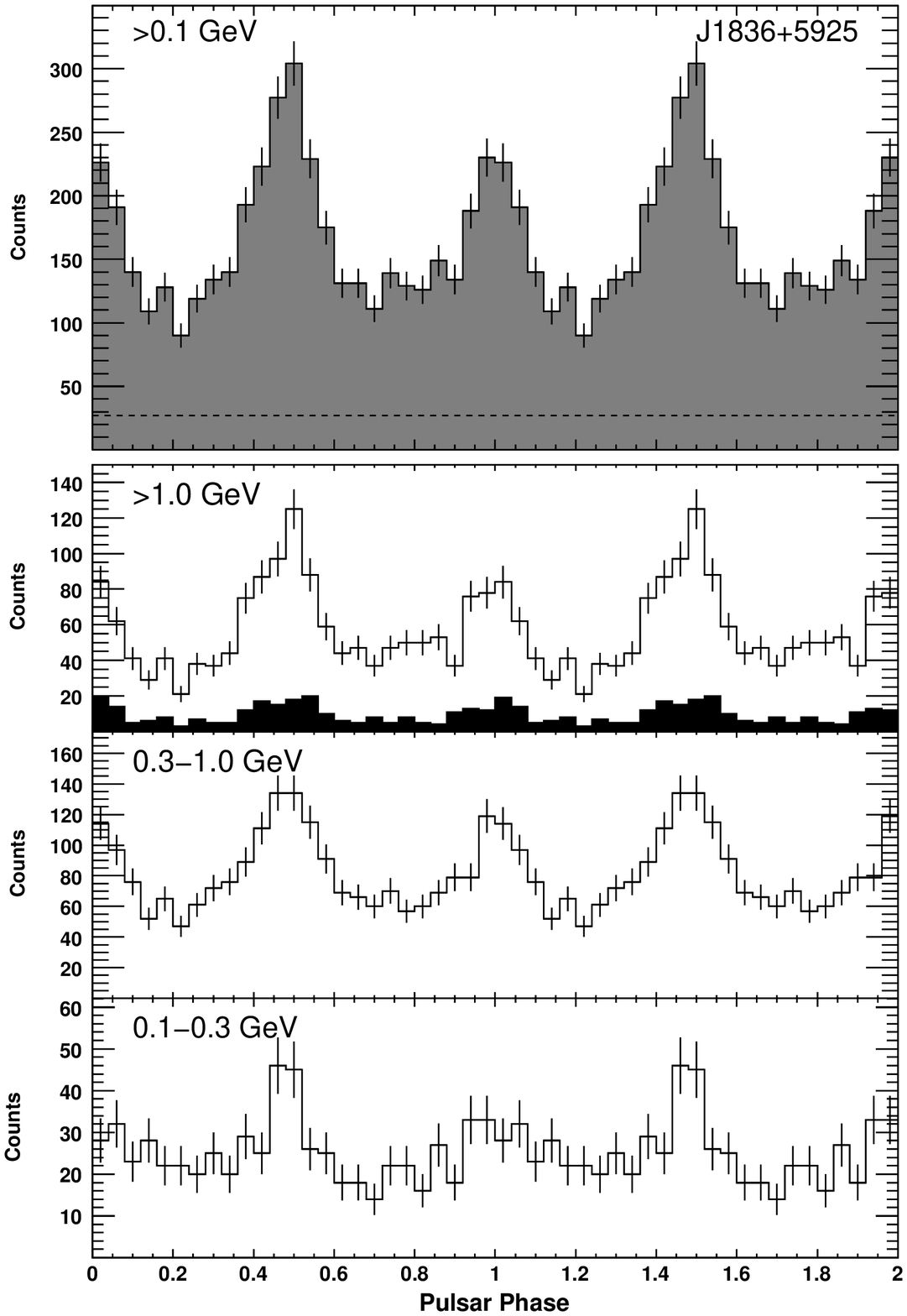}
\caption{Light curves for PSR J1836+5925 ($P=173$\,ms).
\label{fig:J1836p5925_lightcurve}}
\end{minipage}\\
\end{sidewaysfigure}
\clearpage

%----start of new page----
\begin{sidewaysfigure}
\centering
%%----start of figure----
\begin{minipage}[t]{0.45\linewidth}
\centering
\includegraphics[width=\linewidth]{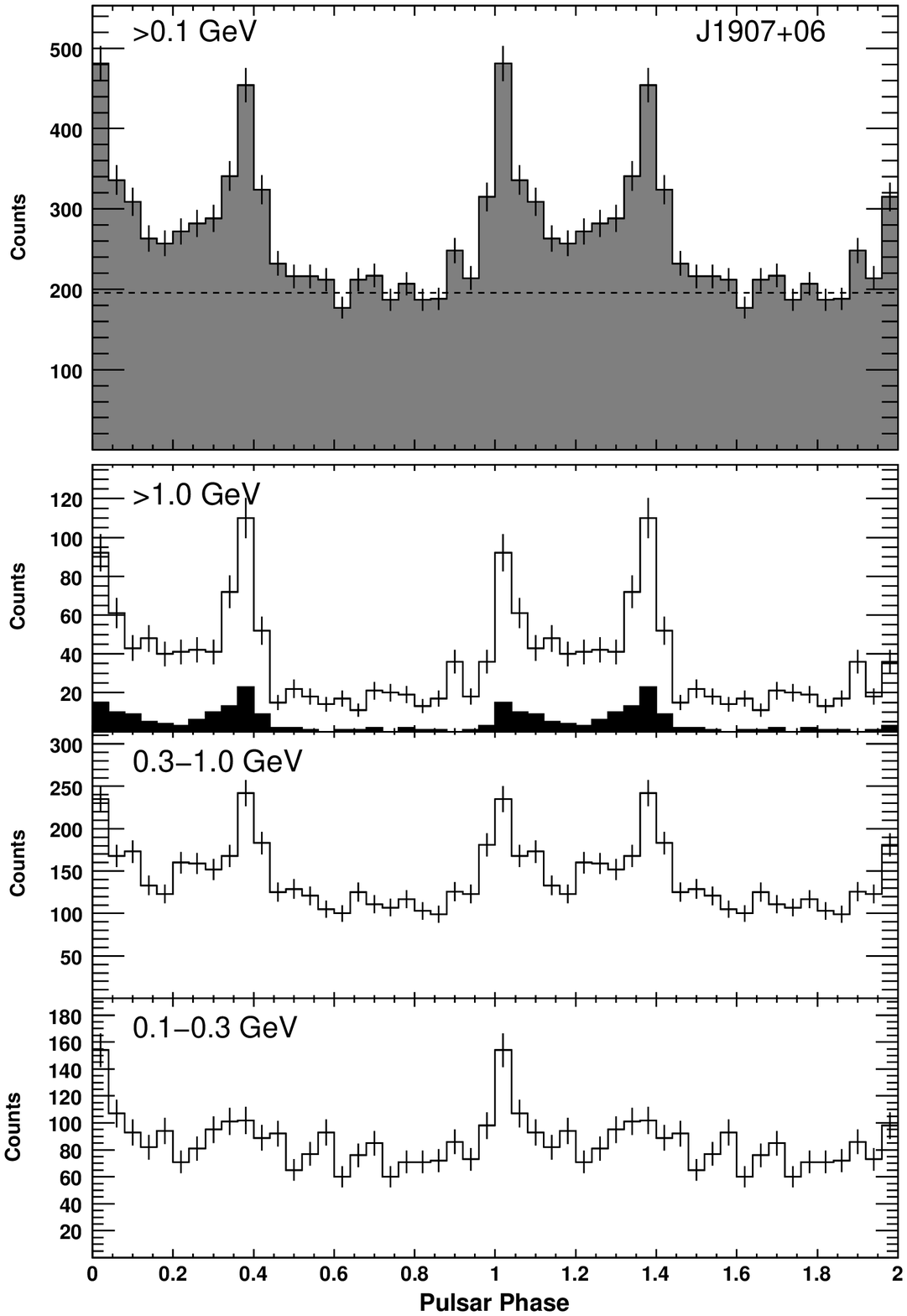}
\caption{Light curves for PSR J1907+06 ($P=107$\,ms).
\label{fig:J1907p06_lightcurve}}
\end{minipage}%
\hspace{1cm}%
%%----start of figure----
\begin{minipage}[t]{0.45\linewidth}
\centering
\includegraphics[width=\linewidth]{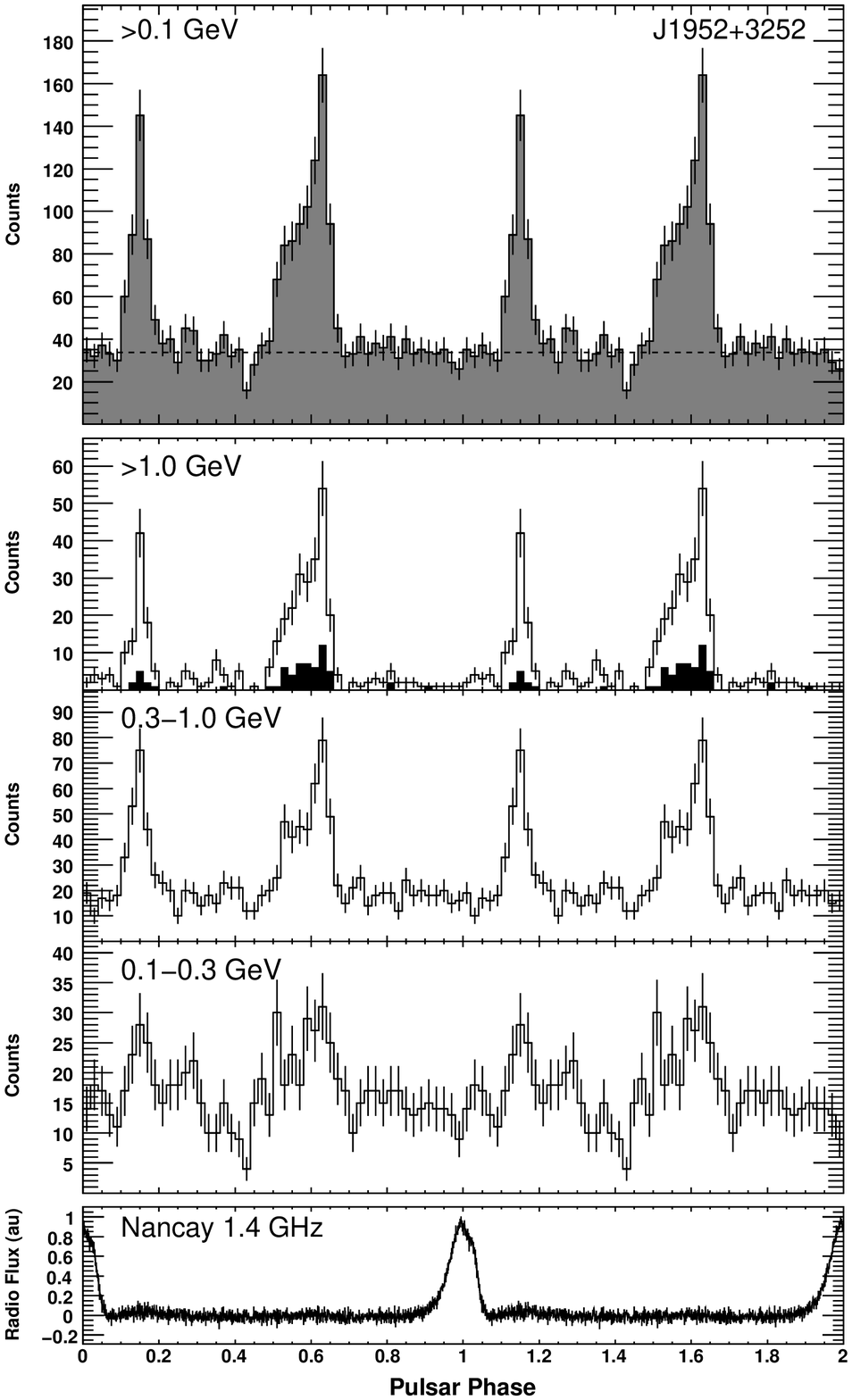}
\caption{Light curves for PSR J1952+3252 ($P=39.5$\,ms, PSR B1951+32).
\label{fig:J1952p3252_lightcurve}}
\end{minipage}\\
\end{sidewaysfigure}
\clearpage

%----start of new page----
\begin{sidewaysfigure}
\centering
%%----start of figure----
\begin{minipage}[t]{0.45\linewidth}
\centering
\includegraphics[width=\linewidth]{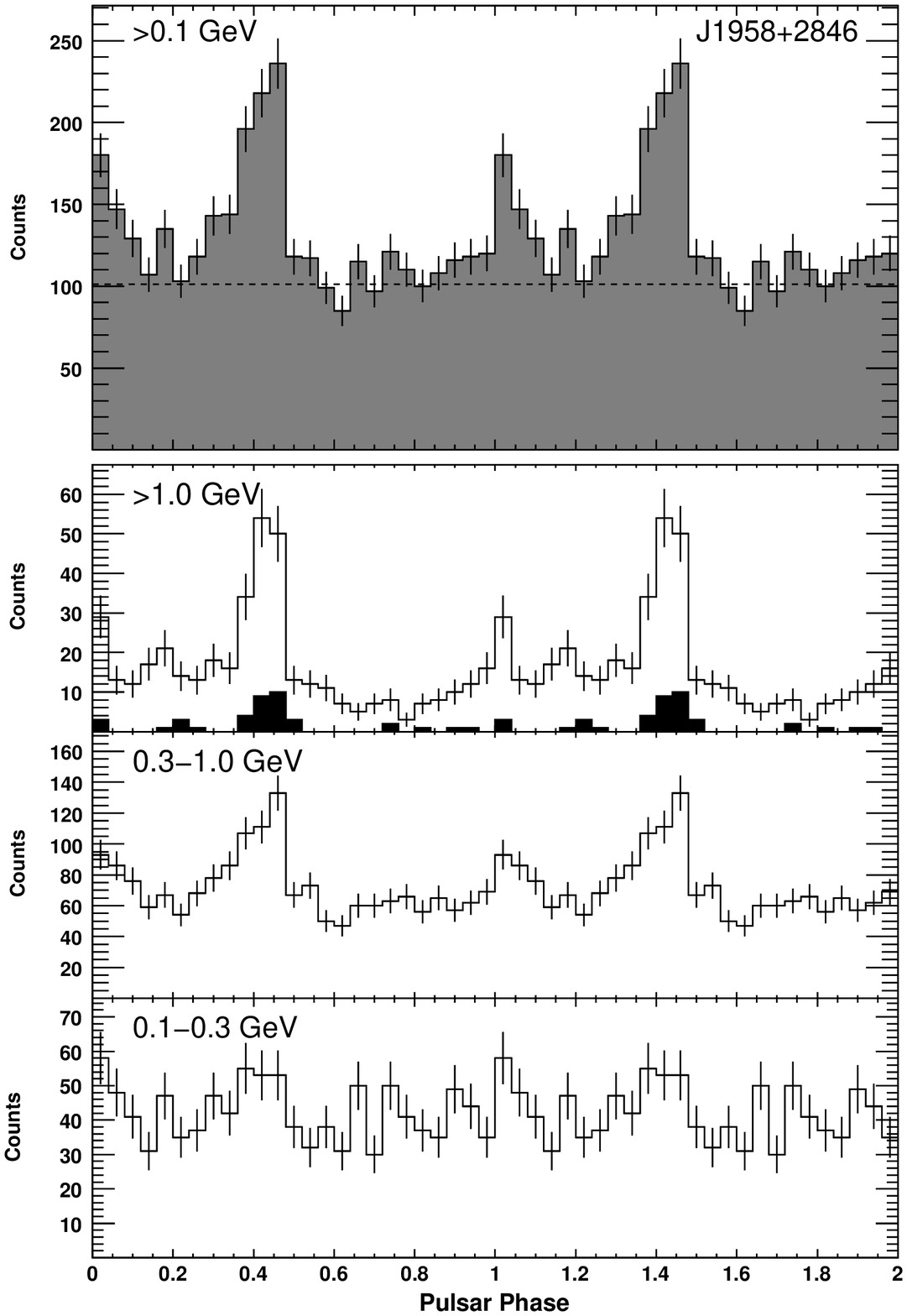}
\caption{Light curves for PSR J1958+2846 ($P=290$\,ms).
\label{fig:J1958p2846_lightcurve}}
\end{minipage}%
\hspace{1cm}%
%%----start of figure----
\begin{minipage}[t]{0.45\linewidth}
\centering
\includegraphics[width=\linewidth]{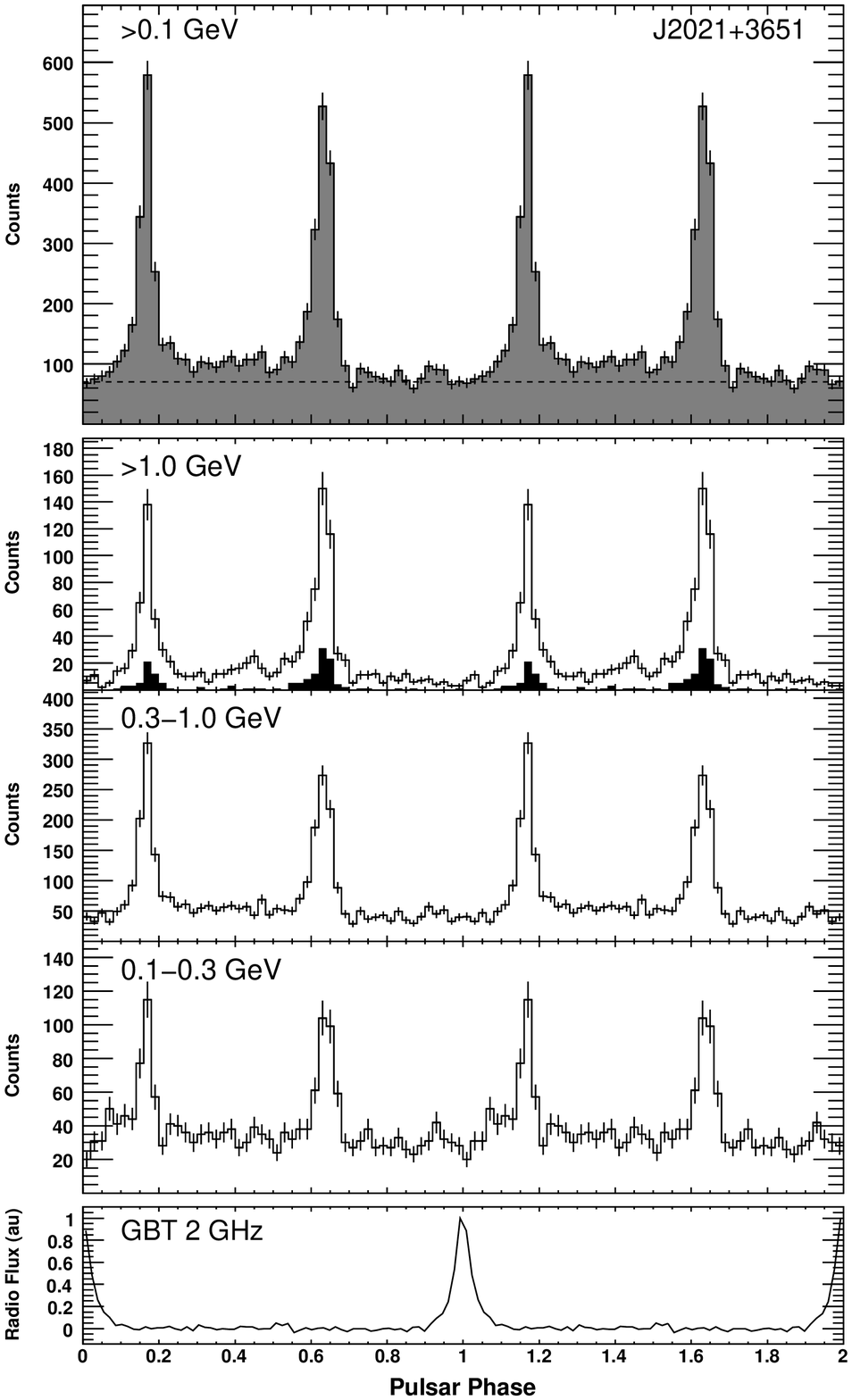}
\caption{Light curves for PSR J2021+3651 ($P=104$\,ms).
\label{fig:J2021p3651_lightcurve}}
\end{minipage}\\
\end{sidewaysfigure}
\clearpage

%----start of new page----
\begin{sidewaysfigure}
\centering
%%----start of figure----
\begin{minipage}[t]{0.45\linewidth}
\centering
\includegraphics[width=\linewidth]{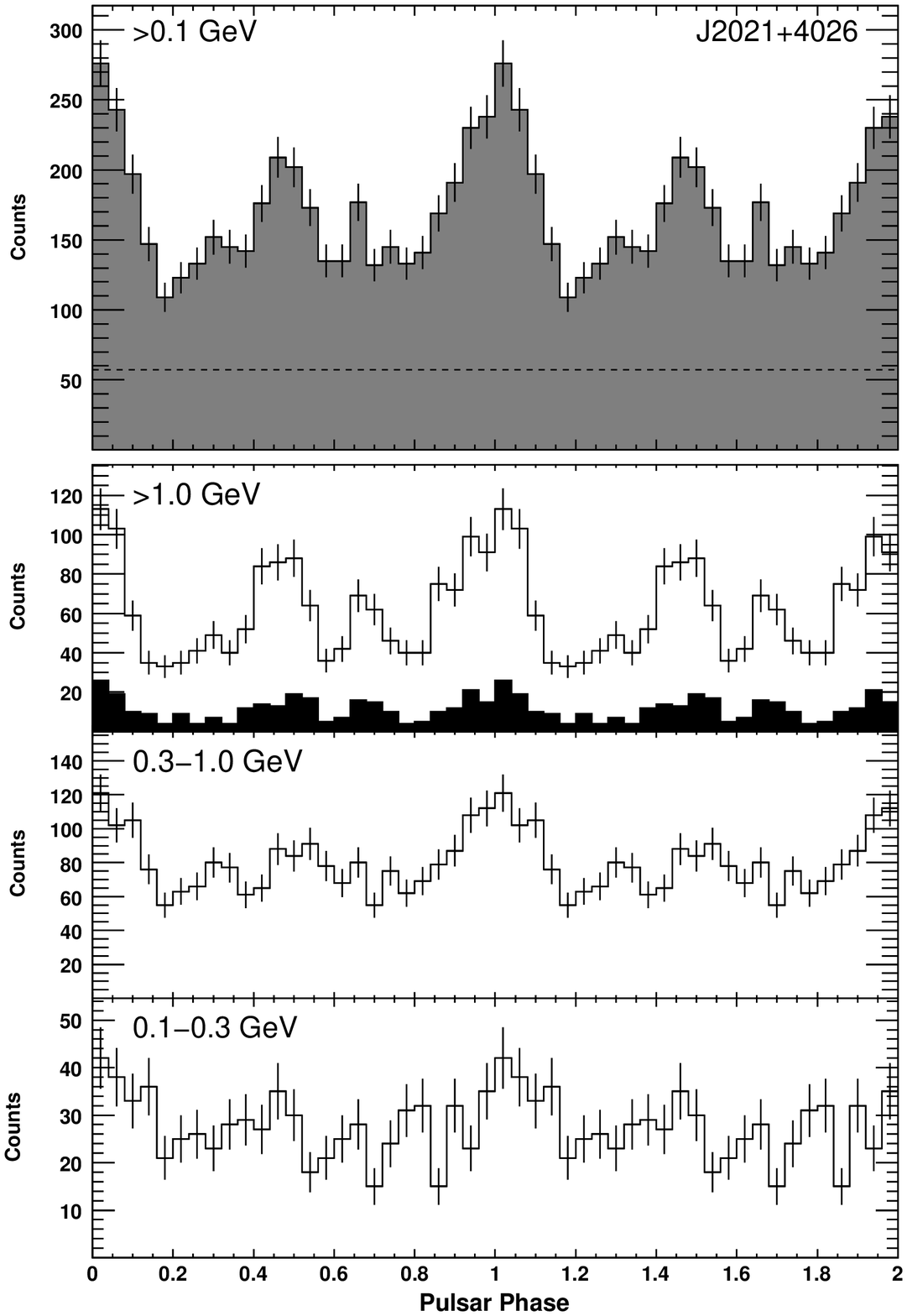}
\caption{Light curves for PSR J2021+4026 ($P=265$\,ms).
\label{fig:J2021p4026_lightcurve}}
\end{minipage}%
\hspace{1cm}%
%%----start of figure----
\begin{minipage}[t]{0.45\linewidth}
\centering
\includegraphics[width=\linewidth]{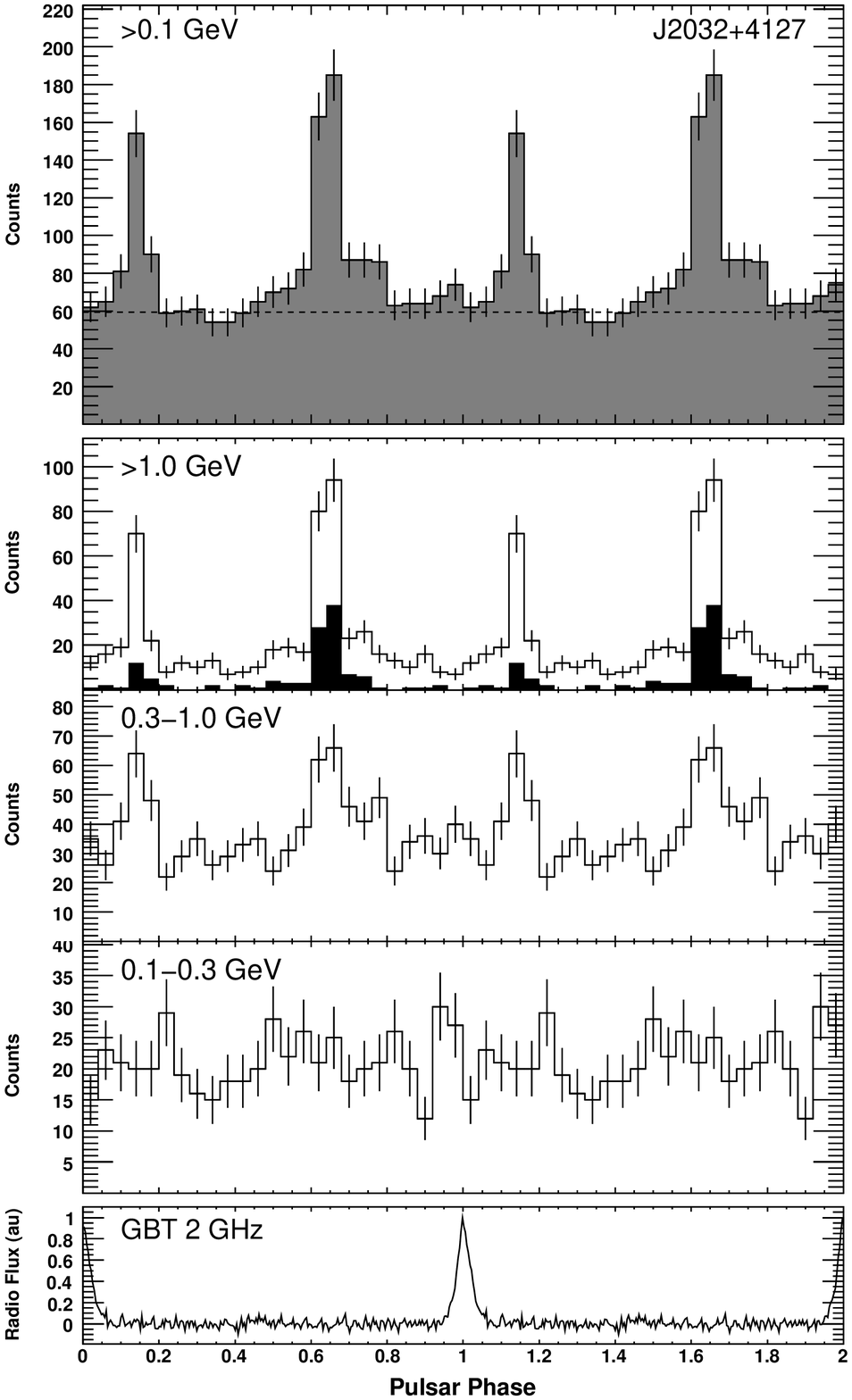}
\caption{Light curves for PSR J2032+4127 ($P=143$\,ms). While this pulsar is detected in the radio \citep{Camilo2009}, it was discovered by the LAT and is considered a gamma-ray-selected pulsar.
\label{fig:J2032p4127_lightcurve}}
\end{minipage}\\
\end{sidewaysfigure}
\clearpage

%----start of new page----
\begin{sidewaysfigure}
\centering
%%----start of figure----
\begin{minipage}[t]{0.45\linewidth}
\centering
\includegraphics[width=\linewidth]{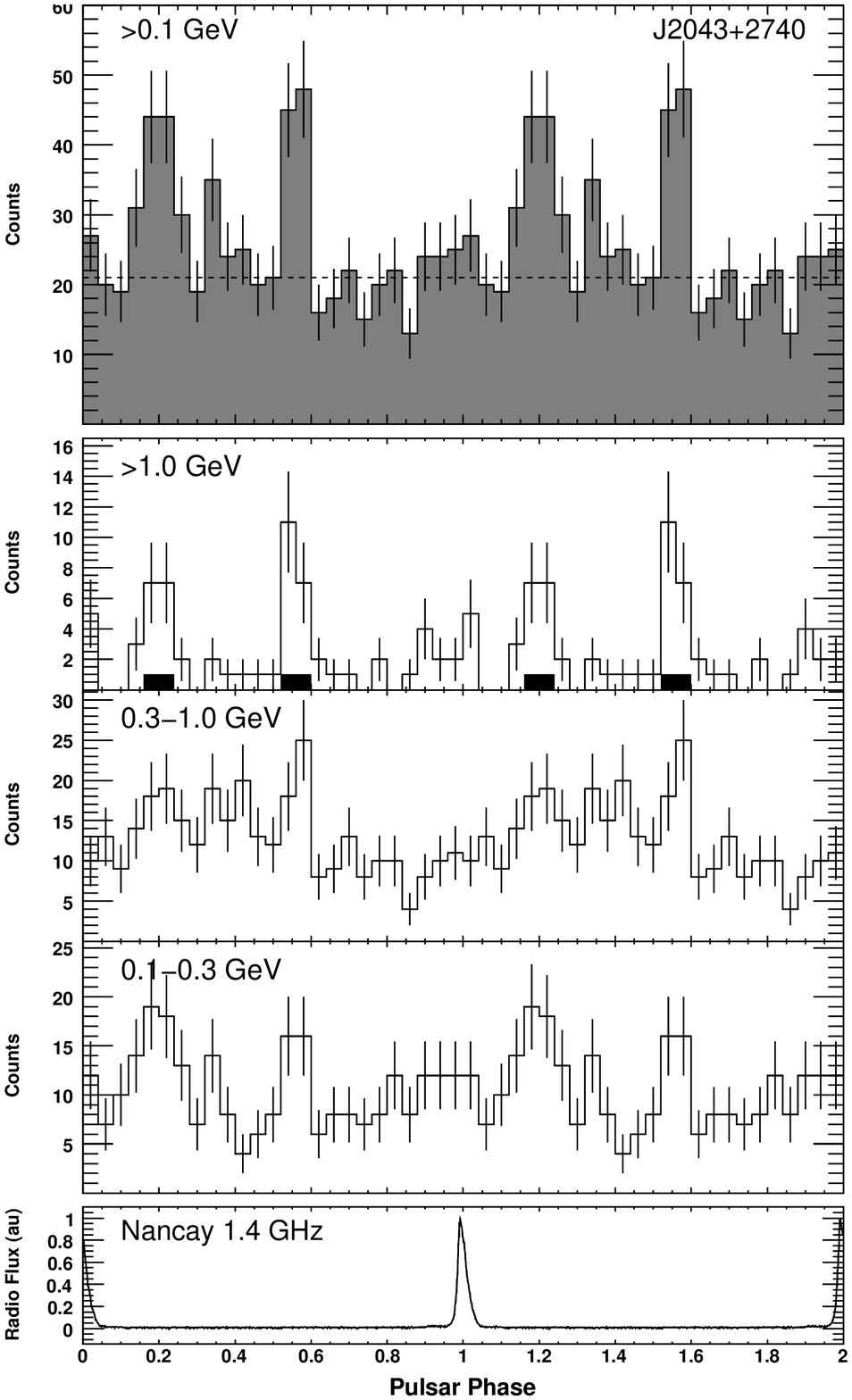}
\caption{Light curves for PSR J2043+2740 ($P=96.1$\,ms).
\label{fig:J2043p2740_lightcurve}}
\end{minipage}%
\hspace{1cm}%
%%----start of figure----
\begin{minipage}[t]{0.45\linewidth}
\centering
\includegraphics[width=\linewidth]{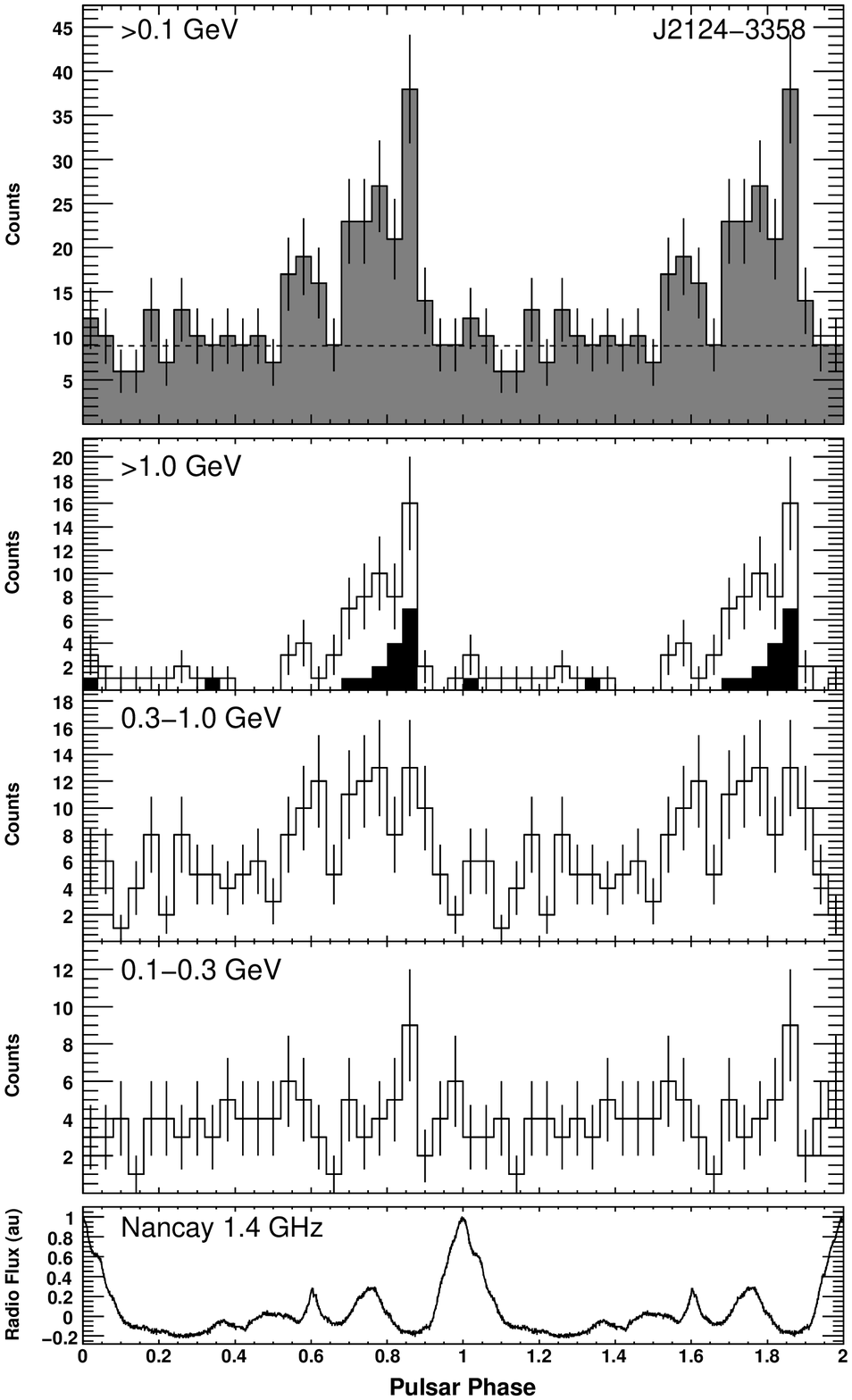}
\caption{Light curves for PSR J2124$-$3358 ($P=4.93$\,ms).
\label{fig:J2124m3358_lightcurve}}
\end{minipage}\\
\end{sidewaysfigure}
\clearpage

%----start of new page----
\begin{sidewaysfigure}
\centering
%%----start of figure----
\begin{minipage}[t]{0.45\linewidth}
\centering
\includegraphics[width=\linewidth]{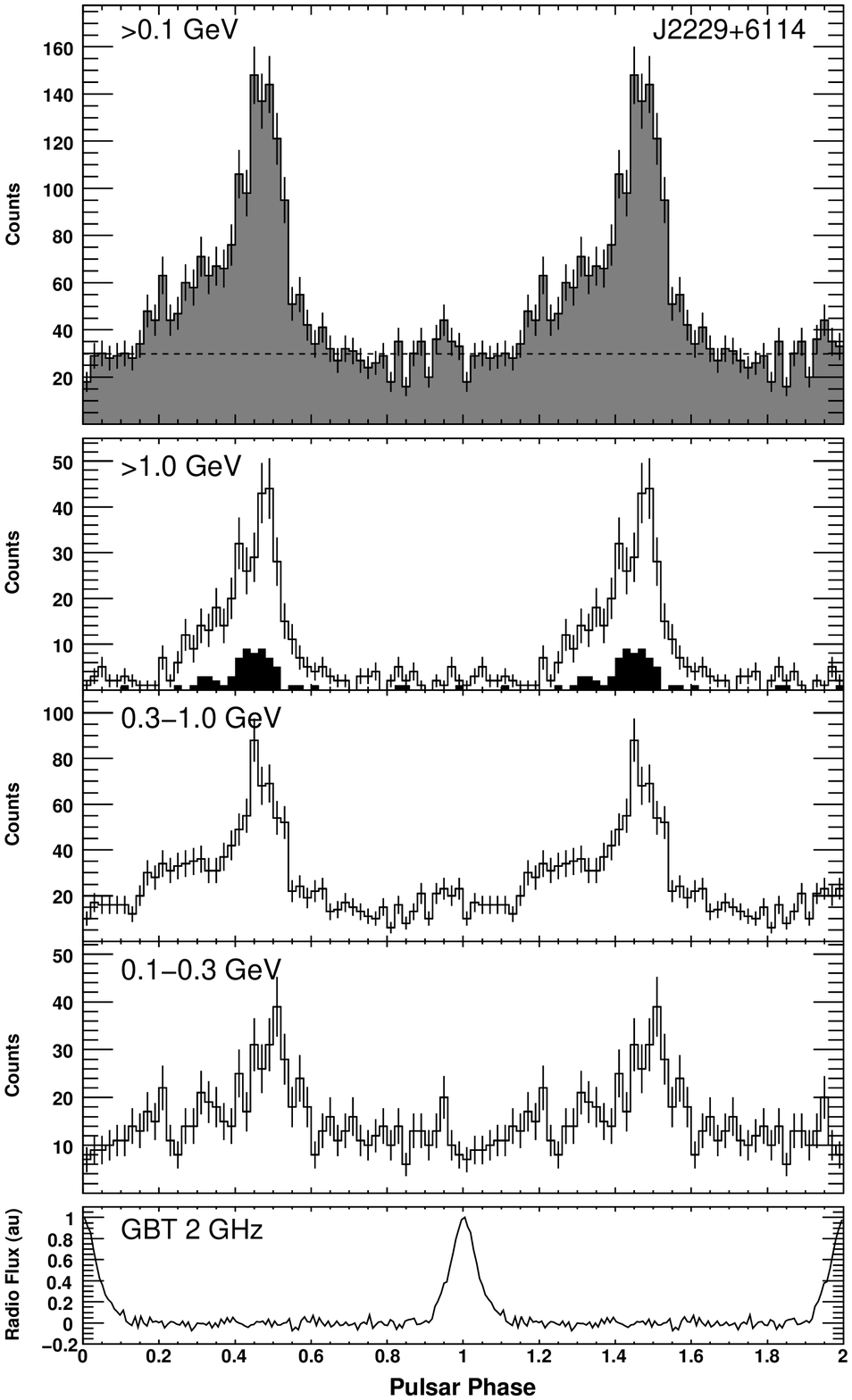}
\caption{Light curves for PSR J2229+6114 ($P=51.6$\,ms).
\label{fig:J2229p6114_lightcurve}}
\end{minipage}%
\hspace{1cm}%
%%----start of figure----
\begin{minipage}[t]{0.45\linewidth}
\centering
\includegraphics[width=\linewidth]{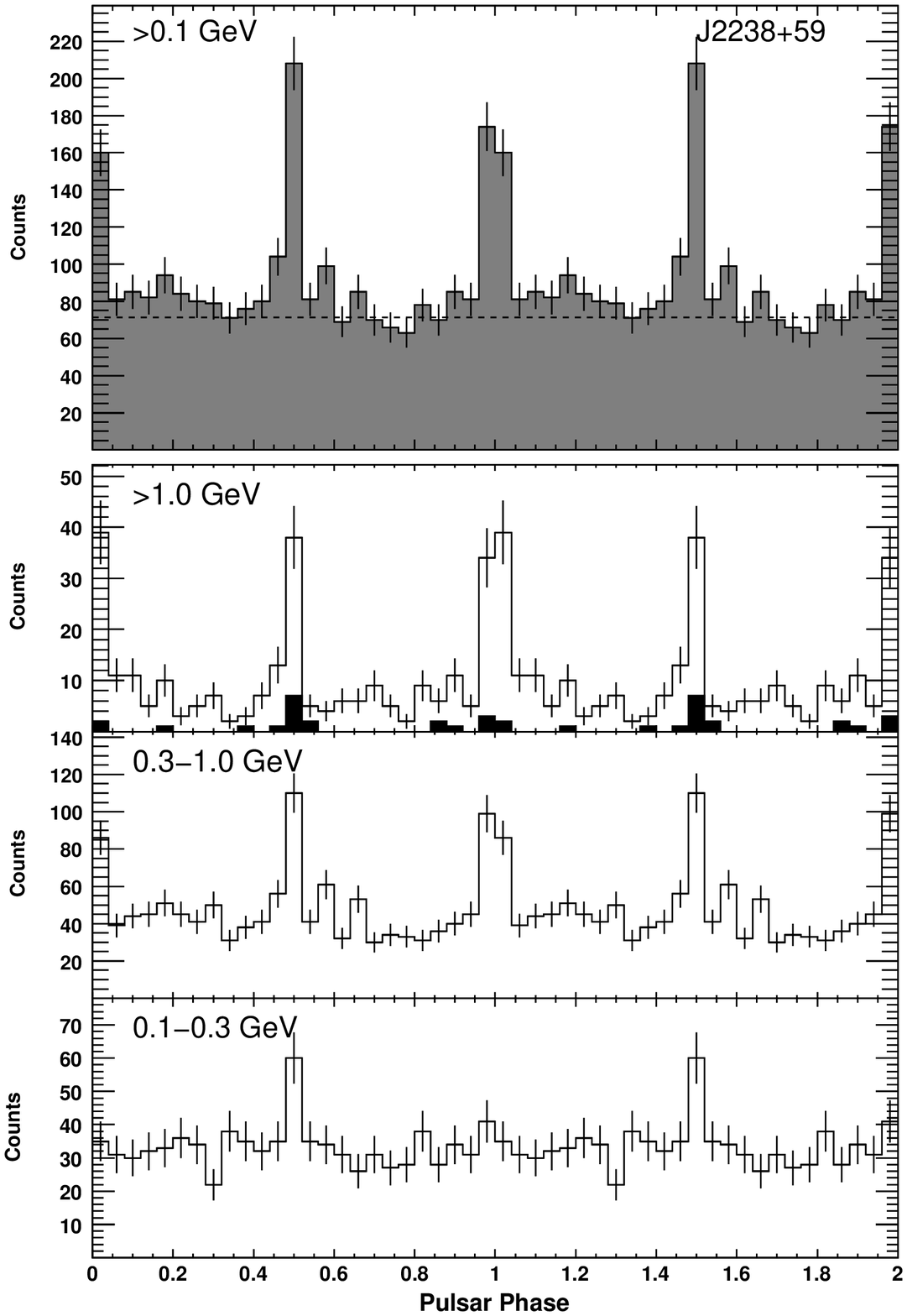}
\caption{Light curves for PSR J2238+59 ($P=163$\,ms).
\label{fig:J2238p59_lightcurve}}
\end{minipage}\\
\end{sidewaysfigure}
\clearpage

%----start of new page----
%\begin{sidewaysfigure}
%\centering
%%----start of figure----
%\begin{minipage}[t]{0.45\linewidth}
%\centering
%\includegraphics[width=\linewidth]{LightCurves/PULSARNAME_catalog_lightcurve.eps}
%\caption{Light curves for PSR J.
%\label{fig:PULSARNAME_lightcurve}}
%\end{minipage}%
%\hspace{1cm}%
%%----start of figure----
%\begin{minipage}[t]{0.45\linewidth}
%\centering
%\includegraphics[width=\linewidth]{LightCurves/PULSARNAME_catalog_lightcurve.eps}
%\caption{Light curves for PSR J
%\label{fig:PULSARNAME_lightcurve}}
%\end{minipage}\\
%\end{sidewaysfigure}
%\clearpage

\end{document}